\documentclass{aa}  

\usepackage{graphicx}
\usepackage{float}
\usepackage{txfonts}
\usepackage{placeins}
\usepackage{enumitem}
\newcommand{\logNH}[1]{$\log N_{\mathrm H}{\mathrm{cm}}^{-2}$}
\newcommand{\NHvalue}[1]{$N_\mathrm{H}={10}^{#1}\mathrm{cm}^{-2}$}
\newcommand{\FeKa}{Fe K$\alpha$ }

\begin{document} 
   \title{XZ: Deriving redshifts from X-ray spectra of obscured AGN}

   \author{C. Simmonds\inst{1} \and  J. Buchner\inst{1,2,3} 
          \and  M. Salvato\inst{4}, L.-T. Hsu\inst{5} 
          \and F.~E. Bauer\inst{1,2}
   }
          
   \institute{Pontificia Universidad Católica de Chile, Instituto de Astrofísica, Casilla 306, Santiago 22, Chile
         \and 
         Millenium Institute of Astrophysics, Vicu\~{n}a. MacKenna 4860, 7820436 Macul, Santiago, Chile
         \and
         Excellence Cluster Universe, Boltzmannstr. 2, D-85748, Garching, Germany
         \and
         Max Planck Institut für Extraterrestrische Physik Giessenbachstrasse, 85748 Garching, Germany
         \and
         Institute of Astronomy \& Astrophysics, Academia Sinica, Taipei, Taiwan
             }


 
  \abstract
   {Redshifts are fundamental for our understanding of extragalactic X-ray sources. Ambiguous counterpart associations, expensive optical spectroscopy and/or multimission multiwavelength coverage to resolve degeneracies make estimation often difficult in practice.}
   {We attempt to constrain redshifts of obscured Active Galactic Nuclei (AGN) using only low-resolution X-ray spectra.}
   {Our XZ method fits AGN X-ray spectra with a moderately complex spectral model incorporating a corona, torus obscurer and warm mirror. Using the Bayesian X-ray Astronomy (BXA) package, we constrain redshift, column density, photon index and luminosity simultaneously. The redshift information primarily comes from absorption edges in Compton-thin AGN, and from the Fe K$\alpha$ fluorescent line in heavily obscured AGN. A new generic background fitting method allows us to extract more information from limited numbers of source counts.}
   {We derive redshift constraints for 74/321 hard-band detected sources in the {\it Chandra} deep field South. Comparing with spectroscopic redshifts, we find an outlier fraction of 8\%, indicating that our model assumptions are valid. For three {\it Chandra} deep fields, we release our XZ redshift estimates. 
   }  
   {
The independent XZ estimate is easy to apply and effective for a large fraction of obscured AGN in todays deep surveys without the need for any additional data.
Comparing to different redshift estimation methods, XZ can resolve degeneracies in photometric redshifts, help to detect potential association problems and confirm uncertain single-line spectroscopic redshifts. With high spectral resolution and large collecting area, this technique will be highly effective for {\it Athena}/WFI  observations.
}

   \maketitle
%

\section{Introduction}
Measuring redshifts for distant Active Galactic Nuclei (AGN) is notoriously difficult, yet crucial for constraining the accretion history of the Universe. Spectroscopic redshifts (hereafter specz's) are desirable but costly, and remain impossible for targets which are faint in the optical and near-infrared bands. Thus it is common practice to rely on photometric redshifts (hereafter photoz's), which when given a sufficient number of photometric bands can effectively provide very low resolution spectra \citep[e.g.,][]{Baum1962,Koo1985,Pello1996}.

For galaxies, photoz values can be extremely accurate when the Spectral Energy Distributions (SEDs) of various types have distinct sharp features (e.g., Lyman and Balmer breaks, or high equivalent width, EW, emission lines). This is complicated for AGN because (1) the AGN component dilutes or hides the host galaxy features and has a number of high EW lines, leading to multiple, degenerate redshift solutions \citep{Salvato2009} and (2) the intrinsic variability of these sources makes the photometry gathered over years very uncertain \citep{Simm2015}. 

Additionally, AGN selected in one waveband (here: X-rays) have thus far had to be associated with counterparts to obtain photometry or spectroscopy. However, finding the correct counterpart when the X-ray position uncertainties are large can be ambiguous, especially in deep X-ray observations \citep[see e.g.,][]{HsuCDFSz2013,Salvato2017}. Misassociations can affect both photoz and specz and are difficult to detect.

In this work we demonstrate that for obscured AGN there is sufficient information in the X-ray spectrum to compute a reliable redshift. Although typical redshift uncertainties are usually substantial ($\sigma(z)\sim 0.2$), they remain sufficiently accurate for verifying the association and checking for photoz outliers. The presented technique can be applied already with the low-resolution X-ray imaging spectrometers in current AGN surveys.

We describe in $\S$2 the redshift information that is present in low-resolution, relatively low-count X-ray spectra. In $\S$3 we explain our method in detail and quantifies the redshift information gain. In $\S$4 we test the reliability of our method with real survey data and provide redshift catalogues (additional fields in Appendix~\ref{moretables}). Synergies with ambiguous specz and photoz information are discussed in $\S$5.
Simulations are presented in $\S$6 highlighting the number of counts needed for current instruments ({\it Chandra, XMM-Newton, NuSTAR, Swift}) and future surveys with {\it eROSITA} and {\it Athena}.

\section{Data}
To demonstrate and validate our method, a meaningful test sample of obscured AGN with X-ray spectra and previously obtained specz and photoz information is needed. We primarily use data from 
the Chandra Deep Field-South \citep[CDF-S;][]{Xue2011,Rangel2013}, which has extensive spectroscopic and multi-wavelength photometric coverage and thanks to its deep exposure features a large fraction of obscured sources.

The CDF-S survey (464.5 arcmin$^2$) was reduced following the procedure of \cite{Laird2009}. Here we work with the 326 hard (2-7 keV) detected X-ray sources in the 4Ms data from \cite{Rangel2013}, whose X-ray identification numbers (IDs) we use throughout. Three Ms of additional X-ray data has been obtained recently \citep{Luo2017}. However, 4Ms spectra and associated data products are available from our previous efforts and these are sufficient for our purposes here, where we seek to generically demonstrate the utility of the method.
Multi-wavelength counterparts were found by \cite{HsuCDFSz2013} who also computed photoz's and collated specz's into a large catalogue. Because our goal is to apply a X-ray spectral model suitable for AGN, we discard sources whose photometry is best fitted with a stellar template in \cite{HsuCDFSz2013}. X-ray spectra and related data products\footnote{auxiliary response files (\texttt{arf}), redistribution matrix files (\texttt{rmf}) and background spectra (\texttt{bkg})}
were generated with ACIS EXTRACT \citep{BroosACISEXTRACT2010} by \cite{Brightman2013}. Our final X-ray sample, previously analysed also in \cite{Buchner2015}, consists of 321 AGN, 264 of them obscured (N$_{\rm H} \geq$10$^{22}$cm$^{-2}$). We have optical redshift measurements for the entire sample (100\% photoz; 56\% specz). For photoz, we use a state-of-the-art photoz technique optimized for AGN \citep{HsuCDFSz2013} and throughout use redshift probability distributions.

To test our technique also in wider fields, we later also report results from the All Wavelength Extended Groth strip International Survey \citep[AEGIS-XD,][]{Nandra2015} and the Cosmological evolution Survey \citep[C-COSMOS;][]{Elvis2009}. The X-ray spectra products were derived similarly as above by \cite{Brightman2013}, with multiwavelength associations and redshift determinations by \cite[][AEGIS-XD]{Nandra2015} and \cite[][C-COSMOS]{Salvato2009}. See \citep{Buchner2015} for more details, where those samples were previously analysed.

\subsection{Redshift information in the X-ray spectrum}

\begin{figure}
  \centering
   \includegraphics[scale=0.4, origin=l]{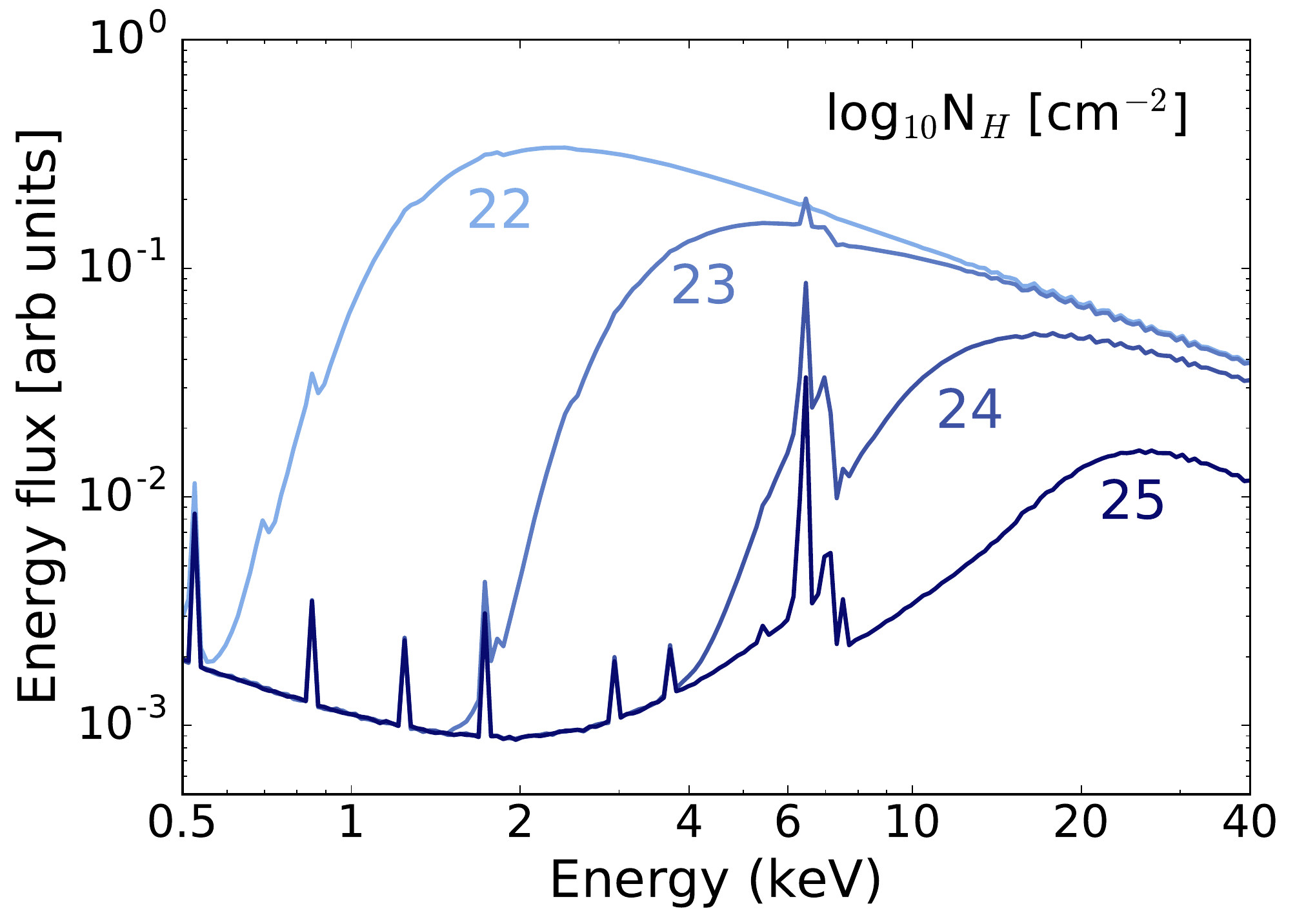} 
   \caption{AGN X-ray spectra from the \texttt{TORUS} model of \cite{Brightman2011a}. The intrinsic powerlaw has a photon index of $\Gamma$ = 1.9. Labeled numbers indicate the obscurer column density \NHvalue{22-25}. At high column densites, the \FeKa line at 6.4 keV becomes prominent. A unobscured powerlaw component normalised to 1\%  of the intrinsic powerlaw has been added. 
   }
              \label{fig:nH}
\end{figure}

\begin{figure*}[ht]
\centering
\includegraphics[width=.43\textwidth]{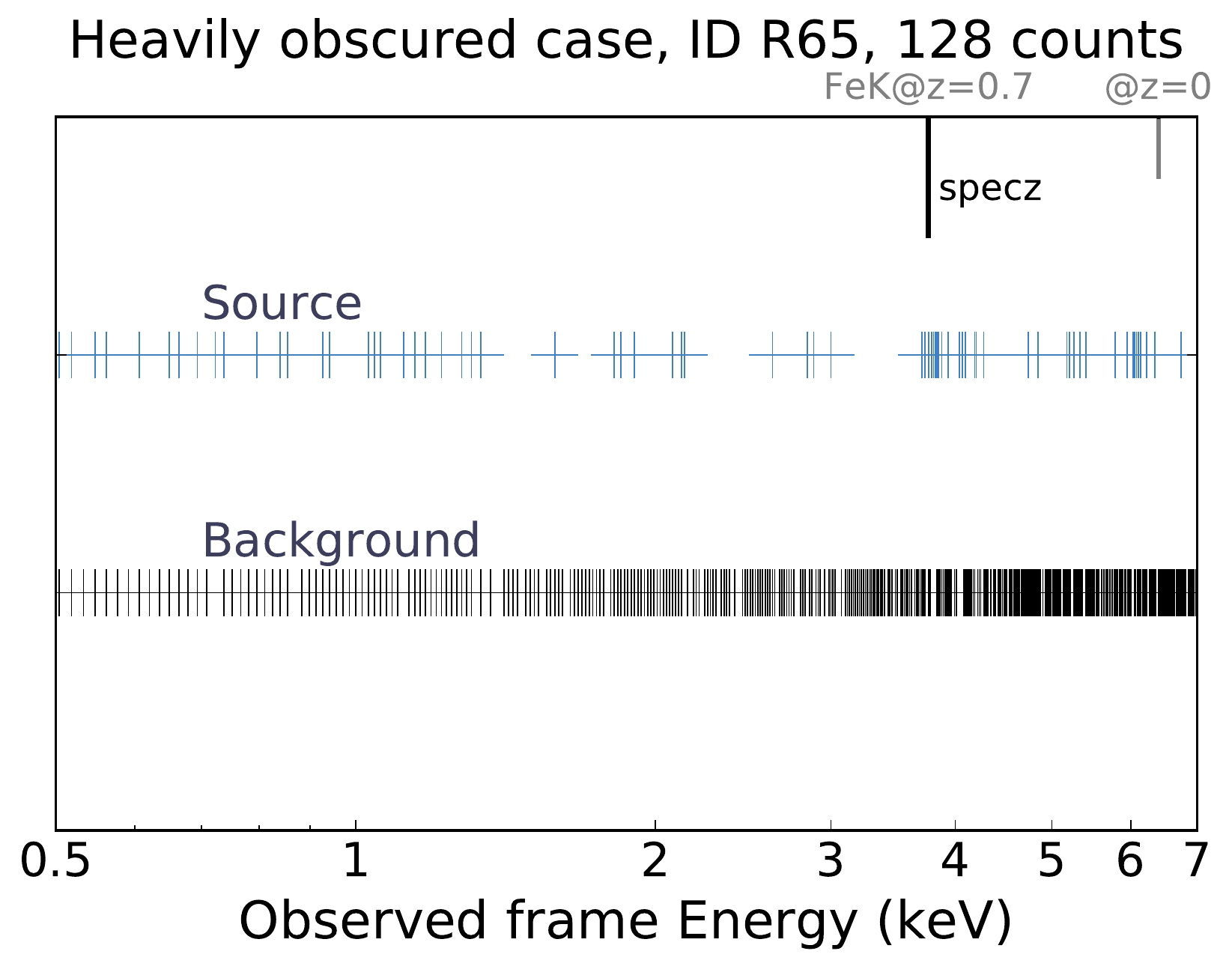}\quad
\includegraphics[width=.43\textwidth]{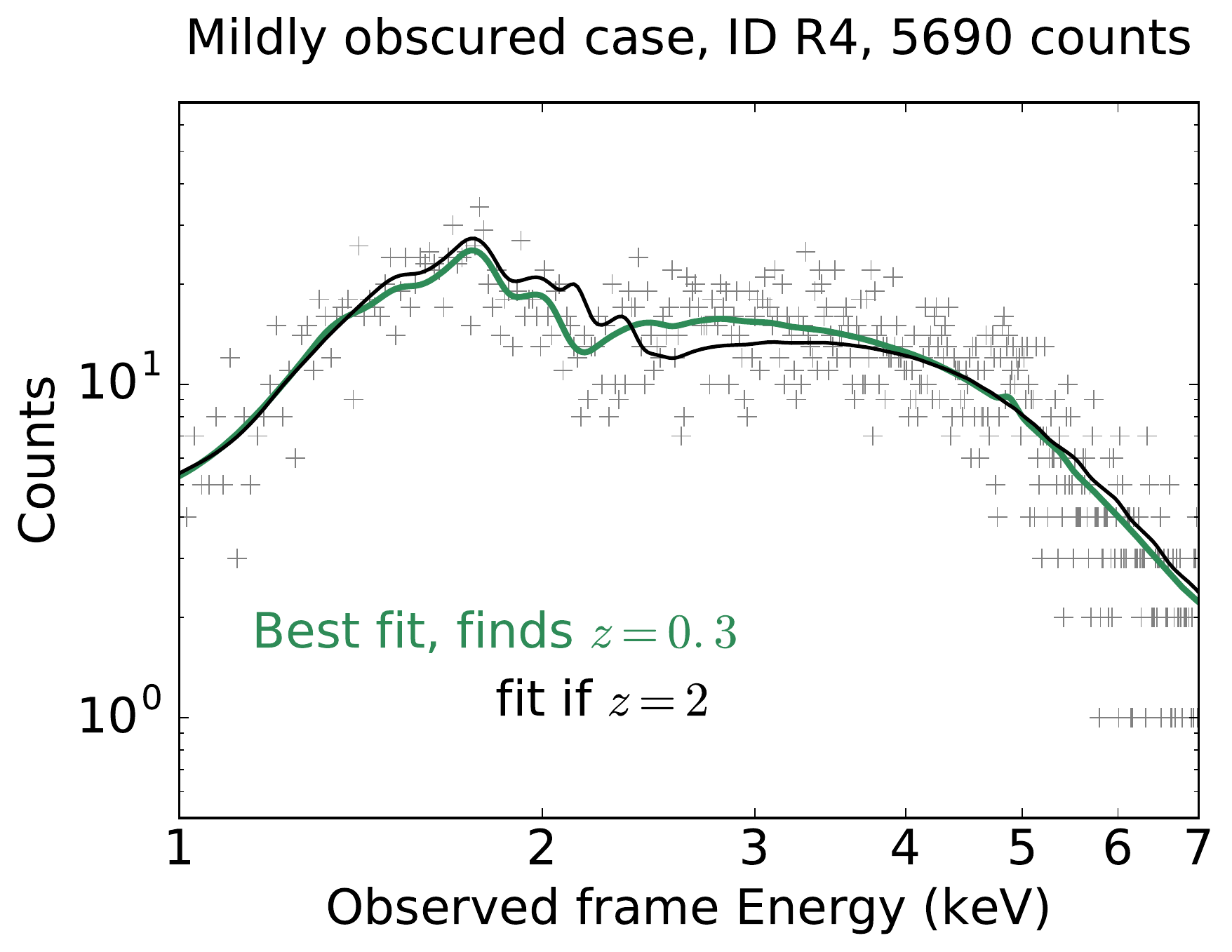}\quad
\caption{Two extreme cases in which XZ provides the correct redshift. \textsl{Left panel:} unbinned X-ray source and background region counts for R65, a heavily obscured AGN ($\log N_\mathrm{H}\simeq 24.4 \mathrm{cm}^{-2}$) with a relatively low number of source counts. The counts are plotted in a horizontal line to show their energy in the observed frame. 
While the background count rate is nearly independent of energy, source counts clearly cluster at 4keV. Assuming that feature corresponds to the rest-frame 6.4keV \FeKa feature, we can compute the corresponding redshift $z = 6.4 keV/E_{obs} - 1$ to the energy (top axis). We find $z = 0.69$, which is consistent with the optically derived spectroscopic redshift. \textsl{Right panel:} binned X-ray spectrum of source R4, a mildly obscured AGN with a high number of counts. The green curve is the best fit, yielding $\log N_\mathrm{H} \simeq 22.6 \mathrm{cm}^{-2}$ and $\Gamma \sim 1.9$ and finds the spectroscopic redshift $z=0.31$ (see Figure~\ref{fig:examples2}). The black curve shows the fit if the redshift is fixed to $z = 2$, which does not follow the data points as well.}
\label{fig:examples}
\end{figure*}

\begin{figure}[ht]
\centering
\includegraphics[width=.5\textwidth]{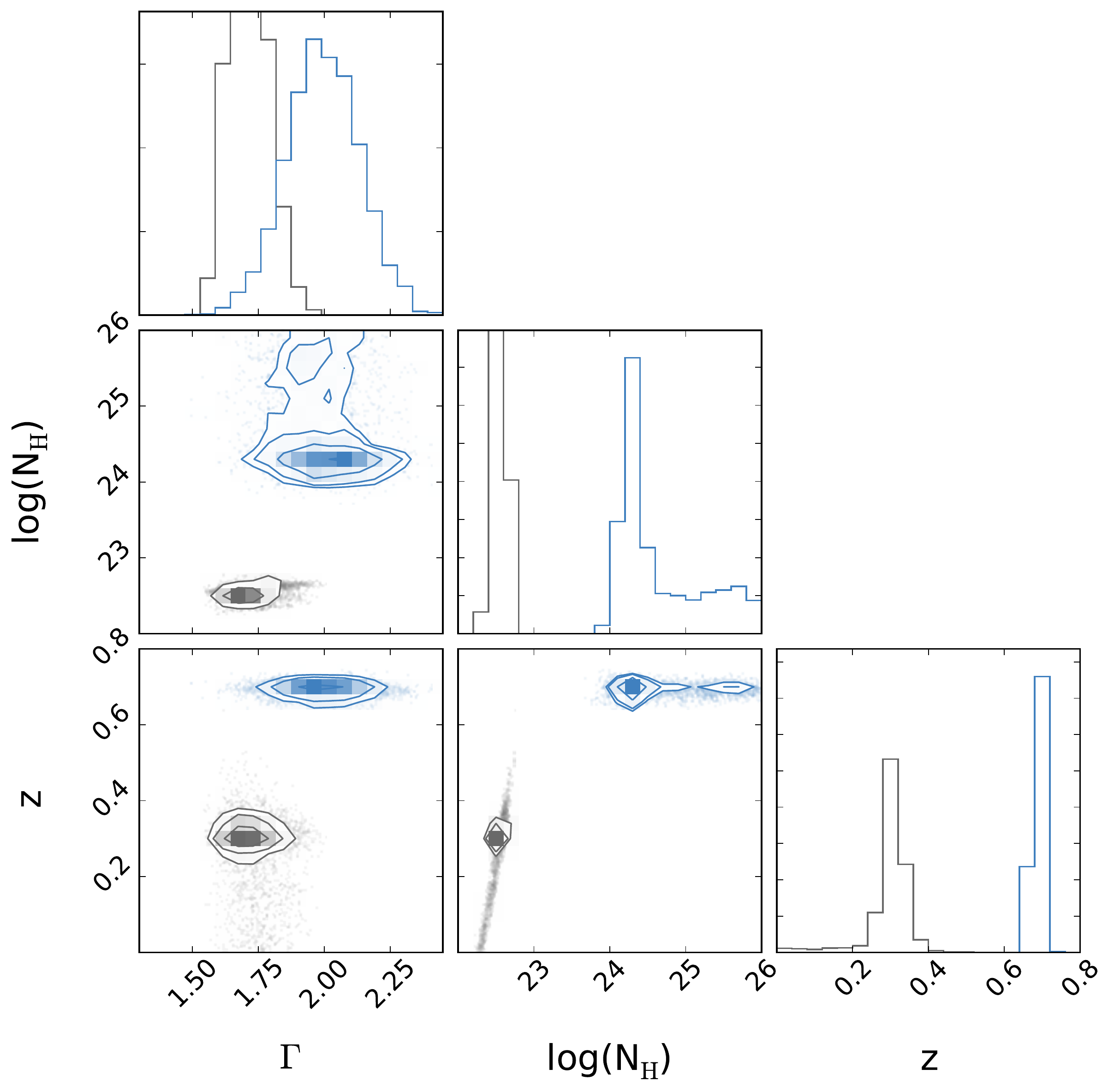}\quad
\caption{Corner plot showing the parameter space over photon index, column density and redshift for the heavily obscured AGN R65 (blue) and the mildly obscured AGN R4 (gray), see Figure~\ref{fig:examples}. 
Contours indicate 1, 2, and 3$sigma$ uncertainties. Any posterior samples outside are plotted as gray dots. The bottom right panel shows the marginalised redshift probability distribution. In the bottom middle panel a modest degeneracy between column density and redshift is present for R4.}
\label{fig:examples2}
\end{figure}

Our technique of determining redshifts from the X-ray spectrum (hereafter XZ) relies on fitting physically motivated templates commonly applied in obscured AGN. The intrinsic X-ray photon spectrum of AGN is to first order a powerlaw $E^{-\Gamma}$ with photon-index \citep[typically $\Gamma = 1.9 \pm 0.15$;][]{Nandra1994}. Most AGN additionally feature photo-electric absorption (due to O, Fe, H and He atoms along the line-of-sight) as well as Compton scattering \citep[e.g.,][]{Rivers2013,Buchner2014}. Figure~\ref{fig:nH} illustrates the resulting spectral shapes. Weak soft powerlaw emission is also frequently observed \citep[e.g.,][]{Brightman2014,Buchner2014}, which can be explained by Thomson scattering of the intrinsic powerlaw by a warm mirror which may bypass the obscurer. 

In unobscured sources, the spectrum is essentially a featureless\footnote{The equivalent width of the \FeKa line is below 100 eV \citep{Turner1998}, which is below typical spectral resolution of most X-ray CCDs operating today.} powerlaw, and little to no redshift information can be gained. However, in obscured sources the obscurer imprints absorption edges at characteristic energies. In heavily obscured AGN ($N_\mathrm{H}\gtrsim{10}^{24}\mathrm{cm}^{-2}$), the \FeKa line is prominent \citep{Turner1998} and can
pinpoint the redshift. 
We find in this work that in mildly obscured AGN, given enough photons, the absorption edge depths and turn-over energies can uniquely constrain the redshift, obscurer column density and photon index simultaneously, even using current CCD spectral resolution data. 

Figure~\ref{fig:examples} shows the spectra of two illustrative examples in which the method provides precise and correct redshift estimates. In the heavily obscured case (left panel; R65), the redshift determination relies on identification of the \FeKa feature, and the redshift can be fit more or less by eye. 

In the mildly obscured case (right panel in Figure~\ref{fig:examples}; R4), the \FeKa feature is not the source of redshift information. Instead, the location of the absorption edges rule out high redshift solutions, as can be seen when comparing the two spectral fit curves. Because of the non-smooth instrument response, the edges are difficult to identify by eye, and full forward modelling is needed to explore the possible redshifts, column densities and spectral slopes. Full parameter space exploration with forward modelling also permits going to lower count spectra. 

Figure~\ref{fig:examples2} shows the redshift, photon index and column density parameter space of the two examples. For R4 (mildly obscured; gray), both the column density ($\log$N$_{\rm{H}}$ = 22.57$^{+0.04}_{0.05}$ cm$^{-2}$) and redshift (XZ = 0.31$\pm$0.02) of the source are well constrained. The redshift uncertainty also agrees with specz (0.31). However, a degeneracy between redshift and column density (bottom middle panel) is visible.

The turnover shape in the model and the limited number of photons detected produces a degeneracy between redshift (decreasing the turn-over energy) and column density (increasing the turn-over energy). If this degeneracy is not resolved, large uncertainties remain on the redshift. However, given enough photon counts, the detection of absorption edges allows a relatively unique determination of redshift and column density. Furthermore, towards the Compton-thick limit of $N_\mathrm{H}\sim 10^{24}\mathrm{cm}^{-2}$ the model shapes change (see Figure~\ref{fig:nH}), as the \FeKa absorption edge becomes stronger. Because of this behaviour, very heavily obscured solutions (and associated low redshifts) can be excluded in the degeneracy (see gray "streak" of degeneracy in the lower middle panel of Figure~\ref{fig:examples2}).

The two cases illustrate the redshift precision that can be attained. The mildly obscured source has 5,690 source counts while the heavily obscured case (R65; blue) has only 128 source counts. Nonetheless, the redshift error for the latter is slightly smaller (XZ = 0.69$\pm$0.01), with a unique column density solution ($\log$N$_{\rm{H}}$ = 24.39$^{+0.02}_{-0.19}$ cm$^{-2}$). We note that photon counts, source obscuration level and source redshift all contribute to the redshift uncertainties.

\section{Methodology}
This section outlines our spectral model assumptions and parameter exploration algorithms used, as well as how redshift probability distributions are computed from X-ray spectra.

\subsection{Spectral model assumptions}

To fit the source we use the torus model from \citet[]{Brightman2011a}, which models the circumnuclear material around the X-ray emitting corona and computes the spectrum with photo-electric absorption, Compton scattering and fluorescent line emission (most importantly Fe K$\alpha$). We combine this model with a soft scattering powerlaw with a normalisation of up to 10\%. We also model the galactic absorption towards the source, such that:
\begin{center}
\texttt{model = tbabs * (torus + powerlaw)}
\end{center}
This is the \texttt{torus+scattering} model described in \cite{Buchner2014}, which fits the CDF-S field data reasonably well. We also tested replacing the torus model with a simple photo-electric absorber model (\texttt{wabs}) and find comparable but slightly worse results for Compton-thin sources. This indicates that the \FeKa line is not necessary to determine redshifts in these sources.

Non-informative priors are assumed for all parameters (uniform on $\log N_{\rm H}=20-26$ and $z=0-7$; log-uniform for normalisations), except for the photon index which has an informative Gaussian prior with mean $1.9$ and standard deviation $0.15$ \citep[e.g.,][]{Nandra1994}.

Solar abundances are assumed throughout. In obscured AGN, XZ is primarily measuring the total column density of oxygen and iron, and XZ succeeds only when the column density is high (e.g., $N_\mathrm{H} > {10}^{22}$). Extremely sub-solar oxygen to iron abundance ratios could hide edges and lead to poor or invalid redshift estimates. In general, if sources do not adhere to our model assumptions we should see a large outlier fraction in our test sample.

To model the contribution of background spectra, we use a new automatic fitting technique, \textit{BStat}. 
Modelling the background carefully, incorporating prior knowledge of background shapes seen in other observations, permits working also with very few source counts. To create empirical background models with just a few parameters we use a simple machine learning method on archival observations (described in the Appendix~\ref{autobackground}). We carry this out for several detectors including those on {\it Chandra}, {\it XMM} and {\it NuSTAR}. To reduce the number of parameters, we fit the background model to the background spectrum first and freeze its shape parameters. The source parameters and background normalisation are then fitted jointly. In low count sources, our approach extracts more information from the X-ray spectrum, because the machine-learned background model exploits instrument-specific prior information.

\subsection{Quantifying redshift information}
\begin{figure}
  \centering
   \includegraphics[width=1.0\columnwidth,trim={0 0 3cm 3cm}]{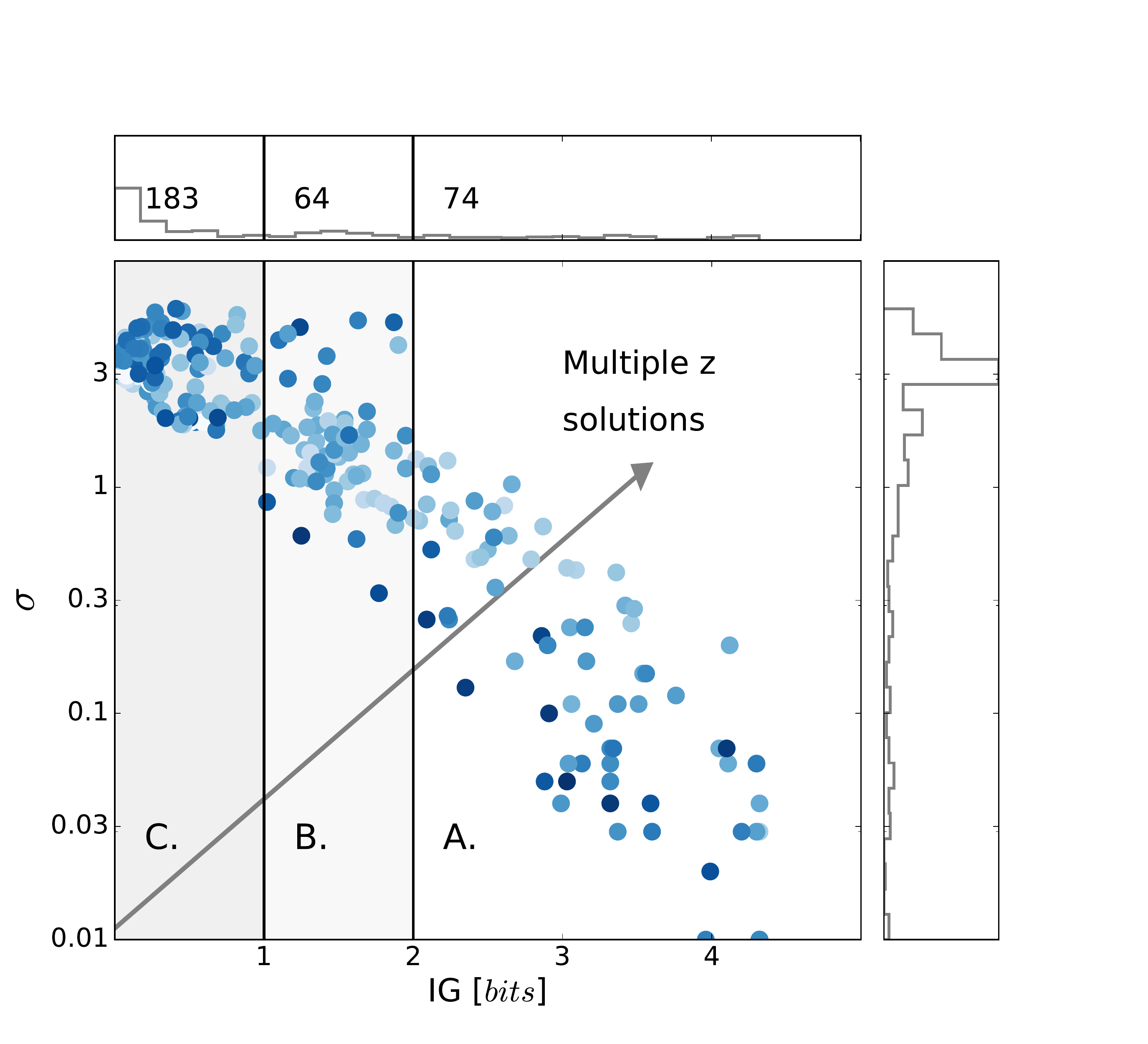}
   \caption{Comparison of two redshift uncertainty measures, the information gain (IG; x-axis) and the  standard deviation $\sigma(z)$ (y-axis). Generally, a larger $\sigma(z)$ implies a smaller IG. However, the IG also captures multiple solutions (examples are shown in subsequent figures) which can inflate the simple $\sigma(z)$ measure (symbolized by the gray arrow). The color-coding indicates the obscuration (median of uncertainties), darker colors represent higher column densities. Heavily obscured sources typically have only a single solution and are at lower $\sigma$. The vertical lines delimit three regions, $\rm{A}$ (IG $\geq$ 2 bits, 'constrained redshift'),  $\rm{B}$ (1 $\leq$ IG < 2 bits, 'has redshift information') and  $\rm{C}$ (IG < 1 bit, 'no redshift information'). The top and right panels show normalised histograms.
   }
   \label{fig:igrelation}
\end{figure}

Full exploration of the parameter space is crucial but not trivial. The parameter space of AGN obscuration presents multiple solutions due to the different spectral behaviour in the Compton-thin and Compton-thick regimes (see Figure~\ref{fig:nH}) and due to the spectral turn-over degeneracy involving redshift, column density, photon index and source flux normalisation. This presents  difficulties for commonly employed minimization algorithms or Monte Carlo Markov Chain techniques, and necessitates a global search algorithm \cite[]{Buchner2014}. We use the analysis software BXA \cite[]{Buchner2014}, which connects the Monte Carlo nested sampling algorithm MultiNest \cite[]{Feroz2009} with the fitting environment CIAO/Sherpa \cite[]{Fruscione2006}. BXA yields probability distributions on all physical parameters left free in the fit, including the photon index, $\Gamma$, the line-of-sight column density $N_\mathrm{H}$, and, most importantly, the redshift $z$. These three parameters are often degenerate and their uncertainties large when the X-ray spectrum has insufficient information. To avoid overconfidence in the best-fit, throughout we use the full posterior probability distribution for XZ. As we will later show, the combination of XZ redshift probability distributions with those from photoz can be very powerful even in the presence of large uncertainties.

A first useful way to characterise the redshift constraints is via the mean and standard deviation of the probability distribution. If there is a single peak and the standard deviation is, say, $\sigma(z)<0.1$, we may say we have constrained the redshift relatively well. However, the situation is less clear when there are multiple redshift solutions and upper or lower redshift bounds. 

Thus we define a redshift uncertainty measure that can deal with multiple redshift solutions. How much redshift information was gained from a measurement can be expressed by how different the posterior is from the (uninformative) prior. If at least some redshift intervals have zero posterior probability (i.e., are excluded), we have learned something. We use the Kullback-Leibler definition of the information gain (IG):
\begin{equation*}
\mathrm{IG} = \int \mathrm{Posterior}(z) \times \log_2\frac{\mathrm{Posterior}(z)}{\mathrm{Prior}(z)} dz ~ \mathrm{bits}
\end{equation*}
A value of $IG=0~\mathrm{bits}$ means there was no information gained, i.e., the posterior is equal to the prior. A high value of IG means a larger amount of information gained. We define three classes with simplifying labels:
\begin{enumerate}[label=(\Alph*)]
\item $IG>2$~bits: Redshift constrained
\item $IG=1-2$~bits: Some redshift information
\item $IG<1$~bit: Little to no redshift information in the X-ray spectrum.
\end{enumerate}
The borders at 1 and 2 bits crudely correspond to redshift uncertainties of $\sigma(z)=0.85$ and $0.4$, respectively, if there is only a single (Gaussian) solution.

\section{Results}

   \begin{table}
   \tiny
   \begin{center}
   \caption[]{Sample statistics}
       \label{summary}
       \begin{tabular}{lccc}
           	 & CDF-S & AEGIS-XD  & COSMOS \\
           	\hline
           	\hline
           	\noalign{\smallskip}
           	X-ray hard-band detected & 326 & 574 & 1016 \\
            Stars removed & 5 & 11 & 6 \\
            Total$^a$ & 321 & 563 & 1010 \\
            \noalign{\smallskip}
            \hline
            \noalign{\smallskip}
            XZ: have redshift information & 20\% & 14\% & 6\% \\  
            XZ: redshift constrained & 23\% & 11\% & 3\% \\ 
            Have photoz & 100\% & 53\% & 98\% \\
            Have specz & 56\% & 17\% & 49\% \\
            Obscured ($\log$N$_{\rm{H}} \geq 22$ cm$^{-2}$)& 82\% & 34\% & 33\% \\

			\noalign{\smallskip}
          	\hline
       	\end{tabular}
     \tablefoot{Sources which have redshift information are those with 1 $\leq$ IG < 2 $bits$, while sources with constrained redshift are those with IG $\geq$ 2 $bits$.  \\$^a$ Total number of sources selected.}
      \label{table:summary}
   \end{center}
   \end{table}

\begin{table*}
{\tiny
\hfill{}
\tiny
   \begin{center}
   \caption[]{Redshift catalogue of CDF-S sources with IG$\geq$1~bit (excerpt).}
       \label{const}
       \begin{tabular}{ccccccccccc}
           	\hline
           	\noalign{\smallskip}
           	R13 ID & Counts & RA & Dec & XZ & photoz & log$_{10}$N$_{\rm H}$ & specz & Quality & IG [$bits$] & Sample\\
            \noalign{\smallskip}
           	\hline
           	\noalign{\smallskip}
 			4 & 5690 & 53.1621   & -27.95005 & 0.31 (+0.02 ,-0.03) & 0.28 (+0.01 ,-0.01) & 22.57 (+0.04 ,-0.05) & 0.2983 & 0 & 4.32 & I\\
5 & 1112 & 53.05775  & -27.93325 & 1.89 (+1.20 ,-1.46) & 2.88 (+1.98 ,-1.17) & 23.00 (+0.48 ,-0.77) & ---  & --- & 1.27 & III\\
6 & 2030 & 53.10342  & -27.93336 & 0.78 (+0.48 ,-0.48) & 1.06 (+0.10 ,-0.02) & 22.36 (+0.26 ,-0.33) & 1.05 & 2 & 2.41 & II\\
7 & 2321 & 53.17007  & -27.92966 & 3.35 (+0.05 ,-0.05) & 2.97 (+0.08 ,-0.11) & 23.93 (+0.05 ,-0.05) & 1.221  & 2 & 3.32 & II\\
13 & 1085 & 53.19605  & -27.89266 & 1.97 (+0.93 ,-1.36) & 2.60 (+0.16 ,-0.19) & 22.92 (+0.32 ,-0.67) & ---  & --- & 1.50 & III\\
16 & 622 & 53.14668  & -27.88828 & 1.81 (+0.16 ,-1.03) & 1.85 (+0.09 ,-0.09) & 23.22 (+0.10 ,-0.54) & ---  & --- & 2.66 & II\\
			\noalign{\smallskip}
           	\hline
       	\end{tabular}
     \tablefoot{{\it Col. 1:} identification number of \cite{Rangel2013}. {\it Col. 2:} total number of detected source counts. {\it Cols. 3,4:} J2000 right ascension and declination in degrees. {\it Col. 5:} redshift (XZ) {\it Col. 6:} redshift (photoz) {\it Col. 7:} neutral hydrogen column in units of cm$^{-2}$ from XZ fit. {\it Col. 8:} reported specz values from literature compiled by \cite{HsuCDFSz2013}. {\it Col. 9:} quality of specz, a value of 0 (2) represents secure (insecure) specz's. {\it Col. 10:} information gained. {\it Col. 11:} subsample. All quoted errors are $1\sigma$.
     }
\label{table:long}
   \end{center}
   }
   \end{table*}

\begin{figure*}
  \centering
  \includegraphics[width=0.9\textwidth, origin=l]{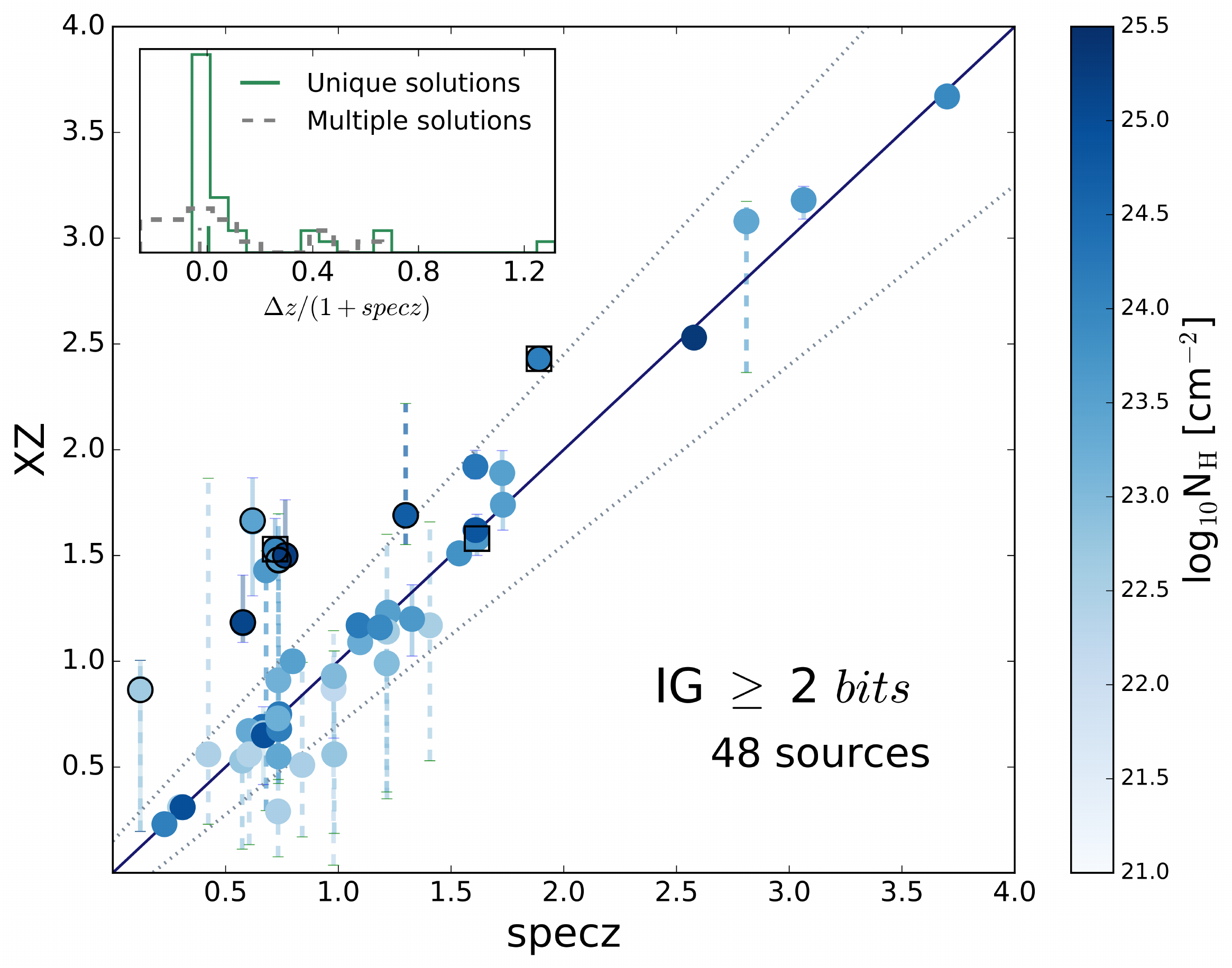} 
  \caption{Comparison of XZ and specz for constrained cases (IG$\geq$2). The filled circles show the median of the XZ probability distribution while the error bars correspond to $1\sigma$-equivalent quantiles and colorcoded by the median of the obscuration level. Dashed error bars indicate XZ found multiple solutions. The black solid line shows XZ = specz, while the gray dotted lines show XZ = specz $\pm$ 0.15(1 + specz). Points outside this region and where the XZ error bars do not cover the specz value are circled in black. Square symbols show insecure specz values \citep{Luo2017}. The inset shows the relative difference between XZ and specz}
  \label{fig:constrained}
\end{figure*}

\begin{figure}
\centering
\includegraphics[width=0.9\columnwidth,origin=l]{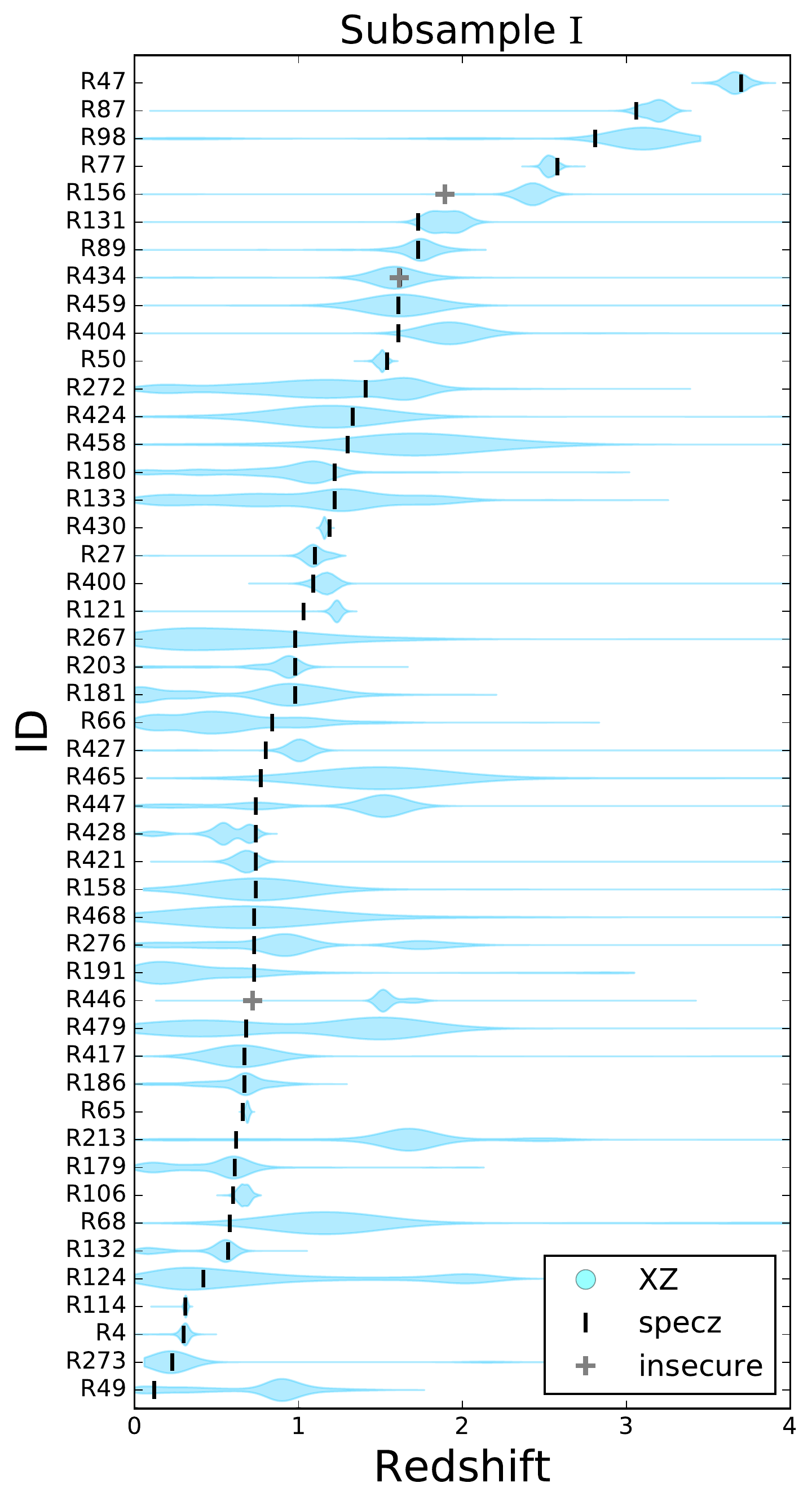}
\caption{XZ probability distributions (cyan violin plots) for subsample~I compared to specz values (black vertical lines). Sources ordered by increasing specz. Insecure specz values are shown as gray crosses.}
   \label{fig:speczPDZ}
\end{figure}

We first present some basic statistics about number and types of sources where XZ is able to extract redshift information from the X-ray data alone. Table \ref{table:summary} shows the sample statistics and the relative effectiveness of XZ, photoz and specz. We find 74 (out of 321) CDF-S sources with constrained redshifts (IG>2) from the X-ray spectra alone, which corresponds to $\sim 23$ percent of the sample. 
Table~\ref{table:long} shows the properties of all 138 CDF-S sources with IG$\geq$1. The XZ method works not only for heavily obscured sources but also in mildly obscured ones, if the count statistics are good enough.

To build a better understanding of the IG quantity, we compare it to the simpler redshift error $\sigma$ in Figure~\ref{fig:igrelation}. In a large fraction of CDF-S sources the IG has a close correspondence to $\sigma(z)$: High $\sigma$ correspond to low IG. As the color-coding in Figure~\ref{fig:igrelation} highlights, high-IG sources tend to be obscured. 
Our three IG classes are delimited by black vertical lines. Nearly all of the sources found in region A (redshift constrained, $IG \geq 2$~bits) have $\sigma \leq 1$. For region~B (some redshift information, $IG=1-2$~bits), there is large scatter in $\sigma$. 
The highest standard deviations here correspond to sources with multiple redshift solutions. In these cases simple error bars do not faithfully represent the uncertainties. Finally, sources in region~C have little or no redshift information in their X-ray spectrum. 

We test XZ against specz and photoz measurements in the following subsections. For this, we define three subsamples: The first two are composed of region A sources with constrained redshifts from XZ, which we compare to specz in subsample~I (48 AGN; §\ref{subsec:subsampleI}), and photoz in subsample~II (26 AGN; §\ref{subsec:subsampleII}). Sources with some XZ redshift information (region B, 1 $\leq$ IG < 2 $bits$) are compared to photoz in subsample~III (64 AGN§\ref{subsec:subsampleIII}). In all cases the specz and photoz information is taken from the compilation of \cite{HsuCDFSz2013}.

\subsection{Constrained XZ vs. specz (subsample I)}
\label{subsec:subsampleI}

\begin{figure}
  \centering
   \includegraphics[width=0.9\columnwidth, origin=l]{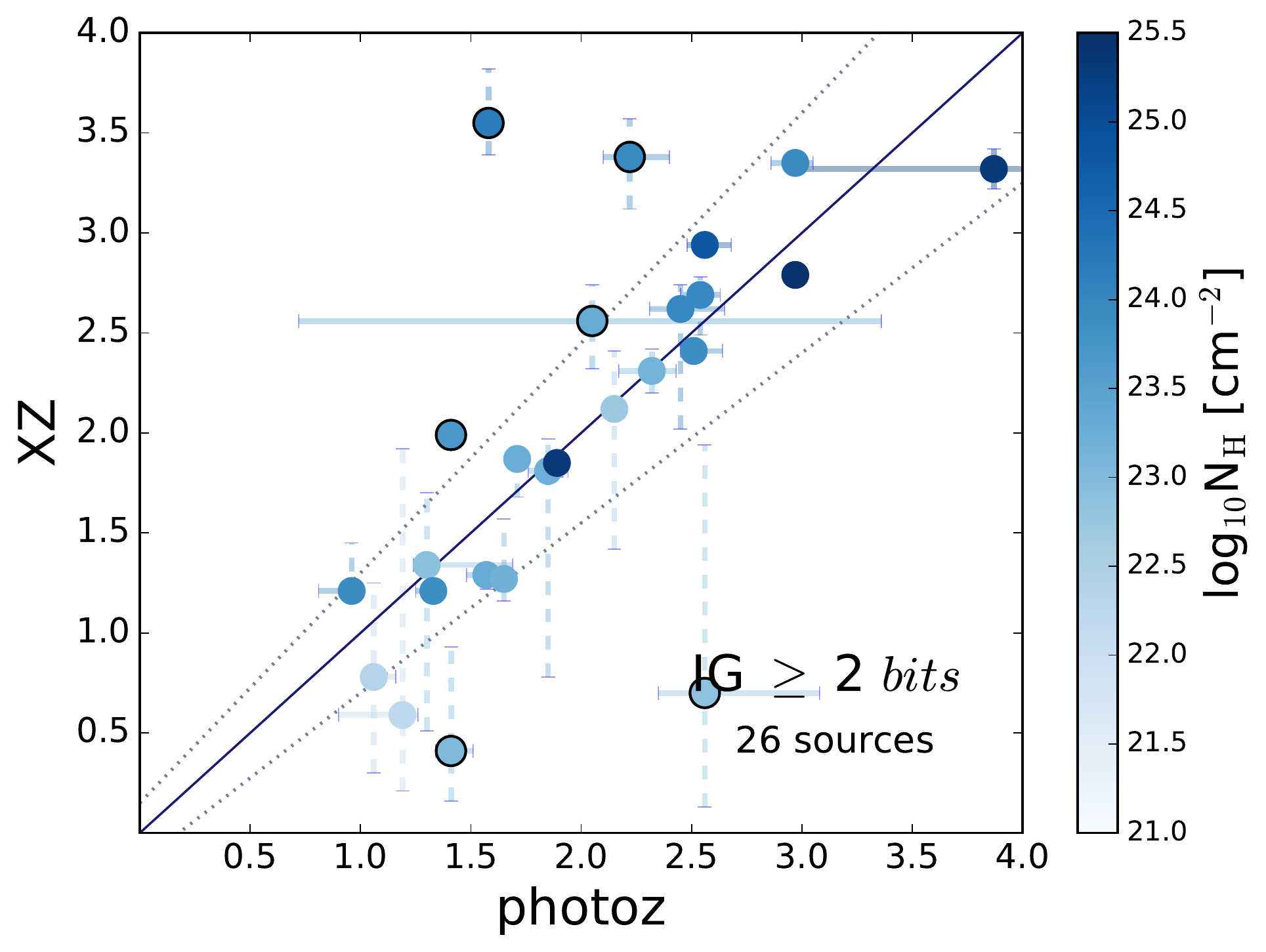} 
   \caption{
   As in Figure~\ref{fig:constrained}, but comparing XZ to photoz (subsample II). }
   \label{fig:constrainedphotoz}
\end{figure}

\begin{figure}
\centering
\includegraphics[width=0.5\textwidth,origin=l]{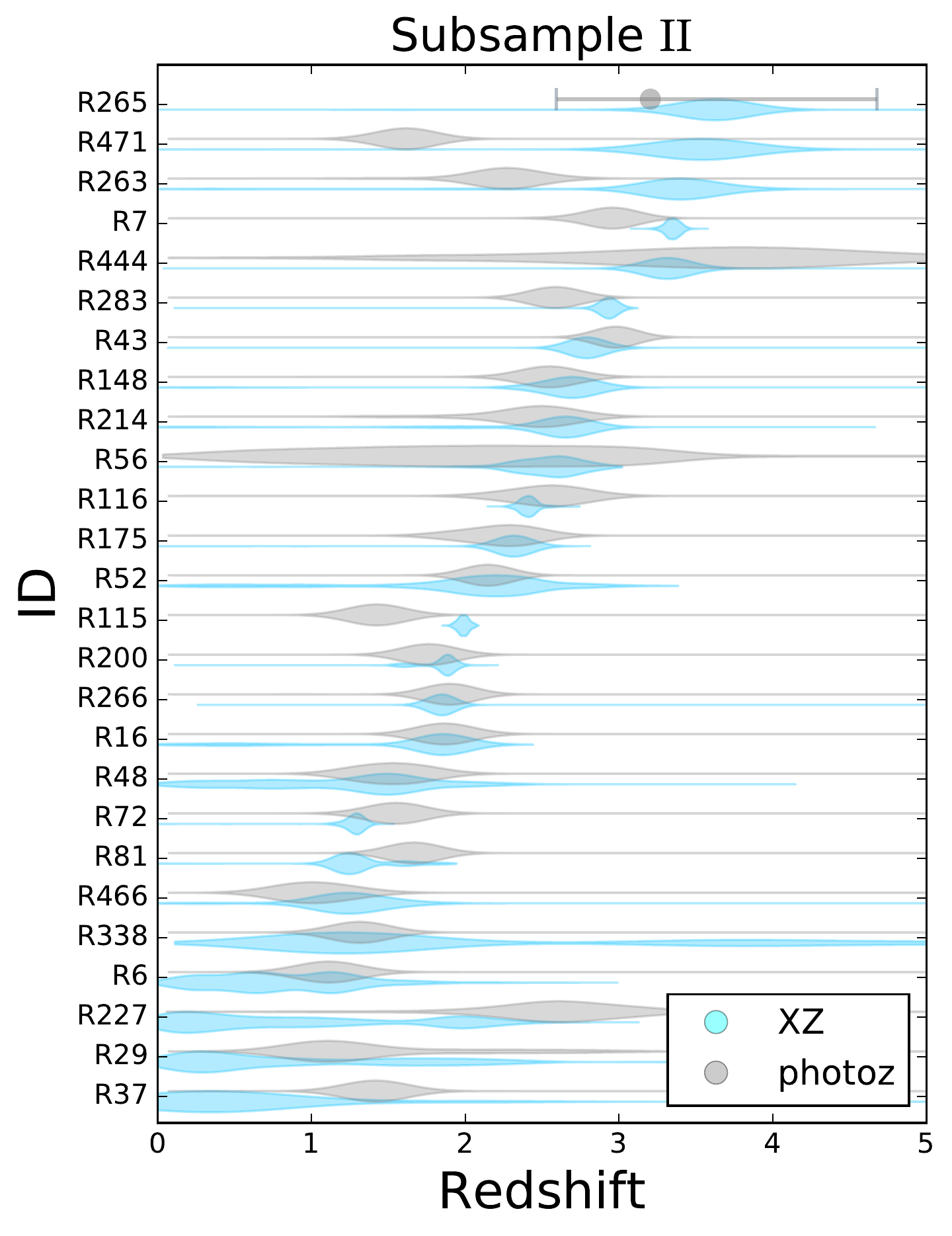}
\caption{XZ (cyan) and photoz (gray) probability distributions for subsample II (IG $\geq$ 2 $bits$, no specz). XZ and photoz can mutually confirm redshifts and narrow uncertainties (e.g., R56, R116, R227, R444) or select between multiple photoz solutions (e.g., R12, R333). Some cases show disagreements (e.g., R115, R273, R471). For R265 the photoz was taken from \cite{Luo2010}. Sources ordered by increasing XZ median values.}
  \label{fig:confirm}
\end{figure}

\begin{figure}
\centering
\includegraphics[width=0.97\columnwidth,origin=l]{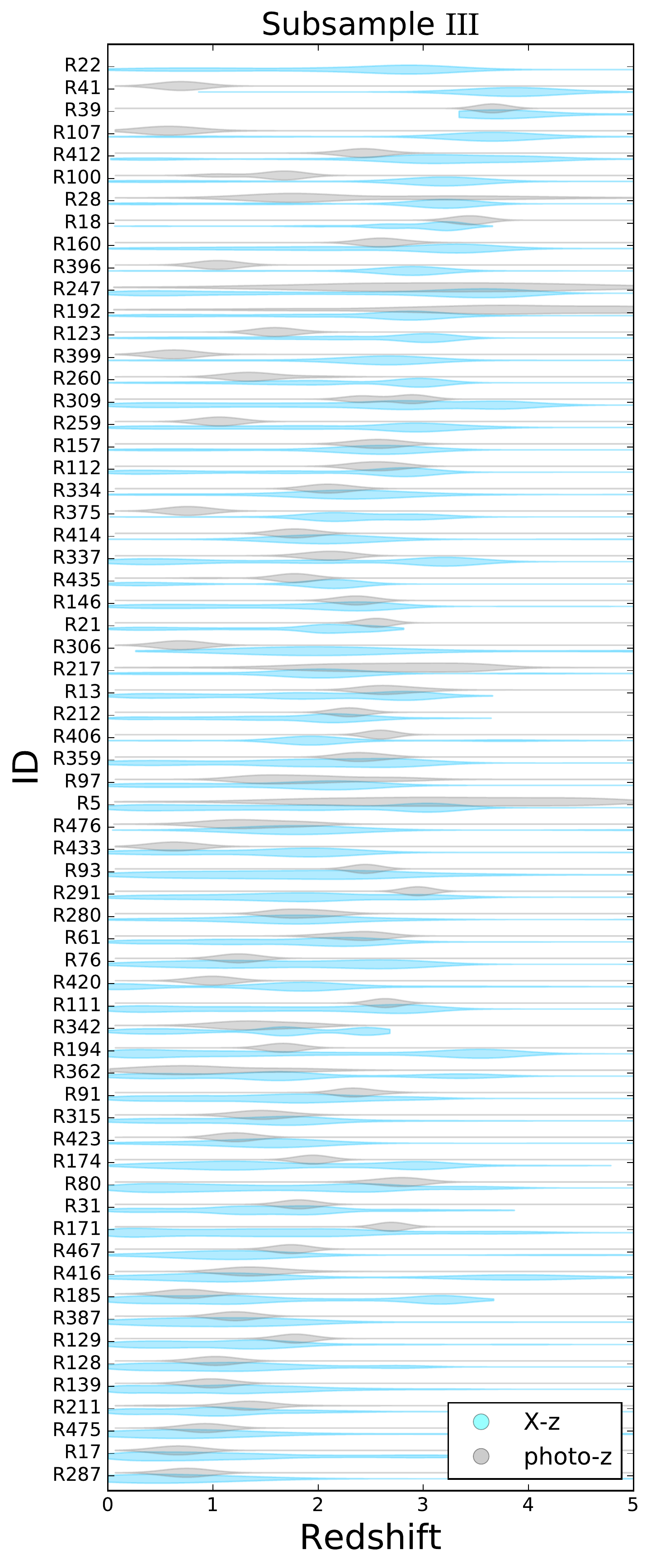}
\caption{XZ (cyan violin plots) and photoz (gray violin plots) probability distributions for subsample III (1 $bits$ $\leq$ IG < 2 $bits$). 
}
  \label{fig:sub3PDZ1}
\end{figure}

We first test the reliability of our XZ method against specz measurements.

Figure~\ref{fig:constrained} plots XZ against specz for subsample I. For $\sim$60 percent, the specz resides within the 68\% quantile of XZ. In many cases, the uncertainties are very small, providing relatively precise redshift measurements. The inset of Figure~\ref{fig:constrained} shows the deviations between the specz and XZ median. For 75 percent, these are within $\pm0.15(1+z)$, with a median absolute deviation of $0.04(1+z)$. Thus, for the vast majority of this sample, XZ provides redshifts consistent with specz. 

The ability to correctly recover specz values implies both that the specz counterparts are correct and that the XZ model assumption are generally valid. We define outliers as those where the 1-$\sigma$ XZ error bars do not cover the specz value and the XZ median is outside specz$\pm$0.15(1+specz). With this criterion, we find seven outliers among the 48 well-constrained sources (R49, R68, R156, R213, R446, R458 and R465). We discuss these in more detail section §\ref{sec:outliers}, where we find that, at least in some cases, the specz has to be corrected.

Dashed error bars in Figure~\ref{fig:constrained} indicate multiple-peaked redshift solutions. In these cases, error bars over-simplify the redshift probability distribution. To show the full distribution, a violin plot visualisation is presented in Figure~\ref{fig:speczPDZ}. There we compare the XZ redshift probability distributions (cyan violin plots) and the specz values (black vertical lines). Multiple solutions can be seen for several sources (e.g., R179, R181, R428). 

\subsection{Constrained XZ vs. photoz (subsample II)}
\label{subsec:subsampleII}

Next, we consider sources without specz measurements, and validate XZ against photoz. This generally probes sources that are fainter in the optical/near-infrared wavelengths. The mutual validation of photoz and XZ can confirm redshifts or point out potentially problematic cases.

Figure \ref{fig:constrainedphotoz} plots XZ against photoz for subsample II. While the majority of the sources show consistent redshifts (XZ$\approx$photoz) within the errors, there are six outliers out of 26 sources (using the same definition as above). This is a higher fraction than in subsample~I. In order to analyze these cases in more detail, we show in Figure~\ref{fig:confirm} the normalised XZ and photoz probability distributions (gray and cyan, respectively). Disagreements are seen e.g., in R115, R263, R471. 
An interesting case is R227, which has two solutions in XZ, one of which overlaps with the photoz uncertainties. The combination of the constraints implies a redshift just above 2, demonstrating the synergy of the two techniques.

\subsection{XZ redshift predictions (subsample III)}
\label{subsec:subsampleIII}

Finally, we consider XZ constraints that are weaker, but can be combined with photoz constraints to distinguish between degenerate or multiple redshift solutions. Figure~\ref{fig:sub3PDZ1} shows the redshift probability distributions for  subsample III (IG=1-2 bits), as before, the gray violin plots represent photoz while the cyan ones show XZ. 
In several cases (e.g., R5, R185, R194, R306) X-ray spectral fitting can confirm results found via photometry and distinguish between multiple photoz redshift solutions.
In one of the cases, R22, no photoz is presented because this X-ray source has an ambiguous association; for one of these opt/NIR counterparts (\citeauthor{HsuCDFSz2013} ID 105068) the photoz is highly uncertain, while the other (61367) is constrained to photoz=$2.11\pm0.06$. Unfortunately, the wide XZ error is not able in this case to provide much clarity.

\section{Discussion}
\label{sec:Discussion}

\begin{figure}
\centering
\includegraphics[trim={0 1.1cm 0 0},width=.48\textwidth]{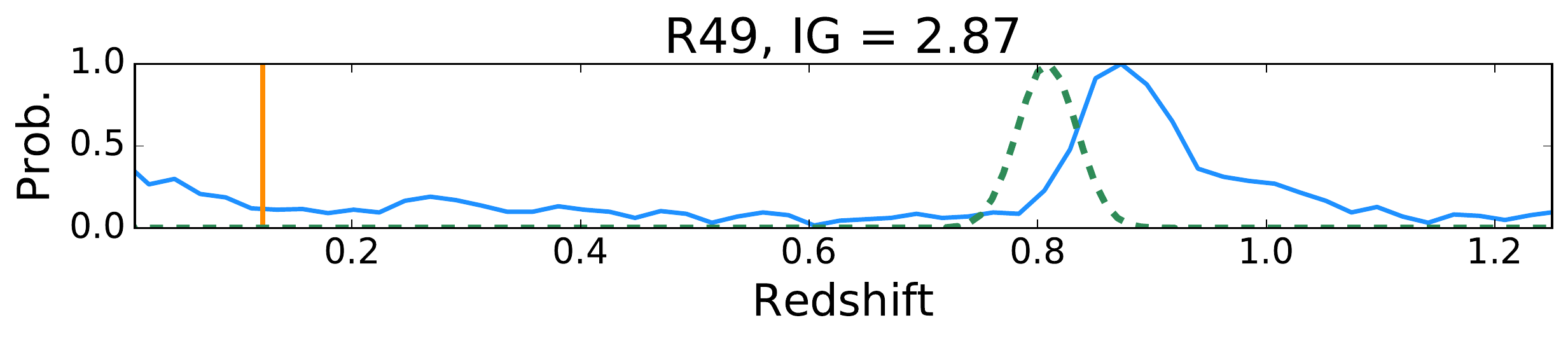}\quad
\includegraphics[trim={0 1.1cm 0 0},width=.48\textwidth]{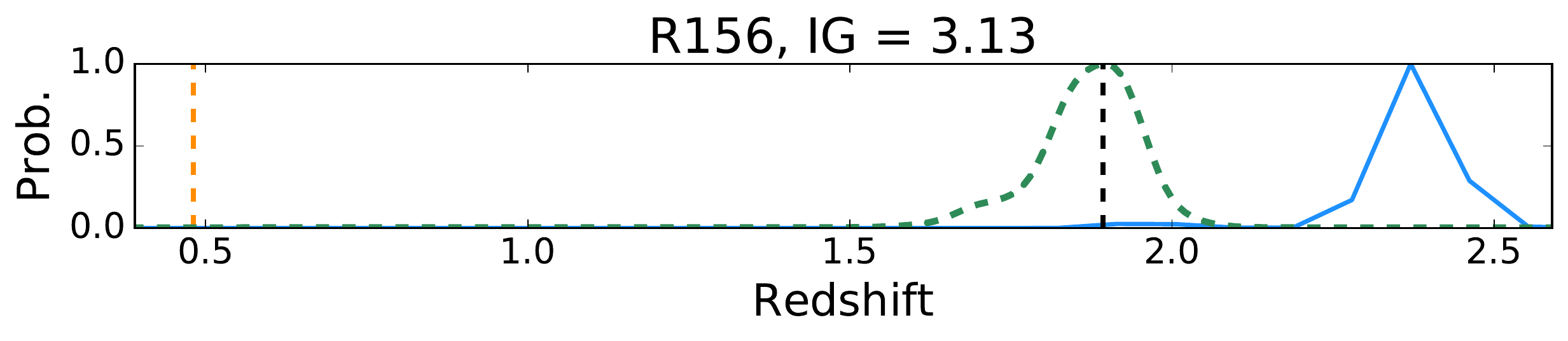}\quad
\includegraphics[trim={0 1.1cm 0 0},width=.48\textwidth]{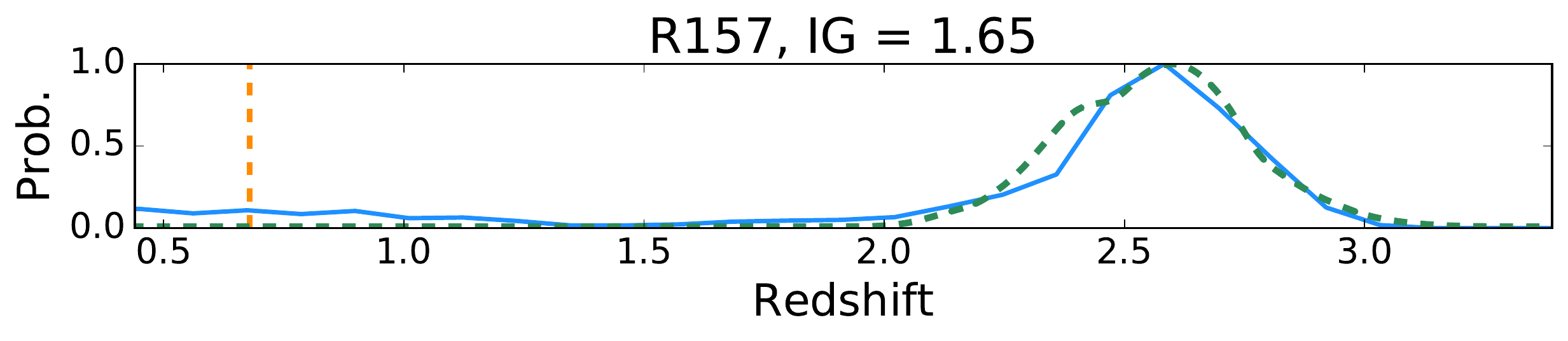}\quad
\includegraphics[trim={0 1.1cm 0 0},width=.48\textwidth]{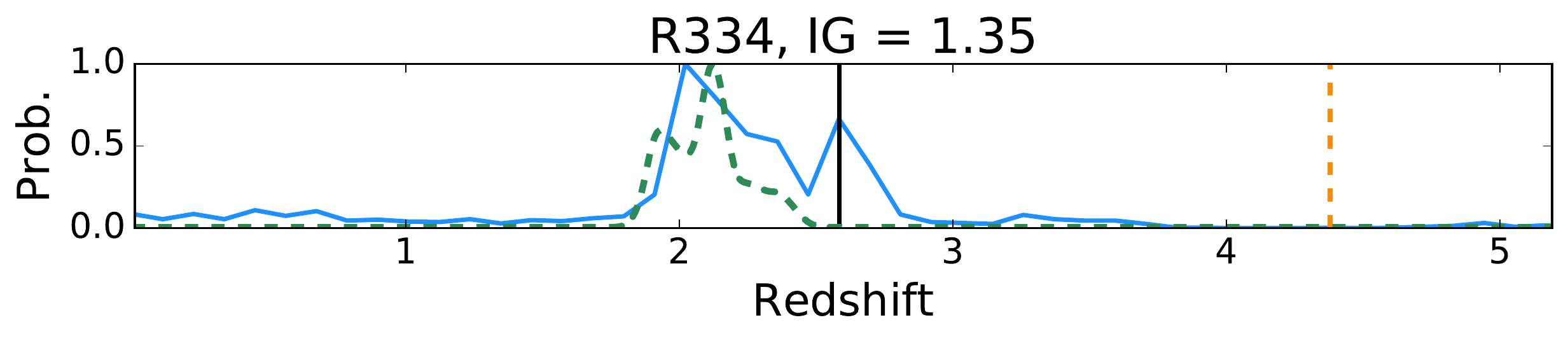}\quad
\includegraphics[width=.48\textwidth]{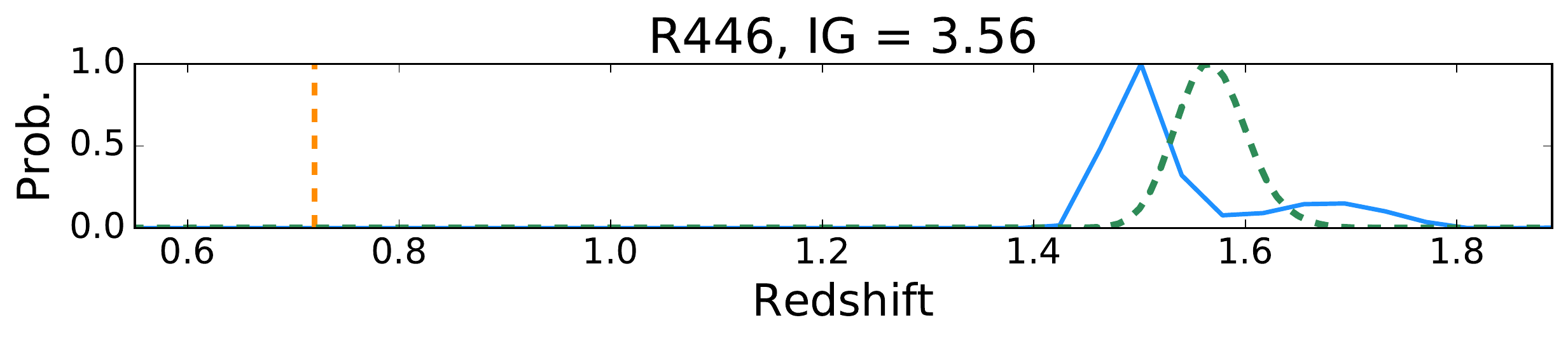}\quad

\caption{Normalised redshift probability distributions where XZ (blue solid) and photoz (green dashed) agree but are different than the specz values (orange vertical lines). Black vertical lines in R156 and R334 show updated specz values from \cite{Balestra2010} and \cite{Barro2014}, respectively, which agree with the photoz and XZ. Dashed vertical lines show insecure specz values.}
\label{fig:PDZ}
\end{figure}

We now discuss differences between XZ, photoz and specz redshift estimates. 

\label{sec:outliers}
We find a few noteworthy cases which we arrange into two distinct categories: (1) cases in which XZ and photoz agree but neither agree with specz, and (2) cases in which photoz and specz agree but neither agree with XZ.

\subsection{Potential specz issues}

Some disagreements with specz may be caused by incorrect {\em spectral line identification}, especially when only a single line has been observed. In some such cases, the redshift is nevertheless classified as 'secure' in published catalogues.

We identified several cases where XZ=photoz$\neq$specz. These include sources with constrained redshifts (R49, R156, R446; from subsample I) and others with more limited redshift information (IG>1 in R157, R334).
For these five sources, we present the (dis)agreements  between XZ, photoz and specz in Figure~\ref{fig:PDZ}. 

\begin{figure}
\centering
\includegraphics[trim={0 0 0 0},clip,width=.45\textwidth]{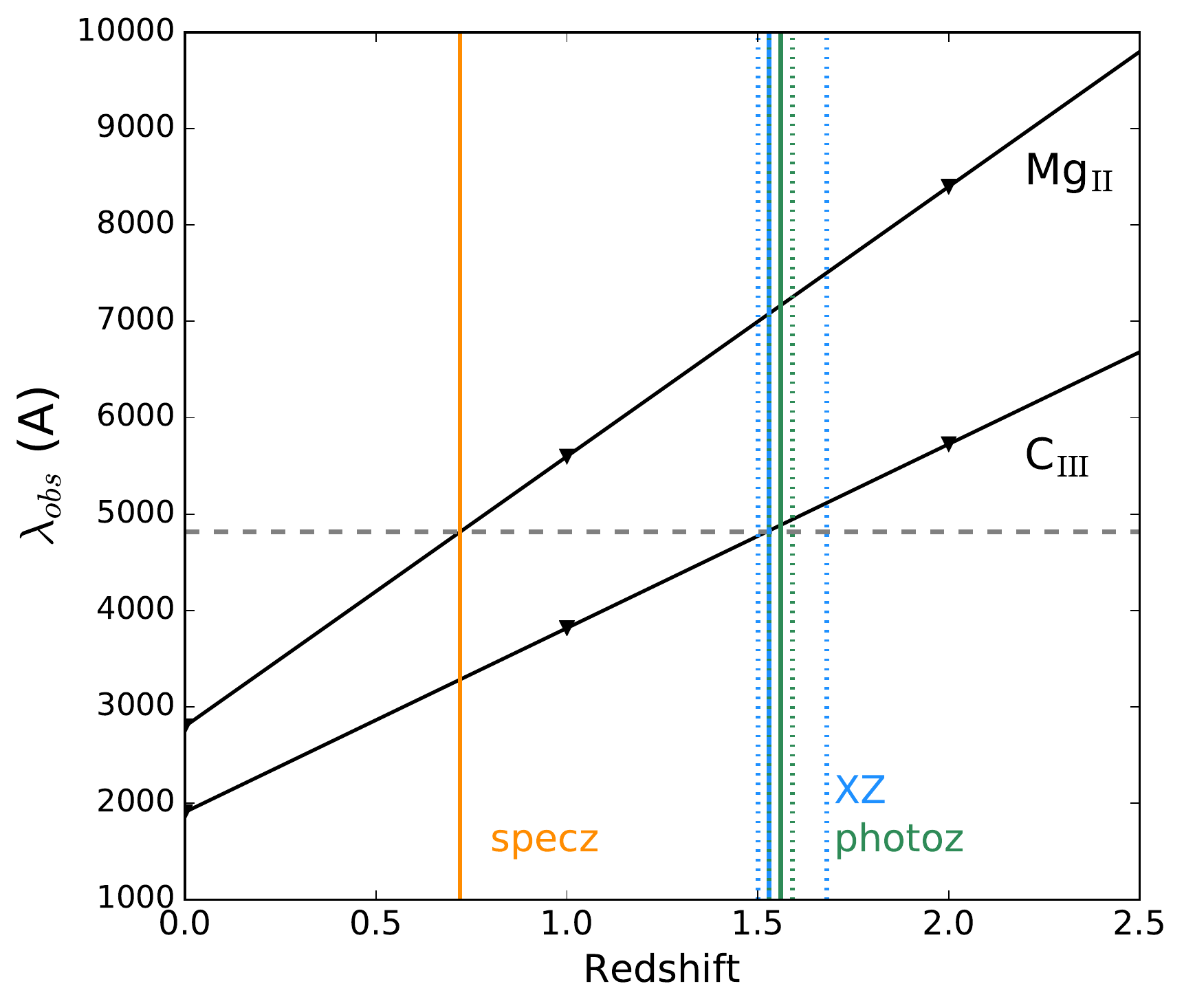}\quad
\caption{Observed wavelength of Mg$_{\rm{II}}$ and C$_{\rm{III}}$ spectroscopic lines as a function of redshift. The horizontal dashed line shows the observed wavelength of a single detected optical spectral line in source R446. The vertical lines indicate the reported specz (orange), photoz (green) and XZ (blue). The dotted vertical lines show the errors of photoz and XZ, respectively. The specz was derived assuming Mg$_{\rm{II}}$, but agrees with XZ and photoz if we assume C$_{\rm{III}}$ instead.}
\label{fig:discussionR446}
\end{figure}

Through a literature search on these sources we found updated specz for two, R156 and R334, in \cite{Balestra2010} and \cite{Barro2014}, respectively. We mark these in Figure~\ref{fig:PDZ} as vertical black lines. These new specz values agree with XZ. This fact makes us question the reliability of specz for the remainder of this group, in particular for R446. 
For R446 we do not have an updated specz, however, the redshift reported in \cite{Szokoly2004} relies on only a single spectral line. 
In Figure~\ref{fig:discussionR446} we show the evolution of the observed wavelength of two spectral lines according to redshift. The specz (0.72) was derived from a spectroscopic line assumed to be Mg$_{\rm{II}}$. However, XZ and photoz indicate $z\sim 1.56$. This would be consistent if the line was C$_{\rm{III}}$.
Given that the literature search provided updated specz values for two sources, which agree with XZ, we caution that  specz can sometimes be incorrect and this can be detected by our method.

\subsection{Potential association issues}

\begin{figure}
\centering
\includegraphics[trim={0 1.1cm 0 0},width=.48\textwidth]{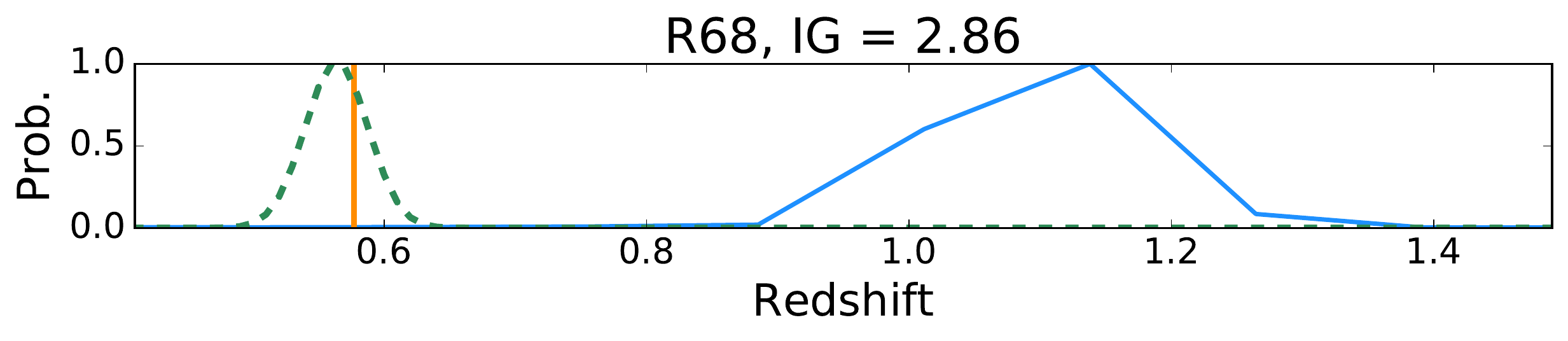}\quad
\includegraphics[trim={0 1.1cm 0 0},width=.48\textwidth]{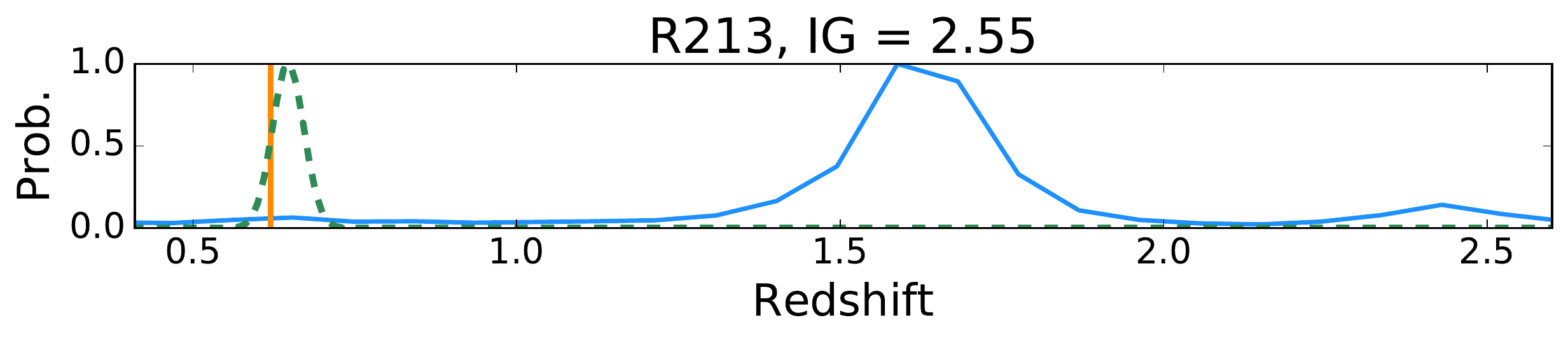}\quad
\includegraphics[trim={0 1.1cm 0 0},width=.48\textwidth]{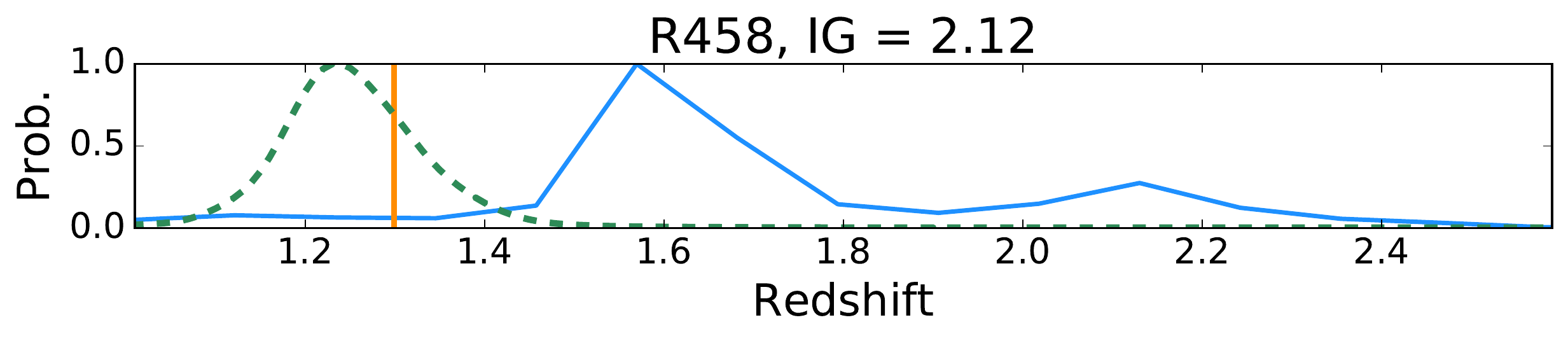}\quad
\includegraphics[width=.48\textwidth]{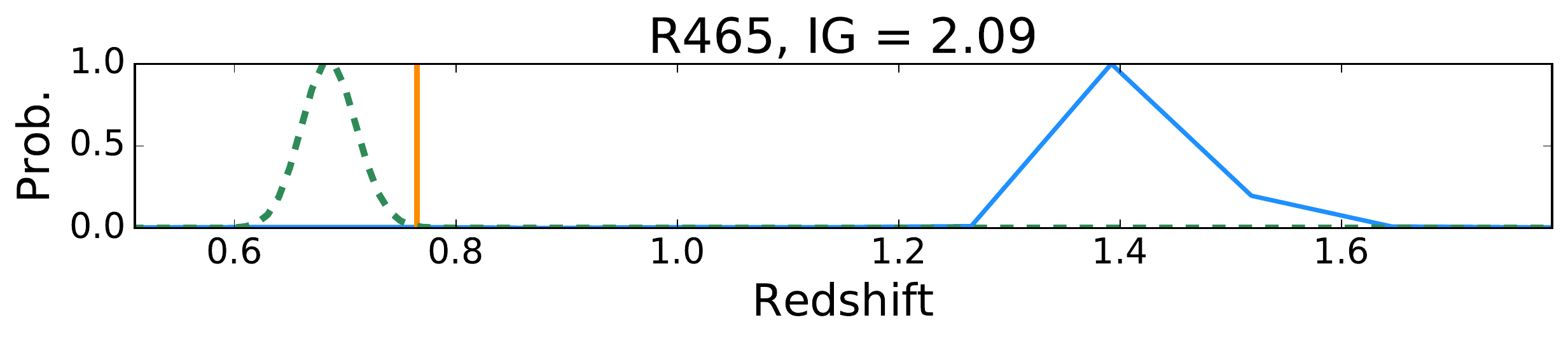}\quad
\caption{Same as Fig.~\ref{fig:PDZ} but for cases where photoz (green dashed) and specz (orange vertical line) agree but are noticeably different than XZ (blue curve).}
\label{fig:PDZ2}
\end{figure}

We find four constrained sources where XZ$\neq$photoz=specz, namely R68, R213, R458 and R465. For these outliers, Figure \ref{fig:PDZ2} shows the redshift probability distributions, along with their associated specz. The disagreement likely indicates that some assumptions of the XZ method are invalid in these cases. From these we estimate that the {\em outlier fraction is $\sim$8\% of the sample}, indicating a high reliability of the XZ method.

For this group, in principle the XZ could be correct if both the specz and photoz are incorrect. This could occur because of incorrect {\em optical counterpart associations}. 
For our sample, this is not very probable because the {\it Chandra} positional error with \~ hundred counts is usually not large enough to allow multiple counterpart associations. 

\subsection{Potential XZ issues}

In six constrained cases XZ disagrees with photoz (from subsample II). These are: R37, R56, R115, R227, R263 and R471, presented in Figure~\ref{fig:confirm}. All of these sources fall within the sensitivity range of the method, with either a high count number or a high level of obscuration. It is necessary to investigate these sources further to confirm the correct redshift since one (or both) methods are failing in these cases. As discussed above, photoz (like specz) can suffer from possible counterpart misassociations while XZ is far less likely to do so (given the X-ray source density). Both photoz and XZ can yield multiple solutions and sometimes the correct solution is outweighed by an incorrect one, leading to wrong redshifts. 

\begin{figure}
\centering
\includegraphics[trim={0 0 0 0},clip,width=.48\textwidth]{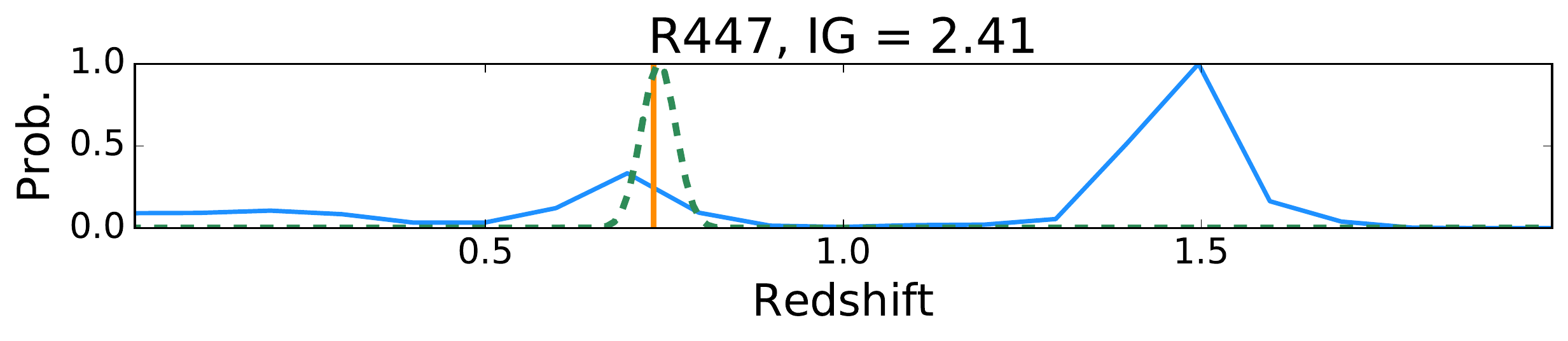}\quad
\caption{Same as Fig.~\ref{fig:PDZ} but for case R447 where XZ (blue) presents multiple solutions, one of which agrees with photoz (green dashed) and specz (orange vertical line).}
\label{fig:discussionR447}
\end{figure}

Figure~\ref{fig:discussionR447} shows an example of the multiple solution issue, R447. 
While the photoz = $0.74\pm0.01$ and specz = 0.735 agree, the XZ (nominally 1.48$^{+0.08}_{-0.87}$) has two solutions, with the less probable solution covering the same range as the photoz and specz.
Thus the most probable redshift solution reported by XZ is not necessarily the correct one, which is why it is {\em advisable to combine the results with photoz in order to get more reliable results.}

In addition, XZ can be affected by AGN variability. This is an open / unclear issue, as variability might either weaken or strengthen features needed for XZ, depending on the sense of the variability (i.e., increasing or decreasing NH or intrinsic continuum).

Finally, there could be fundamental issues with the XZ approach. XZ is model-based and thus invalid model assumptions could lead to incorrect results. The model could be extended to be more realistic, e.g., by allowing different obscurer metallicities, obscurer geometries, multiple AGN components and/or soft energy contaminants. However, the low outlier fraction of \textbf{8\%} when validating against specz indicates that the simple adopted model is adequate in general. 

XZ is most useful when obscured AGN with moderate count statistics make up a sizable fraction of the sample. Because of the observed increase in the obscured fraction at low luminosities and high redshifts \citep[e.g.,][]{Ueda2003,Hasinger2005}, XZ is most useful in deep surveys like the CDF-S. Consistent with this, we find lower percentages of constrained redshifts when testing our method on data from the COSMOS and AEGIS-XD fields (see Table~\ref{table:summary}). The detailed results from these two surveys are presented in Appendix \ref{moretables}, including redshift catalogues and photoz/specz plots. 

\section{Sensitivity Analysis for different X-ray Instruments}
\label{sec:sensitivity}

\subsection{Simulation setup}
   \begin{table}
   \begin{center}
   \caption[]{Simulation Parameter Grid}
       \label{simparams}
       \begin{tabular}{ll}
           	\hline
           	\noalign{\smallskip}
           	Spectral Parameter & Allowed values \\
            \noalign{\smallskip}
           	\hline
           	\noalign{\smallskip}
           	Photon Index $\Gamma$ & 1.9 \\
            Column density $N_\mathrm{H}$ & 10$^{22}$, 10$^{23}$, 10$^{24}$, 10$^{25}$ cm$^{-2}$ \\
            Viewing angle $\theta_\mathrm{op}$ & 45\degr \\
            Opening angle $\theta_\mathrm{view}$ & 87\degr (edge-on) \\
            Scattering fraction $f_\mathrm{scat}$ & 1\% \\
            Redshift $z$ & 0.1 to 5 in steps of 0.1 \\
            Instruments & see Table \ref{table:sims}  \\
			\noalign{\smallskip}
           	\hline
       	\end{tabular}
     \tablefoot{Source fluxes were normalised for each Instrument to target reasonable source count ranges, as listed in Table~\ref{table:sims}.}
      \label{table:simgrid}
   \end{center}
   \end{table}

   \begin{table}
   \footnotesize
   \begin{center}
   \caption[]{Simulation setup for tested instruments}
       \begin{tabular}{ccccc}
           	\hline
           	\noalign{\smallskip}
           	Instrument & Counts & N & E & Exp\\
           	 &  &  & [keV] & [ks]\\
            \noalign{\smallskip}
           	\hline
           	\noalign{\smallskip}
           	Athena/WFI  & 10, 20, ..., 150   & 3,000 & 0.2-10 & 200\\ 
            Chandra/ACIS& 10, 20, ..., 300   & 6,000 & 0.5-8  & 4,000\\
            eROSITA     & 100, 200, ..., 2,000& 4,000& 0.2-10 & 1\\ 
            NuSTAR      & 10, 20, ..., 300   & 6,000 & 3.3-78 & 400\\ 
            Swift/XRT   & 25, 50, ..., 400   & 3,200 & 0.5-5 & 100\\
            XMM/PN+MOS  & 19, 38, ..., 285   & 3,000 & 0.3-8 & 20\\ 
\noalign{\smallskip}
           	\hline
       	\end{tabular}
     \tablefoot{{\it Col. 1:} instrument name.
     {\it Col. 2:} Grid of source net counts used. For {\it XMM}, the counts refer to the number of PN counts, but data from PN, MOS1 and MOS2 are fitted simultaneously.
     {\it Col. 3:} Total number of simulations.
     {\it Col. 4:} Energy range considered.
     {\it Col. 5:} Exposure time.
      }
      \label{table:sims}
   \end{center}
   \end{table}

\begin{figure*}[htp]
\centering
\includegraphics[trim={0.3cm 1cm 0.3cm 0},clip,width=.4\textwidth]{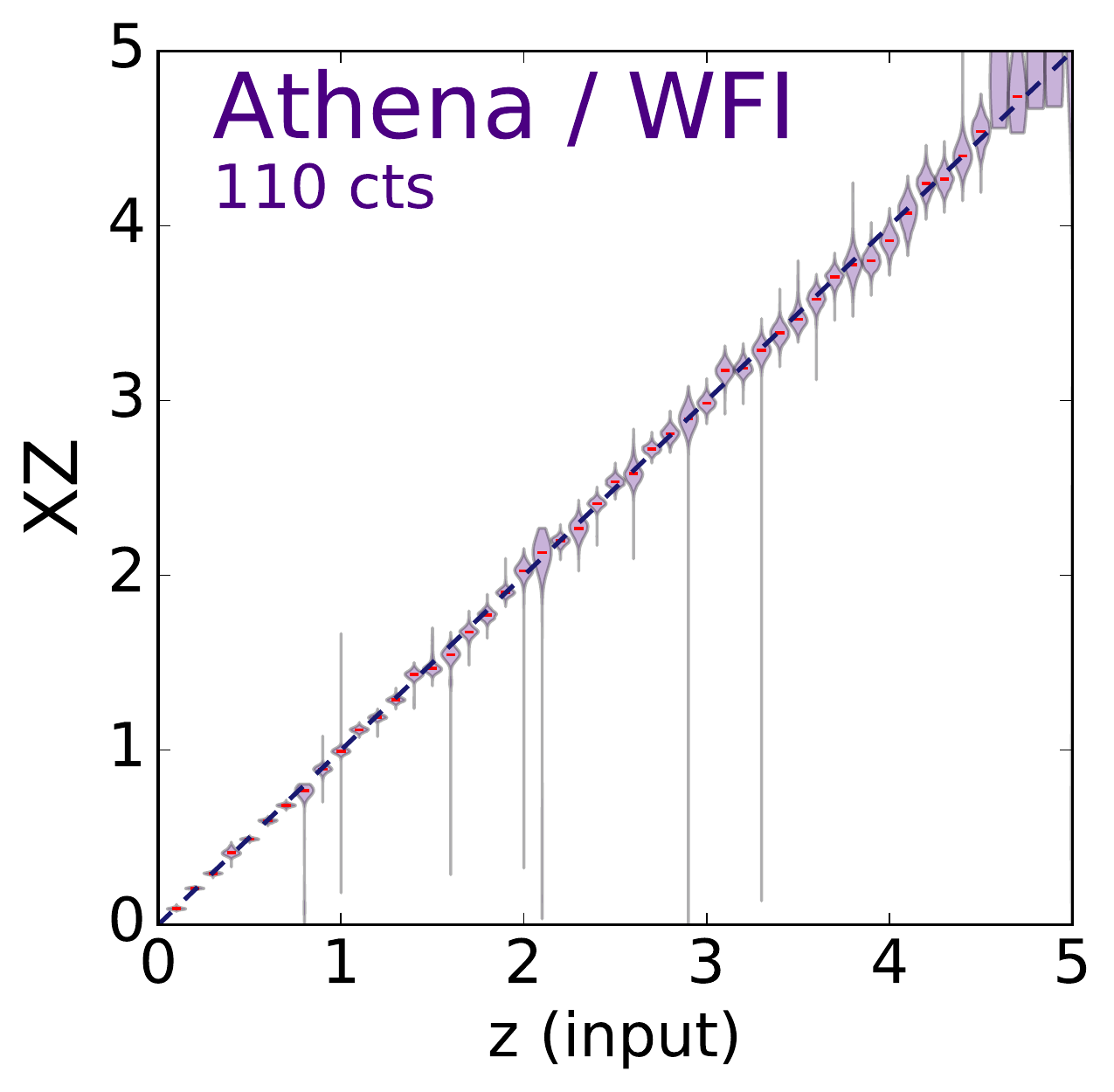}\quad
\includegraphics[trim={0.3cm 1cm 0.3cm 0},clip,width=.4\textwidth]{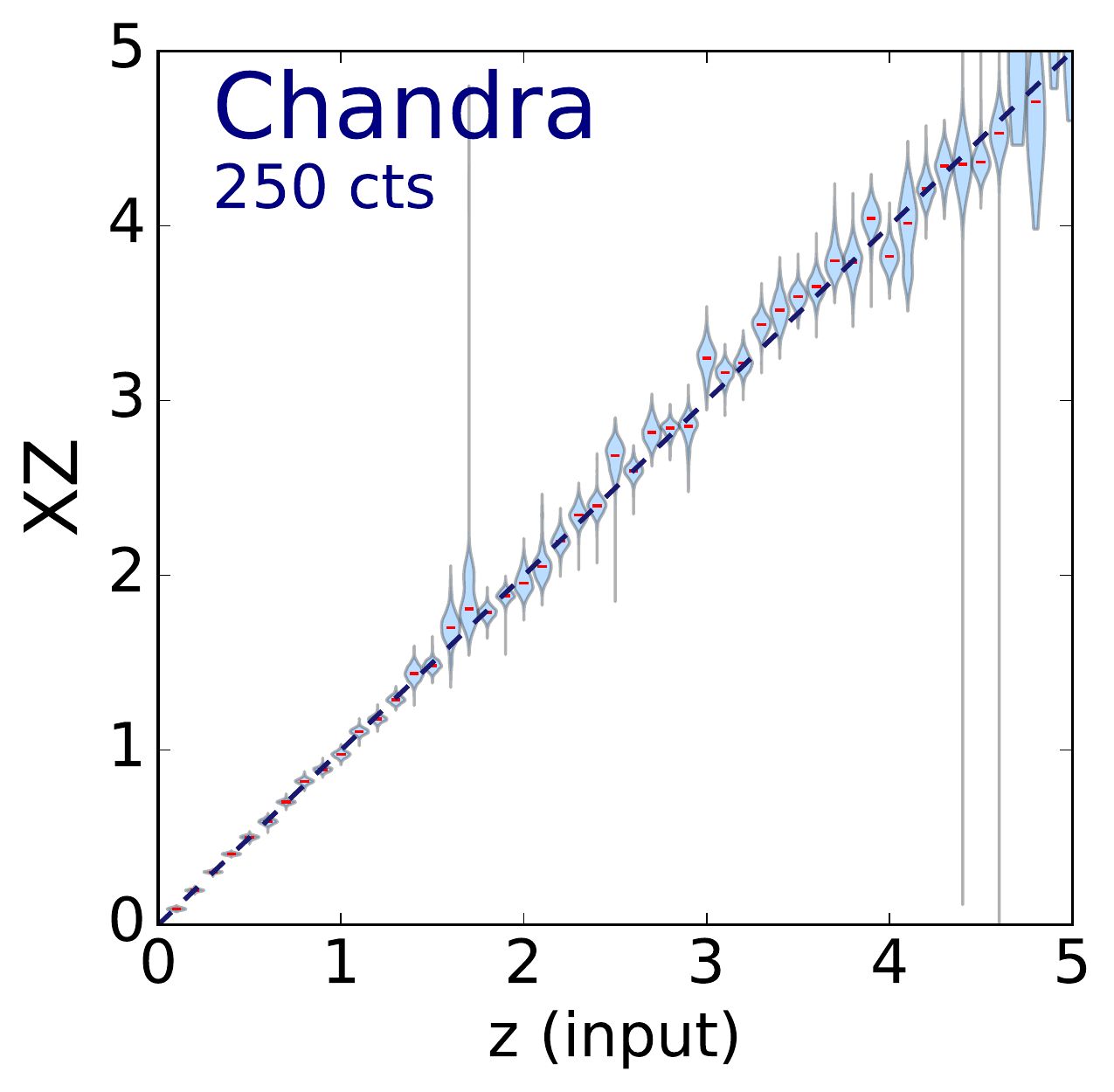}\quad

\includegraphics[trim={0.3cm 1cm 0.3cm 0},clip,width=.4\textwidth]{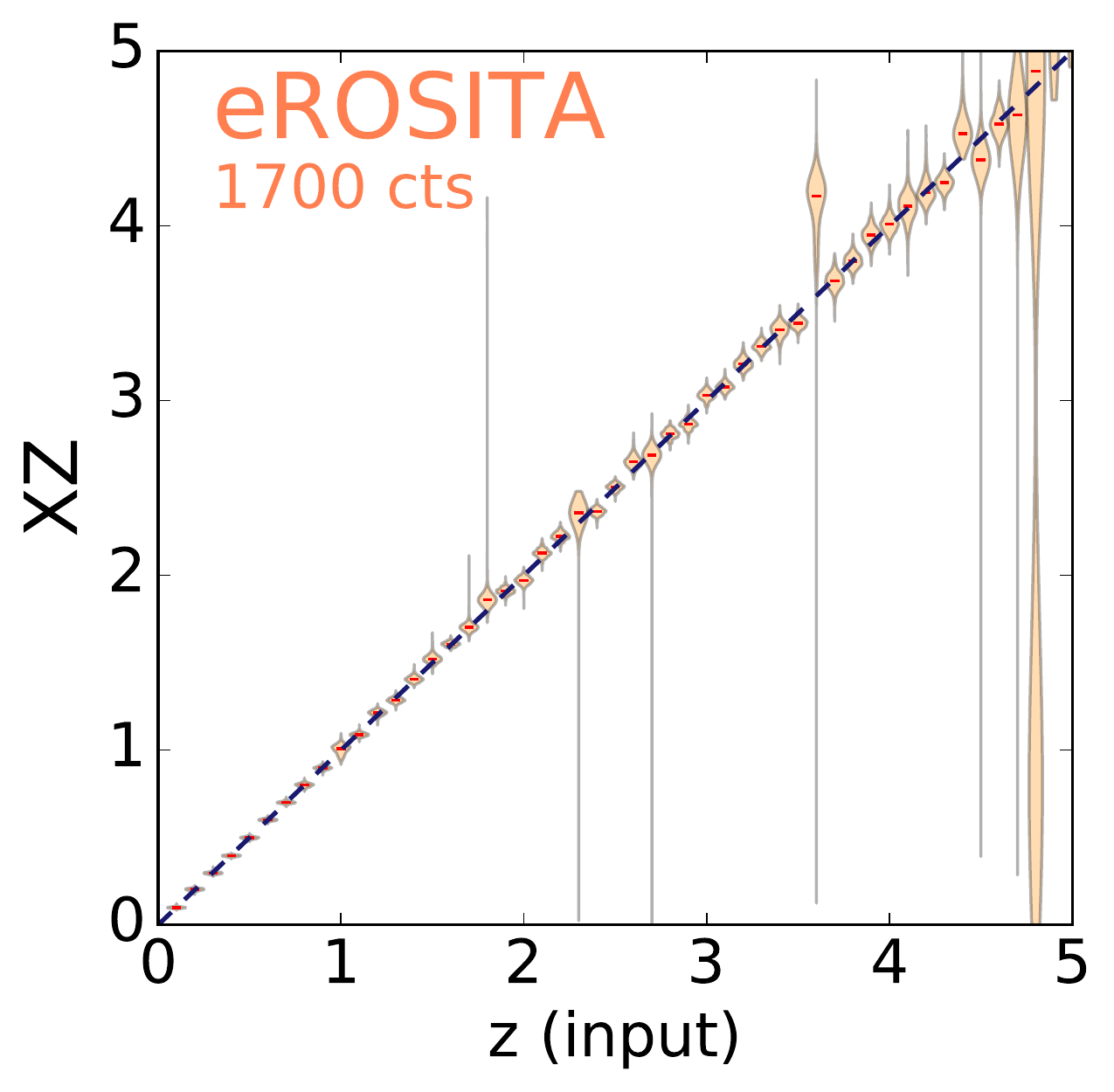}\quad
\includegraphics[trim={0.3cm 1cm 0.3cm 0},clip,width=.4\textwidth]{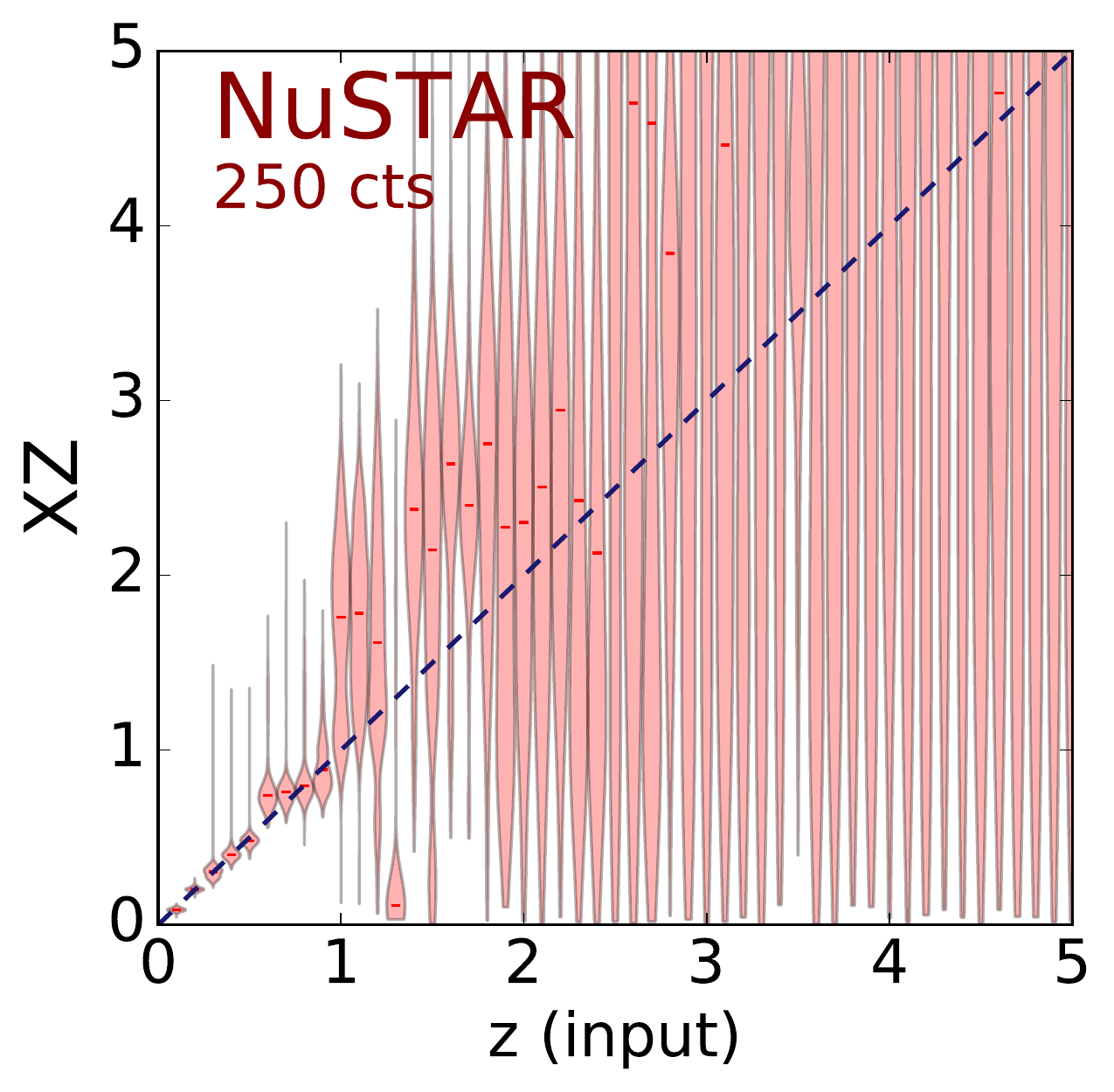}\quad

\includegraphics[trim={0.3cm 0 0.3cm 0},clip,width=.4\textwidth]{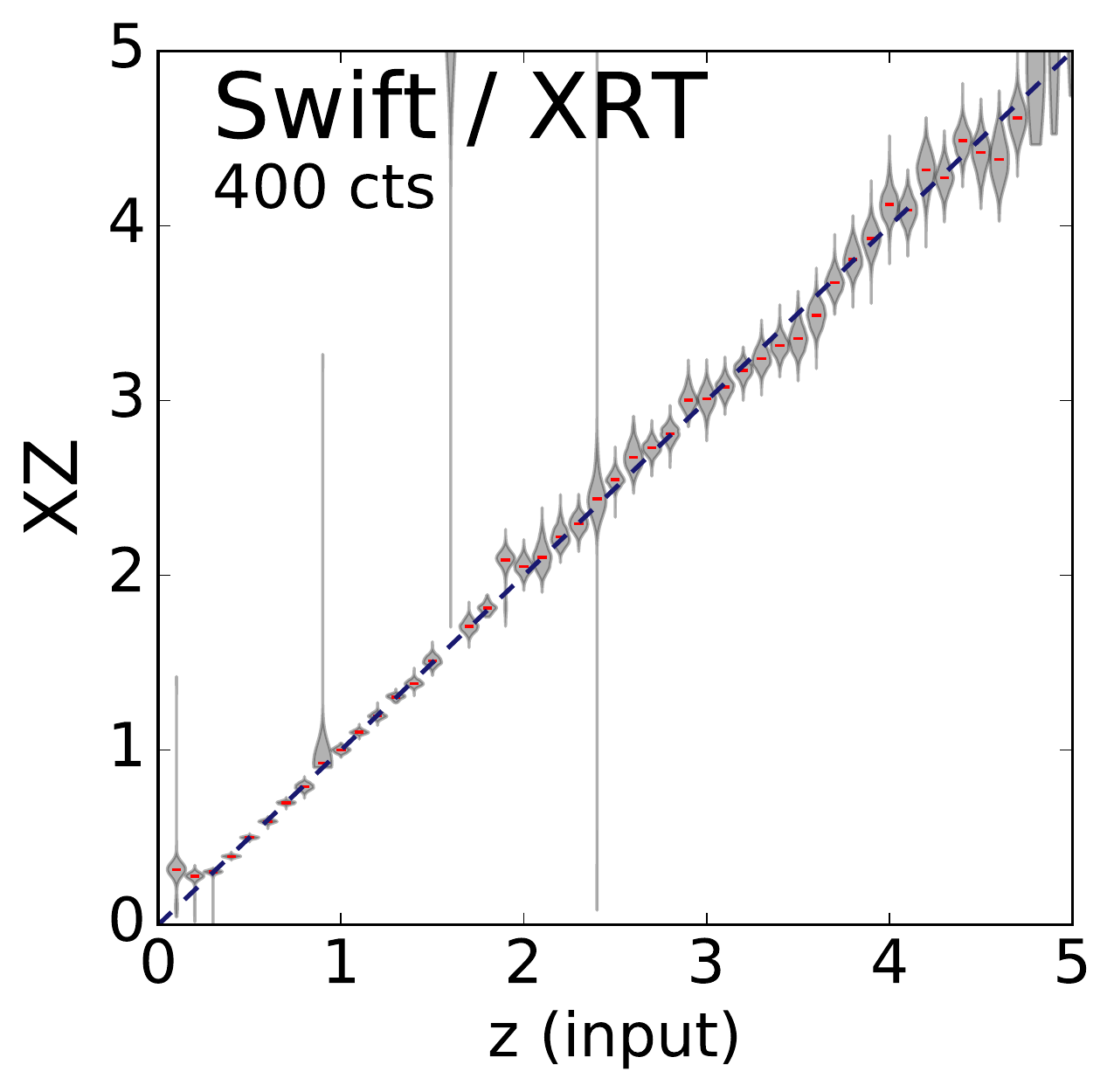}\quad
\includegraphics[trim={0.3cm 0 0.3cm 0},clip,width=.4\textwidth]{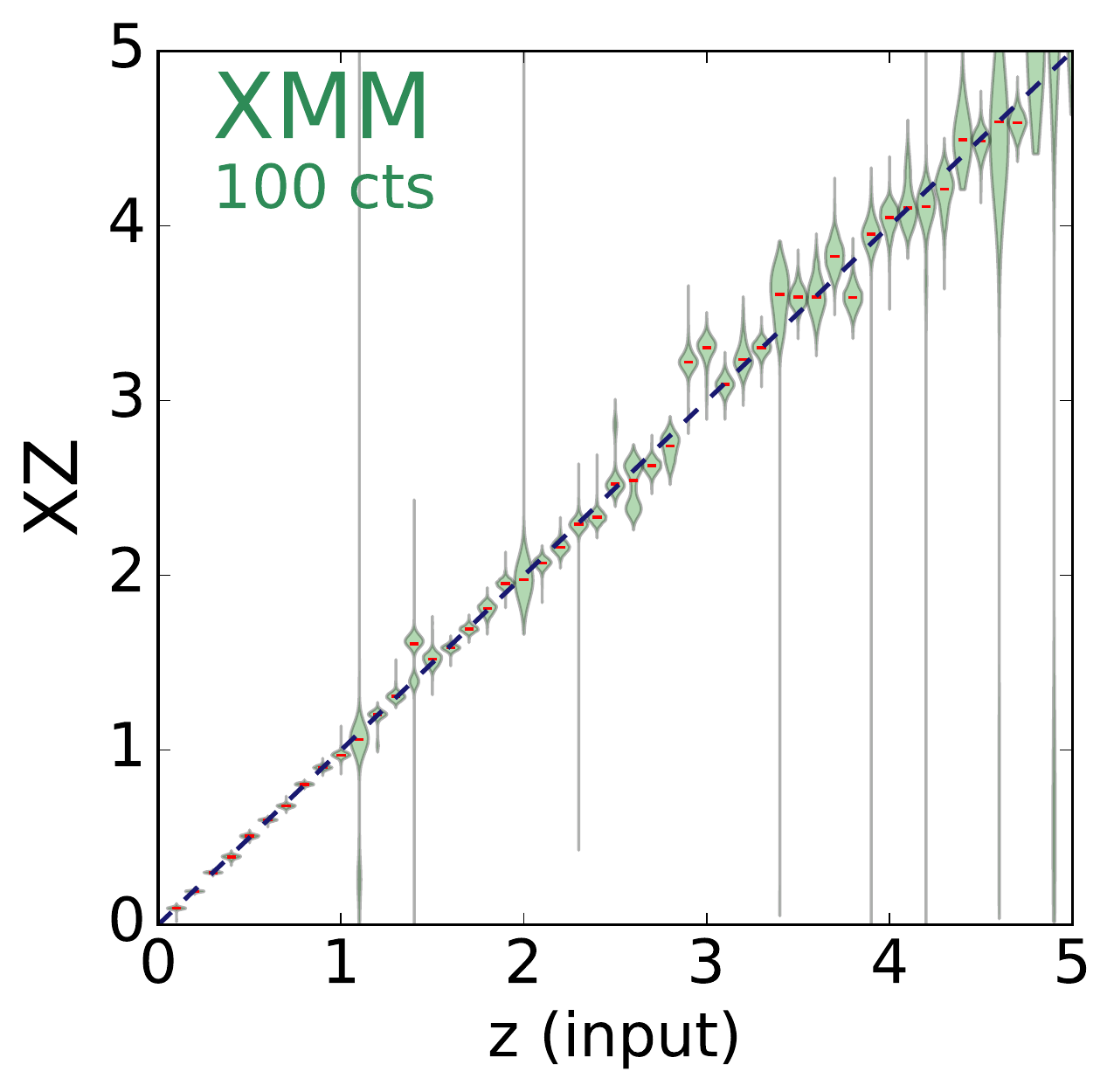}\quad
\caption{Recovery of redshifts by the XZ method. Output XZ redshift probability distributions are shown as violin plots for each input $z$ (x-axis). In all cases \NHvalue{24} is shown. The panels correspond to different instruments, with a total number of source counts chosen at the approximate threshold where the XZ method starts to work optimally.}
\label{fig:violins}
\end{figure*}

\begin{figure*}[htp]
\centering
\includegraphics[trim={0 2.15cm 0 0},clip,width=.89\textwidth]{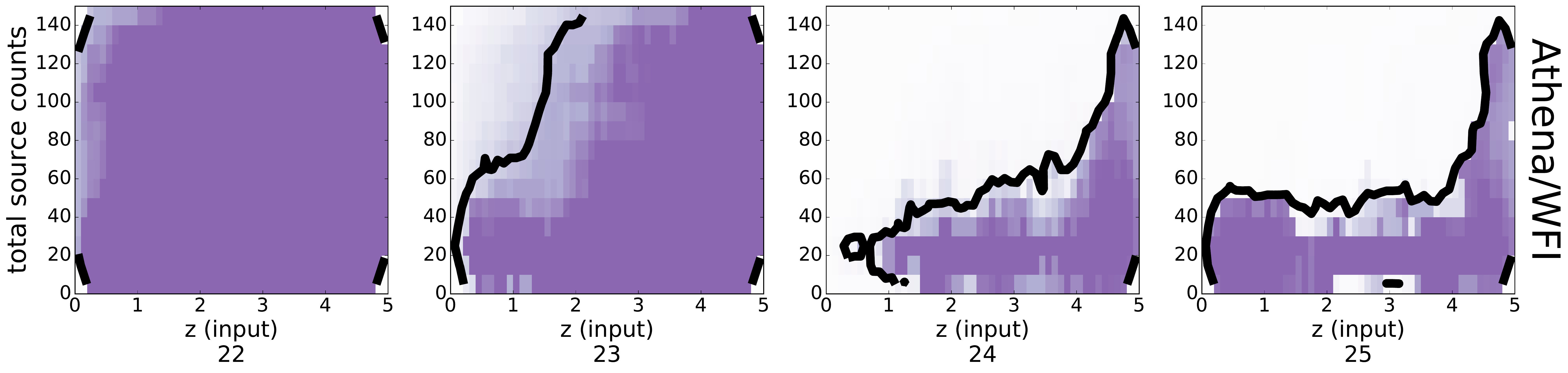}\quad
\includegraphics[trim={0 2.15cm 0 0},clip,width=.89\textwidth]{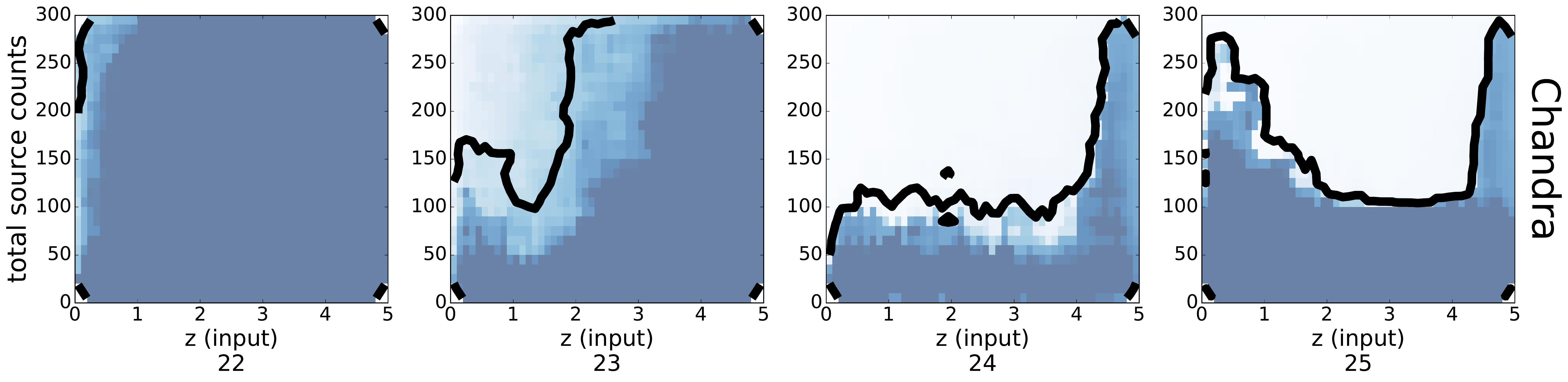}\quad
\includegraphics[trim={0 2.15cm 0 0},clip,width=.89\textwidth]{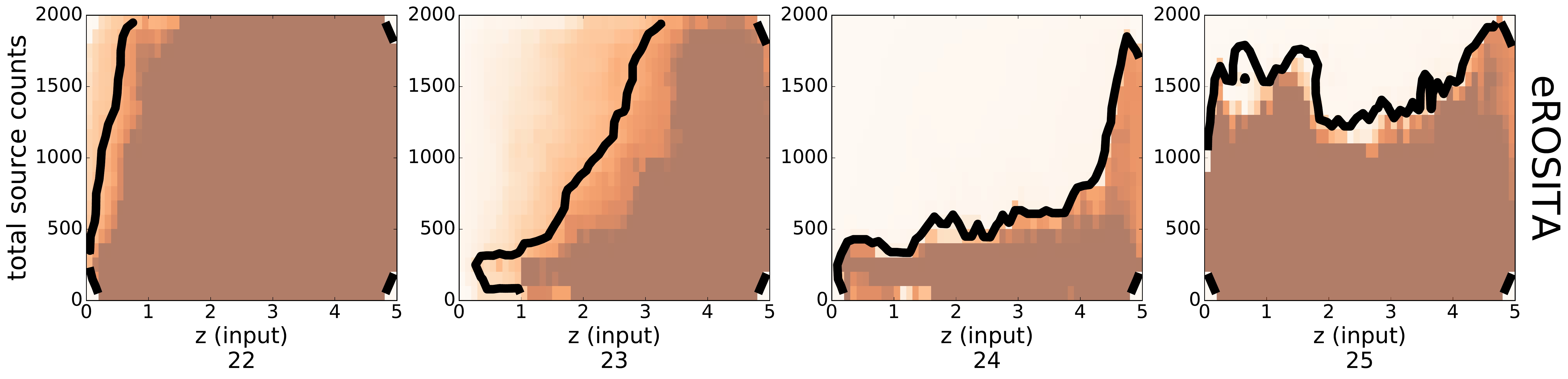}\quad
\includegraphics[trim={0 2.15cm 0 0},clip,width=.89\textwidth]{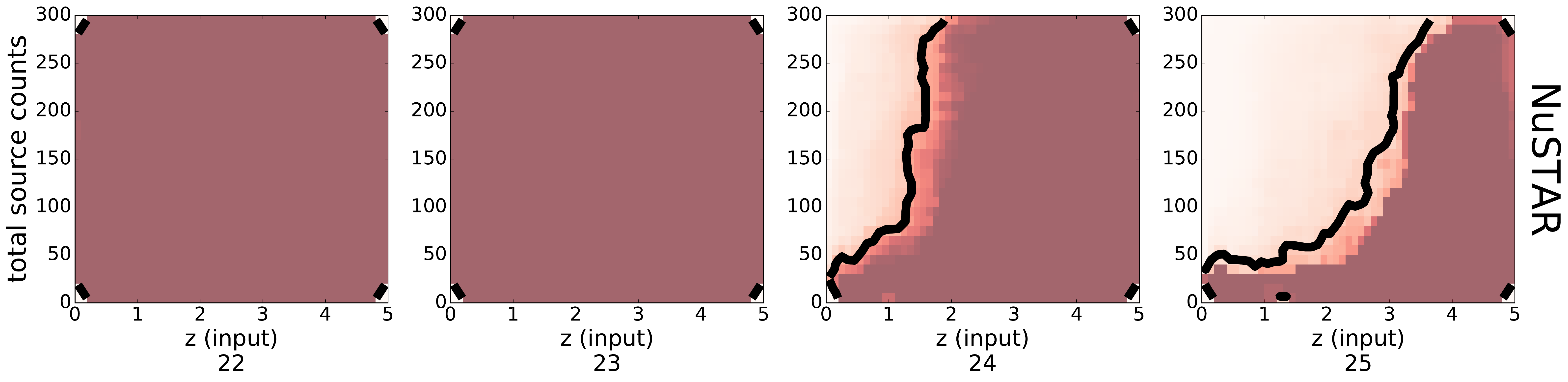}\quad
\includegraphics[trim={0 2.15cm 0 0},clip,width=.89\textwidth]{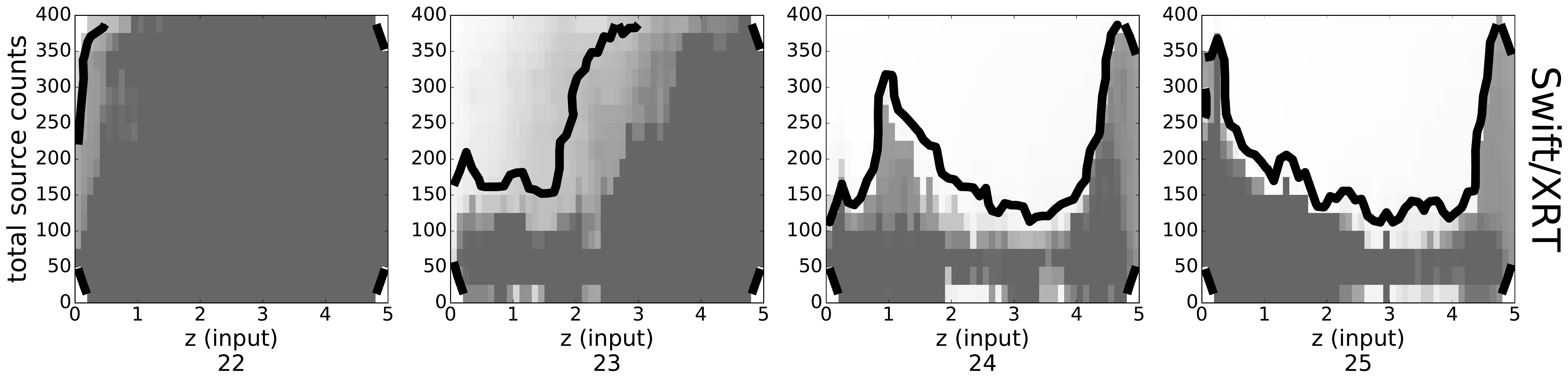}\quad
\includegraphics[width=.89\textwidth]{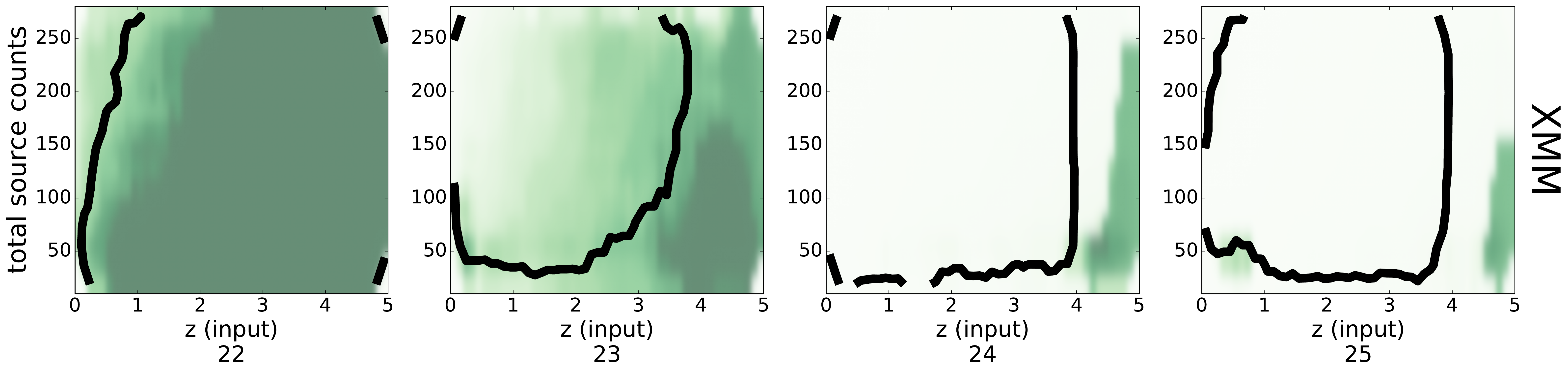}\quad
\caption{Sensitivity heat maps for each instrument. Redshift uncertainty (colors) are plotted vs source redshift (x-axis) and total source counts (y-axis). Above the thick contour lines, IG > 1 $bit$, i.e., the number of counts is sufficient to provide some redshift information. From left to right, the columns show $\log$N$_{\rm H}$ = 22, 23, 24 and 25 cm$^{-2}$. \textsl{XMM} total source counts refer to total counts across PN/MOS1+2. 
}
\label{fig:heat}
\end{figure*}

To answer how many source photon counts are needed for good redshift constraints, we performed extensive realistic simulations for multiple instruments. We generate source and background spectra and fitted them as with the real observations. The recovery is tested across different redshift (z=0-5) and column density regimes ($N_\mathrm{H}=10^{22-25}$ cm$^{-2}$). The simulation parameter grid is listed in Table~\ref{table:simgrid}. 

For each such scenario ($z$, $N_\mathrm{H}$), several instruments are considered, including {\it Chandra, XMM-Newton, Swift, NuSTAR, eROSITA} and {\it Athena}. Importantly, we vary the expected number of source counts to find the relation between number of counts and XZ redshift uncertainty. We also tested different background levels by varying the exposure time, but find that these show small differences (also thanks to the background model fitting), and that $N_\mathrm{H}$ and number of counts remain the most important variables. For e.g., Chandra, we adopted a relatively long exposure time representative of the CDF-S, which gives the worst case of high background levels. 
For each instrument, Table~\ref{table:sims} shows the counts, the exposure time and the total number of simulations created and fitted. 

Simulations were created in {\em{Sherpa}} using the  \texttt{fake\_pha} command. We use the same spectral model for simulation as in the fitting. 
To simulate background spectra, we let \texttt{fake\_pha} resample a representative \texttt{bkg} file for each instrument. For upcoming missions {\it Athena} and {\it eROSITA} we simulate a background following the information provided by the respective mission websites\footnote{\url{http://www.mpe.mpg.de/1341347/eROSITA_background_v12.pdf} for {\it eROSITA}, \url{https://www.cosmos.esa.int/web/athena/resources-by-esa} for {\it Athena}}.
\FloatBarrier 
\subsection{Redshift recovery}
The recovery of input redshifts is presented in Figure~\ref{fig:violins}. The XZ uncertainties are presented for various instruments and redshifts. Here, a total source count number for each instrument representing the approximate threshold where the XZ method starts to be useful was chosen.
Figure~\ref{fig:violins} demonstrates the accuracy and  precision of XZ. The correct retrieval of the redshift(y-axis) at each input redshift (x-axis) is expected because the model used for simulating data is the same as the fitted model. While optimistic, Fig.~\ref{fig:violins}  demonstrates that the parameter space is reliably explored. It also shows that when the X-ray spectrum contains no redshift information, the XZ uncertainties increase dramatically but the results are not incorrect. For instance, at $z>1$ any discriminatory information from the \FeKa line is redshifted out of the {\it NuSTAR} observing window, and the uncertainties increase drastically.

\begin{figure}
\centering
\includegraphics[trim={0.3cm 1cm 0 0},clip,width=.45\textwidth]{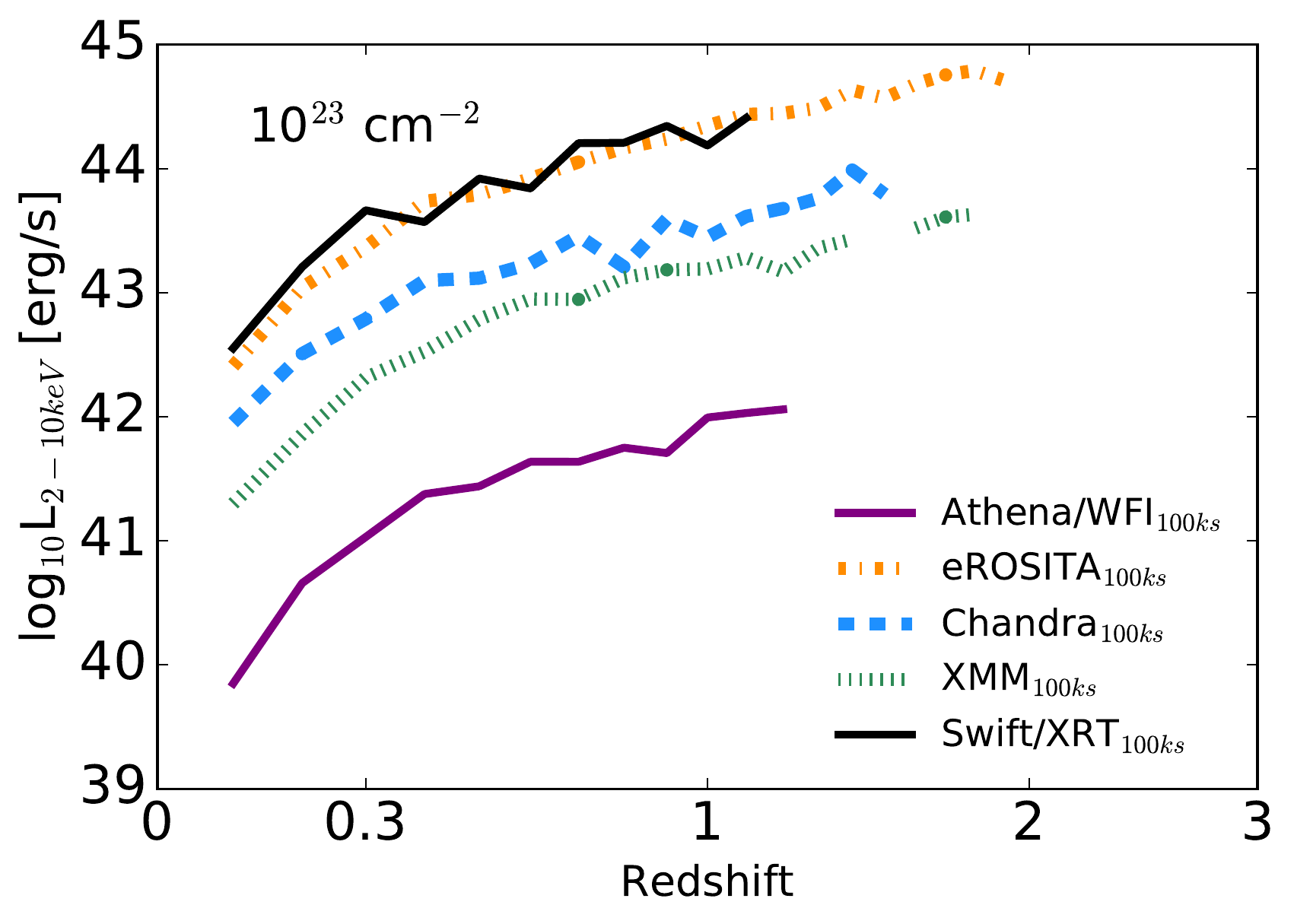}
\includegraphics[trim={0.3cm 1cm 0 0},clip,width=.45\textwidth]{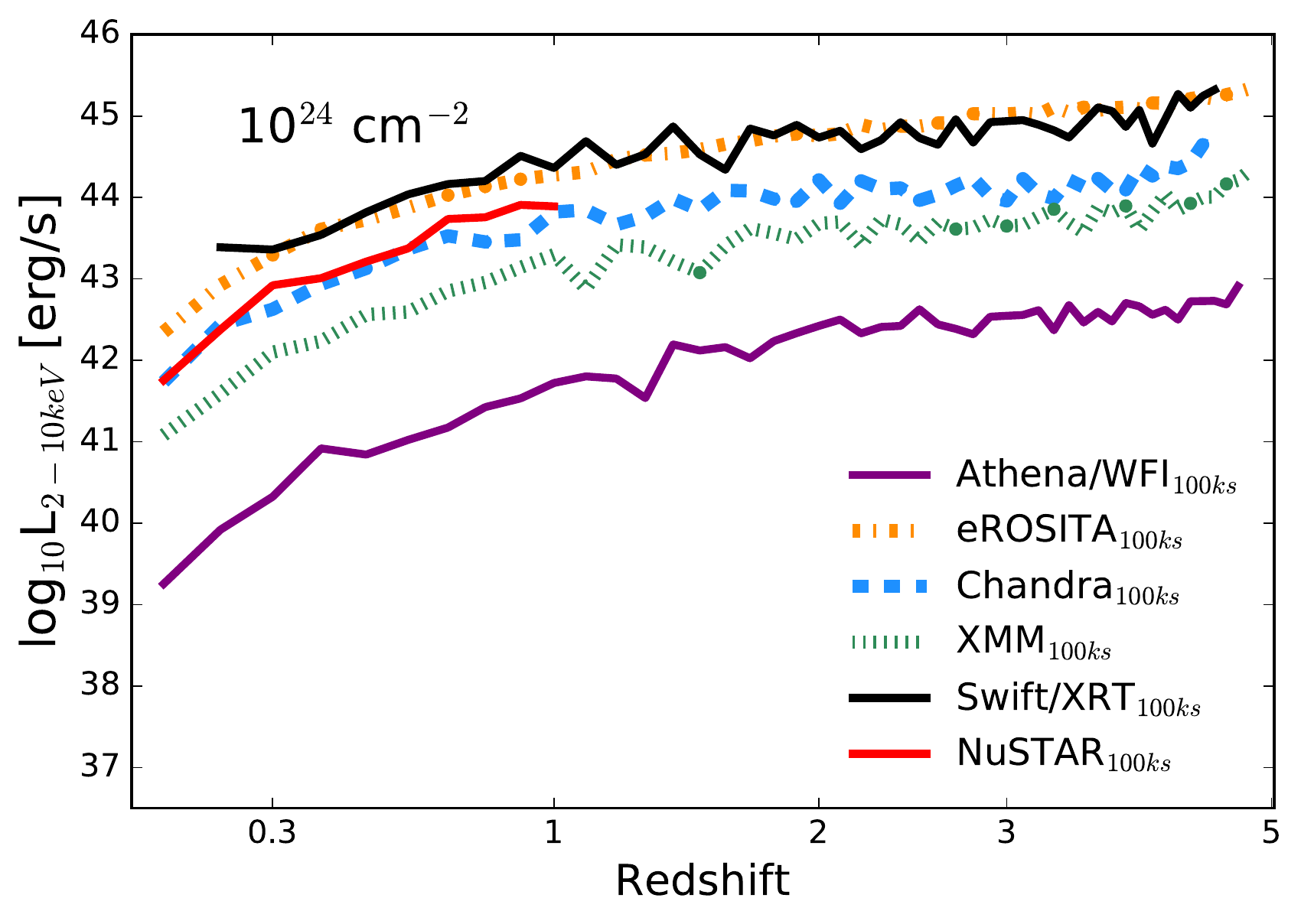}
\includegraphics[trim={0.3cm 0 0 0},clip,width=.45\textwidth]{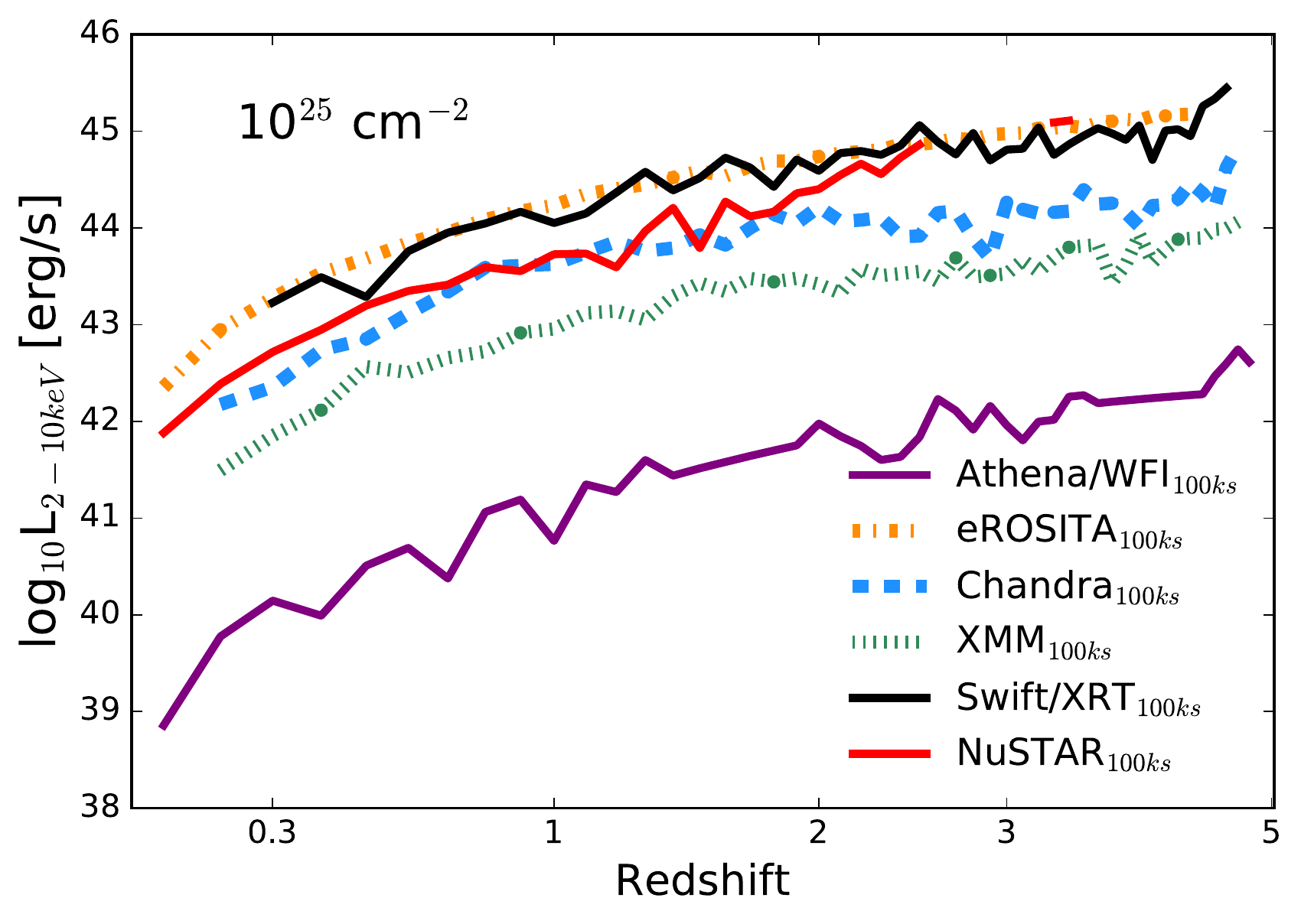}
\caption{2-10 keV luminosity (observed-frame, absorbed) versus redshift plot that shows the sensitivity limit of the XZ method for N$_{\rm H}$ = 10$^{23}$ (top), 10$^{24}$ (middle) and 10$^{25}$ (bottom) cm$^{-2}$. In order to compare the response of XZ for different instruments, we normalize the luminosity values to 100 ks. Both current and future missions are shown, XZ is particularly sensitive to the upcoming mission {\it Athena}.}
\label{fig:lums}
\end{figure}

\subsection{Minimal counts for XZ constraints}
We now investigate the minimum number of counts needed to obtain redshift information (IG$\geq$1). Figure \ref{fig:heat} shows redshift uncertainty (darker: larger errors, white: tightly constrained) as a function of total source counts (y-axis), input source redshift (x-axis), instrument (rows) and column density (columns). The thick black contour line demarks IG=1 bit, i.e., above it (higher counts), some redshift information can be extracted by XZ from the X-ray data alone. Considering {\it Chandra} for instance, we find that the sensitivity of the XZ method depends on both the count number and the neutral hydrogen column density value along the line of sight. In general, the best results are obtained with heavily obscured column densities. The required count number varies from instrument to instrument, from 50 to 500 minimal counts. Compared to {\it Chandra} the number of counts required is higher with {\it eROSITA} and {\it Swift} (both are less sensitive at >4keV). In {\it NuSTAR}, only Compton-thick AGN are usable because the \FeKa line is the primary source of information.

The required number of counts imply minimal exposure times and luminosities for the different instruments. For e.g., {\it Chandra}, we assume a relatively long exposure time (see Table \ref{table:sims}),
which gives the worst case of high background levels. We tested different exposure times, but find that these show small differences (also thanks to the background model fitting), and that N$_{\rm H}$ and number of counts remain the most important variables. To compare the efficiency of the XZ technique for each instrument, we assume 100ks exposures and determine the luminosity limit that can be probed. Figure~\ref{fig:lums} shows this limit as a function redshift for each instrument. The panels correspond to different column densities \NHvalue{23-25}.  At higher luminosities, {\it Swift} and {\it eROSITA} are found to be similarly sensitive. However, the {\it eROSITA} all-sky survey (eRASS) will only reach exposures of several ks \citep{Merloni2012eRositaWhitebook}, thus its luminosity limit will be substantially higher. At these fluxes, only a few AGN are expected across the full sky \citep{Kolodzig2012}, and are likely already identified, e.g., in the WISE all-sky survey \citep{Wright2010}. 

We have also considered applying the XZ method to gamma-ray bursts, which frequently show Compton-thin obscuration up to \NHvalue{23} \citep[e.g.,][]{Buchner2017}. {\it Swift} \citep{Gehrels2004} detects gamma-ray bursts and observes X-ray spectra with XRT. Obtaining redshift constraints is crucial to probe massive star formation in the early Universe. Today, this requires timely follow-up and favorable observing conditions. XZ could improve this process by augmenting photometric redshifts. We tested XZ on a large sample of X-ray afterglow spectra from the {\it Swift} archive \citep[described in][]{Buchner2017}. However, the emission in these sources is generally too short to accumulate the necessary counts for XZ to provide constraints with {\it Swift/XRT}. For some ultra-long gamma-ray bursts the counts are sufficient, but such sources present unusual spectral features that are poorly understood and likely arise from spectral variability \citep{Piro2014,Evans2014}. In any case, these sources are a small sub-population of all gamma-ray bursts. An observatory with higher collecting area is required to obtain XZ constraints for gamma-ray bursts.

The best performance is seen with the next-generation {\it Athena} mission \citep{Nandra2013}, which will feature a mirror with large effective area ($2 \mathrm{m}^2$), wide energy coverage (0.3-12 keV) and good resolution ($<150\mathrm{eV}$) even for the Wide-field Imager \citep[WFI;][]{Meidinger2017}. With this mission, surveys as deep as the CDF-S 4Ms can be performed twenty times faster \citep{Rau2016}, and cover larger areas ($\sim$100$\mathrm{deg}^2$), allowing a deep and complete census of the AGN population at high redshift. 
Previously, \cite{Castello-Mor2011} developed a redshift extraction technique based on finding the \FeKa line in Fourier-transformed spectra. In 100ks exposures, this succeeds in unobscured AGN with luminosities of ${10}^{44}\mathrm{erg/s}$ at $z=1$. In Figure~\ref{fig:lums} we estimate that XZ can measure redshift of obscured AGN, which constitute the bulk of the population in deep observations, down to luminosities as low as ${10}^{41}\mathrm{erg/s}$.
This indicates that for a substantial fraction of sources detected in such a survey, expensive follow-up may be unnecessary, as redshift information can be extracted directly from {\it Athena}/WFI X-ray spectra.

\section{Conclusions}

Deep X-ray fields probe accretion in the early Universe, and are dominated by obscured AGN. The deep observations require time-consuming follow-up campaigns to obtain reliable redshift estimates. In this regime we demonstrated that for a sizable fraction of sources, the X-ray spectrum contains substantial redshift information in the absorption edges and, if Compton-thick, the \FeKa fluorescent line. 
Through simulations, we characterised in detail the obscuration level and number of counts required to find constrained and reliable solutions. Failure to meet such conditions does however not lead to wrong results, only large uncertainties.

Sidestepping association problems, our XZ method is promising for estimating redshifts or confirming/reinforcing photoz and specz estimates. The XZ estimate is easy to apply and does not require additional data. In addition to direct redshift estimates in some high-count sources, it can discriminate between multiple photoz solutions in a large fraction of obscured AGN in todays deep surveys. We demonstrate that disagreements of the independent XZ estimate with photoz/specz estimates can help to flag problematic measurements, e.g., due to incorrectly identified spectroscopic lines, contaminated photometric data due to blending or variability or ambiguous counterparts. With high spectral resolution and large collecting area, this technique will be highly effective for {\em{Athena+}/WFI} observations.

\begin{acknowledgements} 
    We acknowledge support from the CONICYT-Chile grants Basal-CATA PFB-06/2007 (JB, FEB), FONDECYT Regular 1141218 (FEB), FONDECYT Postdoctorados 3160439 (JB), “EMBIGGEN” Anillo ACT1101 (FEB), and the Ministry of Economy, Development, and Tourism's Millennium Science Initiative through grant IC120009, awarded to The Millennium Institute of Astrophysics, MAS (JB, FEB). This research was supported by the DFG cluster of excellence “Origin and Structure of the Universe”.
\end{acknowledgements}

\bibliographystyle{aa}
\bibliography{xz}

\newpage
\clearpage

\begin{appendix}

\section{Automatic Background Fitting}\label{autobackground}
This section describes a new technique, BStat, for characterising spectra of background regions. With principal component analysis (PCA) this technique learns the typical background shape and variations from a large sample. The technique is applied for each telescope and detector separately and forms a general solution to empirical background modelling.

\begin{figure*}[p]
  \centering
  \includegraphics[width=0.99\textwidth]{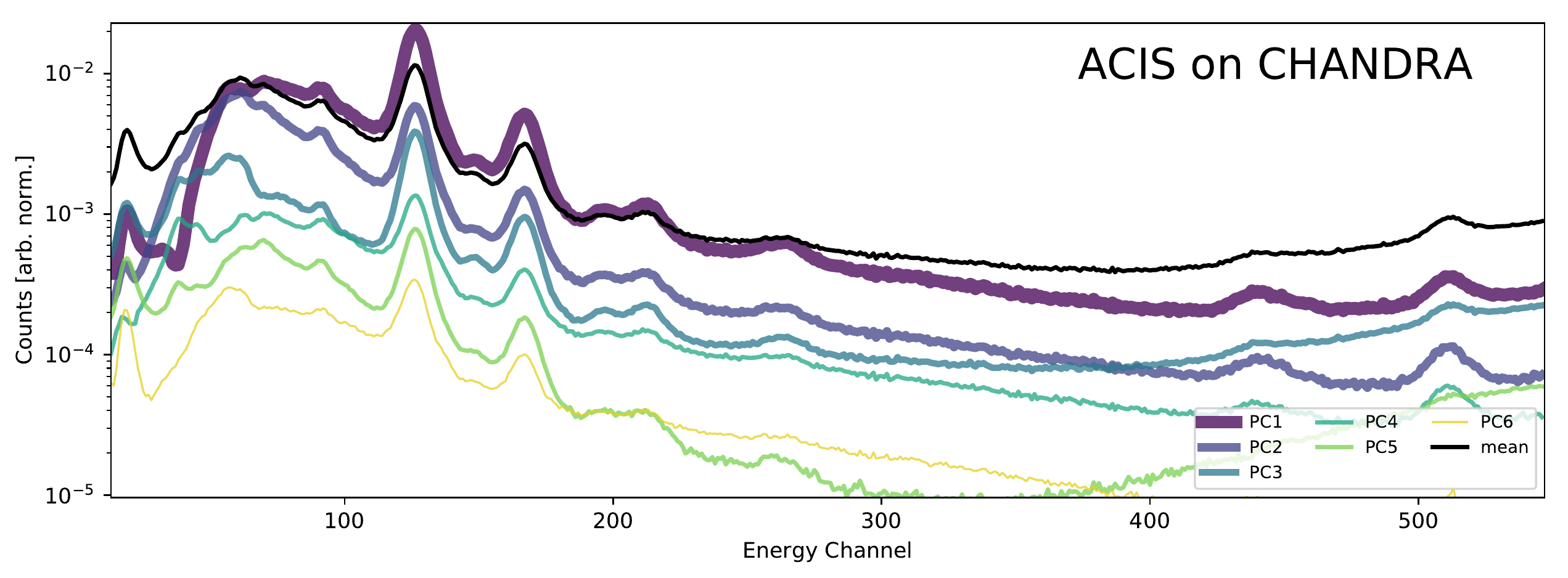}
  \caption{\textit{Chandra}/ACIS background PCA components. The average spectrum plus the selected principle component is plotted, in arbitrary units. All CCDs have been combined as they show broadly similar background shapes.}
  \label{fig:bkgchandra}
\end{figure*}
\begin{figure*}[p]
  \centering
  \includegraphics[width=0.99\textwidth]{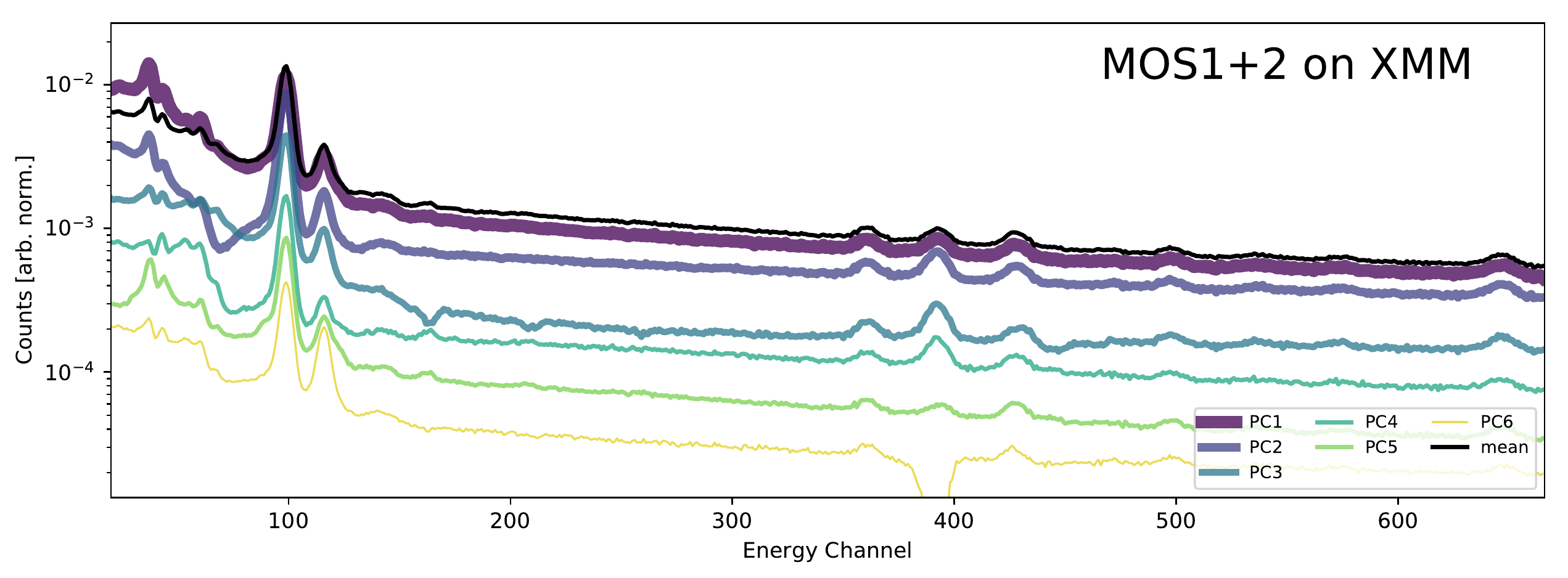}
  \caption{\textit{XMM}/MOS background PCA components. MOS1 and MOS2 have been combined as they show similar background shapes.}
  \label{fig:bkgxmmmos}
\end{figure*}
\begin{figure*}[p]
  \centering
  \includegraphics[width=0.99\textwidth]{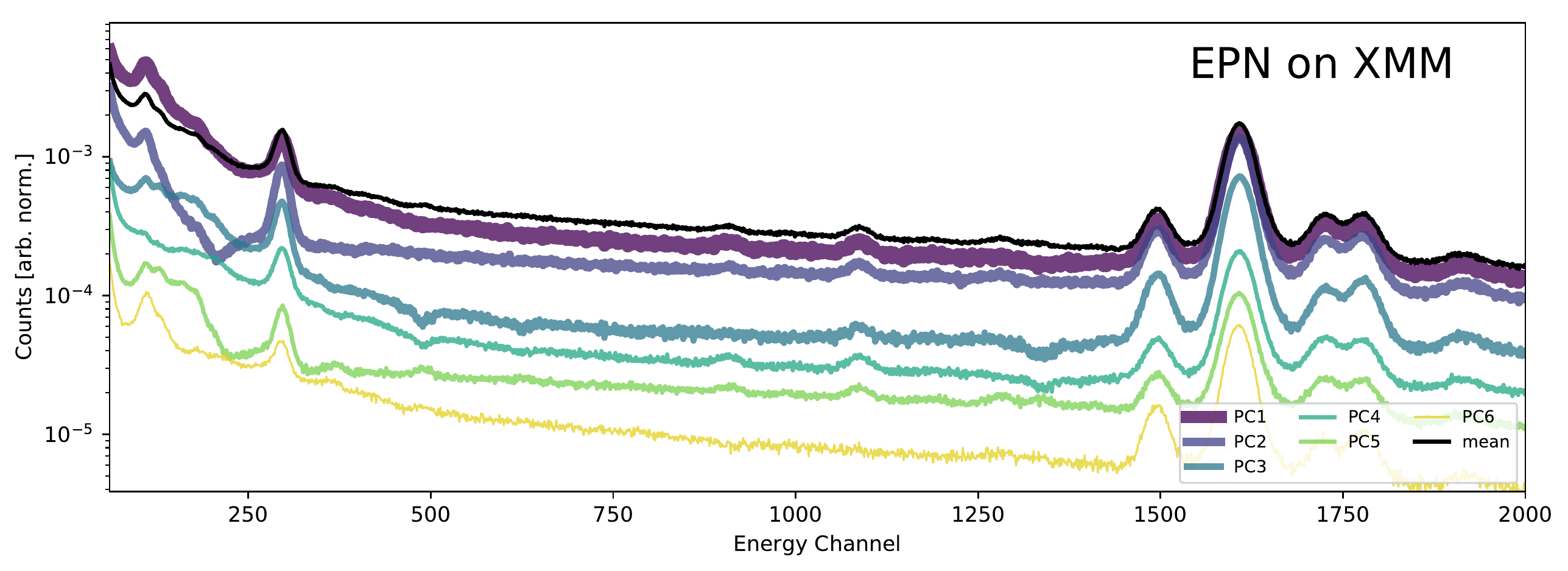}
  \caption{{\it XMM}/PN background PCA components.}
  \label{fig:bkgxmmpn}
\end{figure*}
\begin{figure*}[p]
  \centering
  \includegraphics[width=0.99\textwidth]{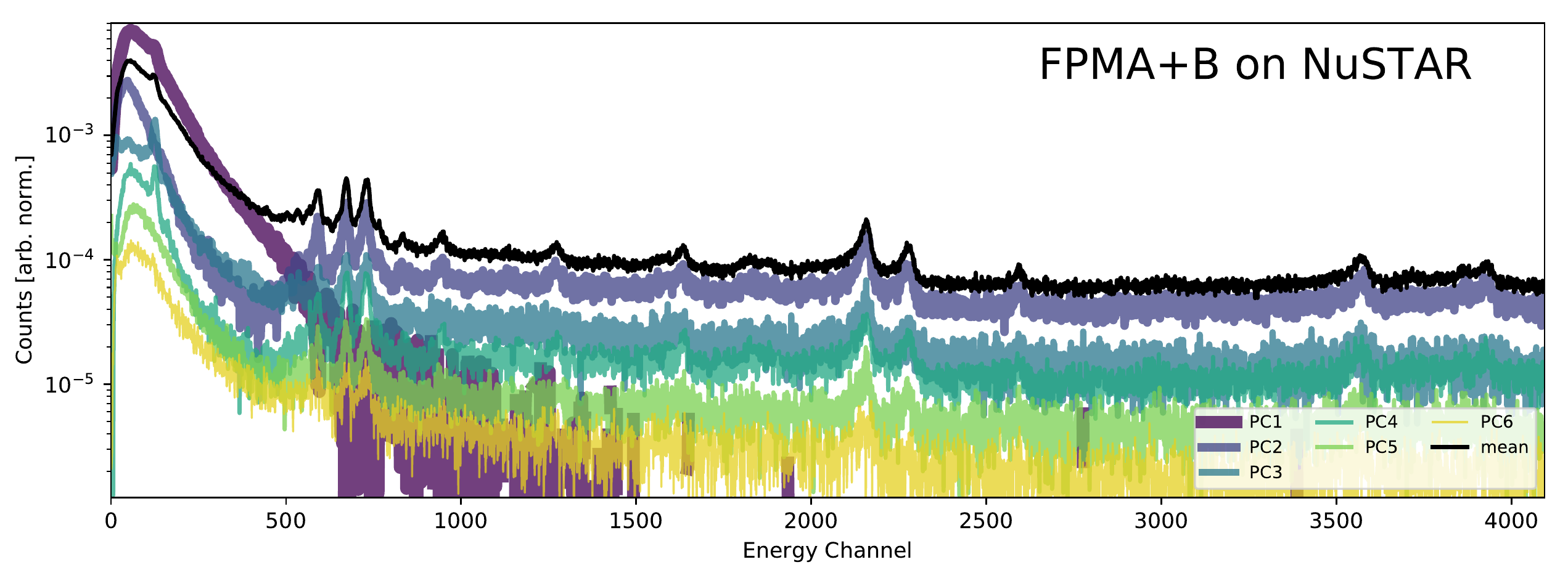}
  \caption{{\it NuSTAR} background PCA components. The FPMA and FPMB detectors have been combined as they show similar background shapes.}
  \label{fig:bkgnustar}
\end{figure*}
\begin{figure*}[p]
  \centering
  \includegraphics[width=0.99\textwidth]{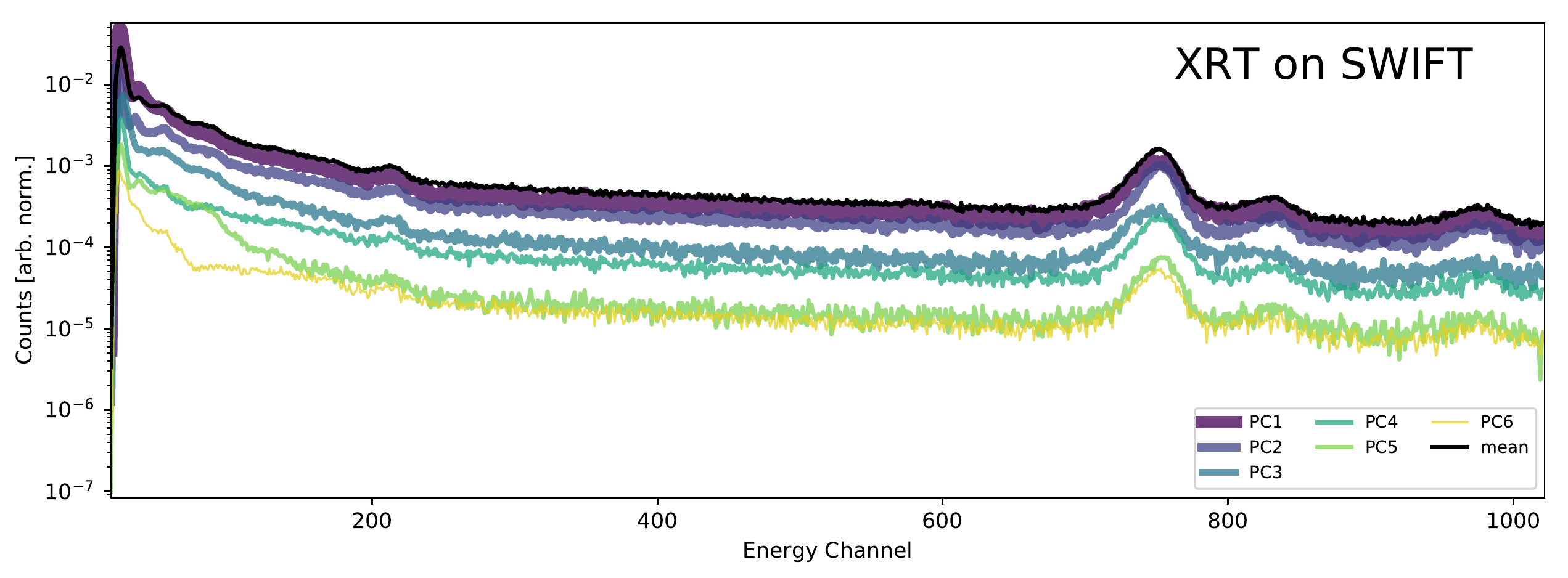}
  \caption{{\it Swift}/XRT background PCA components. WT and PC mode observations have been combined as they show similar background shapes.}
  \label{fig:bkgswiftxrt}
\end{figure*}

To obtain training data, we download background spectra for {\it Chandra} ACIS and {\it XMM-Newton} PN\&MOS from the Xassist website \url{http://xassist.pha.jhu.edu/} \citep{Ptak2003}. We extracted {\it NuSTAR} point source spectra with corresponding backgrounds at the locations of the {\it Chandra} hard-band detected sources in the COSMOS, CDF-S and AEGIS-XD surveys handled in this paper using the standard {\it NuSTAR} pipeline. Most of these sources are undetected. We further obtained {\it Swift} XRT background spectra from http://swift.ac.uk/ \citep[][]{Buchner2017} and the {\it Swift}/BAT survey \citep[][]{Ricci2017}. This yielded a sample of 1895 ({\it Chandra}/ACIS), 575 ({\it NuSTAR}), 414 ({\it Swift}/XRT), 1859 ({\it XMM}/PN) and 1982 ({\it XMM}/MOS) background spectra.

PCA identifies the eigenvectors (in this case, background components) as primary axes of variance (background diversity). We stacked the background spectra so that each background spectrum has at least 10000 total counts. This avoids non-Gaussianity in the Poisson counts. On this sample we apply PCA and extract the mean and the 20 most important features (background spectra vectors). Figures~\ref{fig:bkgchandra} ({\it Chandra}), \ref{fig:bkgxmmmos} \& \ref{fig:bkgxmmpn} ({\it XMM} MOS \& PN), \ref{fig:bkgswiftxrt} ({\it Swift}/XRT) and \ref{fig:bkgnustar} ({\it NuSTAR}) illustrate some examples of these components.

We then use in \texttt{Sherpa} these components to fit the background. Our fitting procedure starts by having all component normalisations free and performing a normal simplex fit to the background spectrum. Then, the least important component is disabled (normalisation set to zero) and another fit is done. This is repeated until only the mean background component remains. For each of these fits we note the Akaike Information criterion \citep[AIC; in this case the the C statistic plus twice the number of fit parameters;][]{AIC,Buchner2014}, and choose the fit with the highest AIC. For fit robustness against local minima, we try to increase the number of components again from there until the AIC does not increase further. Choosing the fit with the highest AIC selects only the $K$ most important components where the background spectrum has enough information to justify activating these components, while the remainder of the spectrum is aligned with the typical (based on the training sample) background spectrum for this detector. Using the AIC thus avoids over-fitting.

Throughout we work with background counts without convolving through the response, as some background components originate on the detectors and this is an empirical model. Working in linear space (on the counts) would be ideal because some background components are additive. However, this can lead to negative backgrounds when we fit. Instead, we transform into logarithmic space with $\log$(counts+1). This has the benefit that all possible combinations of background spectral features are positive counts, because in linear space the (positive) spectral components are multiplied together. However, a drawback is that Gaussian lines of varying strength can not be characterised well. We address this by adding Gaussian lines on top when needed (i.e., again as long as justified by an increase in the AIC). Overall this fitting procedure works robustly and captures the background spectrum well. 

When fitting a science target we add the background model to the source model. We scale the background contribution proportionally to the source and background region sizes. In the source fitting we keep the background shape fixed. However, we allow the background normalisation to vary freely and simultaneously fit the source and background spectra. Freezing also the background normalisation sometimes leads to overly small uncertainties, in particular when few background counts have been detected. 

Our technique decreases the fit uncertainties compared to, e.g., \textit{WStat}, the xspec default behaviour. \textit{WStat} does per-bin estimates of the background, i.e., does not assume any continuity between spectral bins. Additionally, zero or low-count count bins produce biases \citep[e.g.,][]{Willis2005} and require rebinning. In contrast, our technique always works with continous background spectral models across bins, and can handle even zero-count bins. Both in our approach and when using custom-built background models, source fit uncertainties are smaller than in \textit{WStat}, because prior knowledge about typical background shapes is exploited. Our machine-learning approach however does this without requiring the user to hand-craft a model.

Our automatic background fitting technique is released as part of BXA \citep{Buchner2014}. We thank Alex Markowitz and Claudio Ricci for providing example background spectra for XRTE, {\it Suzaku} and {\it Swift}/XRT.


\section{COSMOS and AEGIS-XD results}
\label{moretables}

Similar to Table~\ref{table:long} for CDF-S sources, we present XZ constraints catalogues for the COSMOS and AEGIS fields in Table~\ref{table:COSMOStable} and Table~\ref{table:AEGIS}, respectively. The data products used are described in  \cite{Brightman2013} and \cite{Buchner2015}.

\begin{figure}
\centering
\includegraphics[width=0.5\textwidth,origin=l]{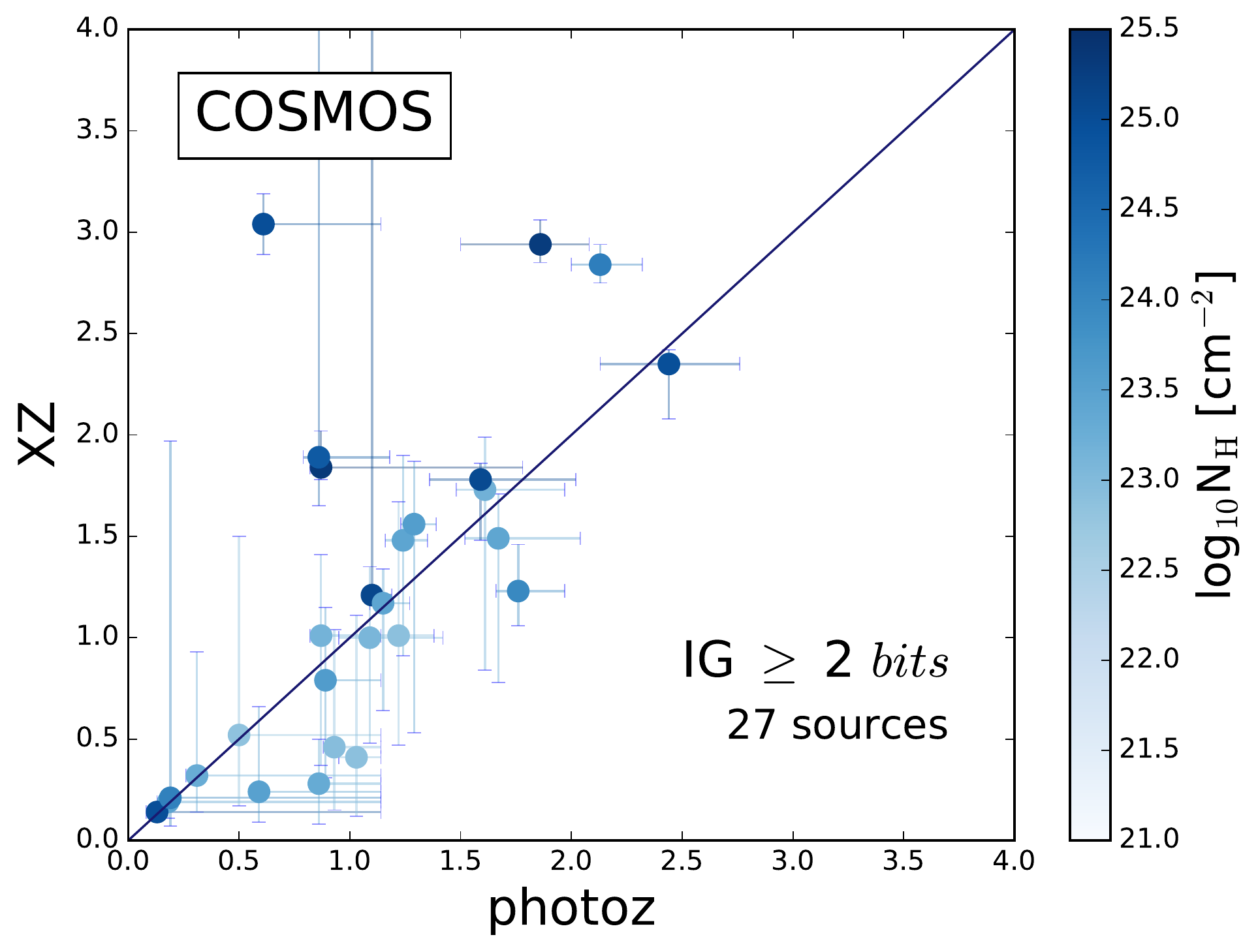}
\caption{Plot of best-fit XZ values versus photoz from \cite{Salvato2009}.}
   \label{fig:cosmosIG}
\end{figure}

\begin{figure}
\centering
\includegraphics[width=0.5\textwidth,origin=l]{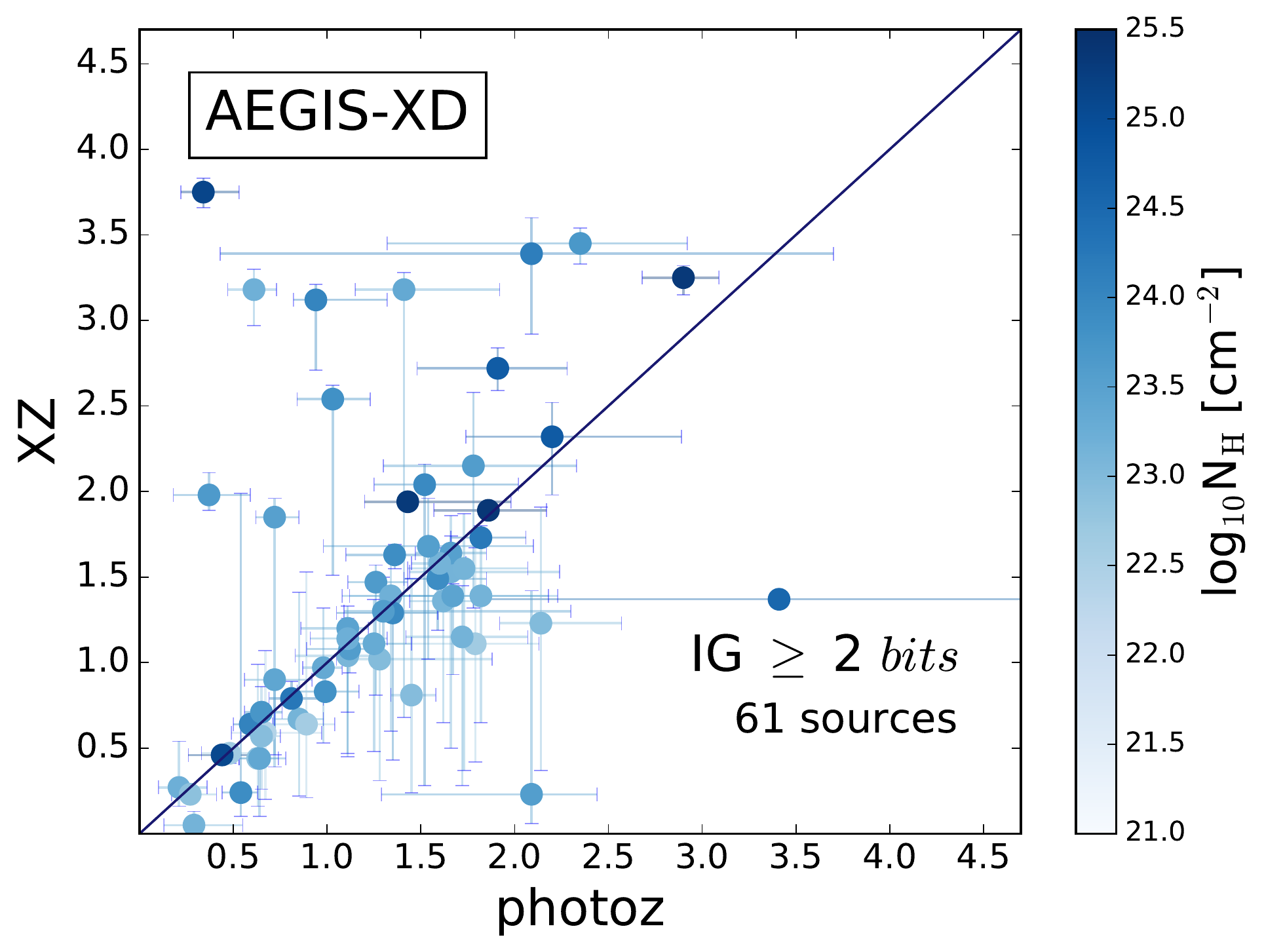}
\caption{Plot of best-fit XZ values versus photoz from \citet{Nandra2015}.}
   \label{fig:aegisIG}
\end{figure}

Figures~\ref{fig:cosmosIG} and \ref{fig:aegisIG} present the performance of COSMOS and AEGIS samples, respectively, where XZ constrained the redshifts. AEGIS, a relatively deep yet wide field has the largest number of constrained redshifts across the three fields.

\begin{figure}
\centering
\includegraphics[width=0.5\textwidth,origin=l]{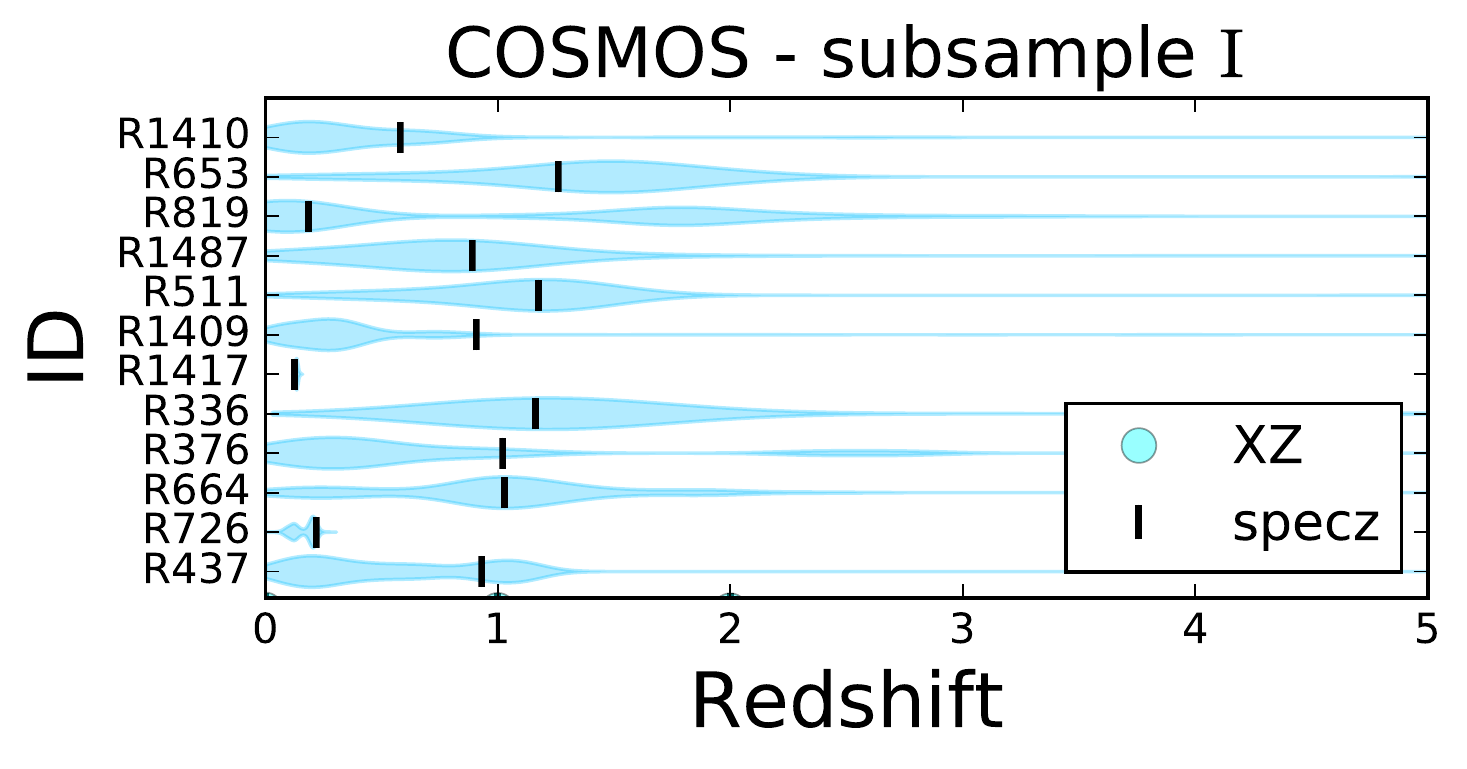}
\caption{XZ (cyan violin plots) and specz (black vertical lines) probability distributions for subsample I (IG $\geq$ 2 $bits$).}
   \label{fig:cosmosI}
\end{figure}

\begin{figure}
\centering
\includegraphics[width=0.5\textwidth,origin=l]{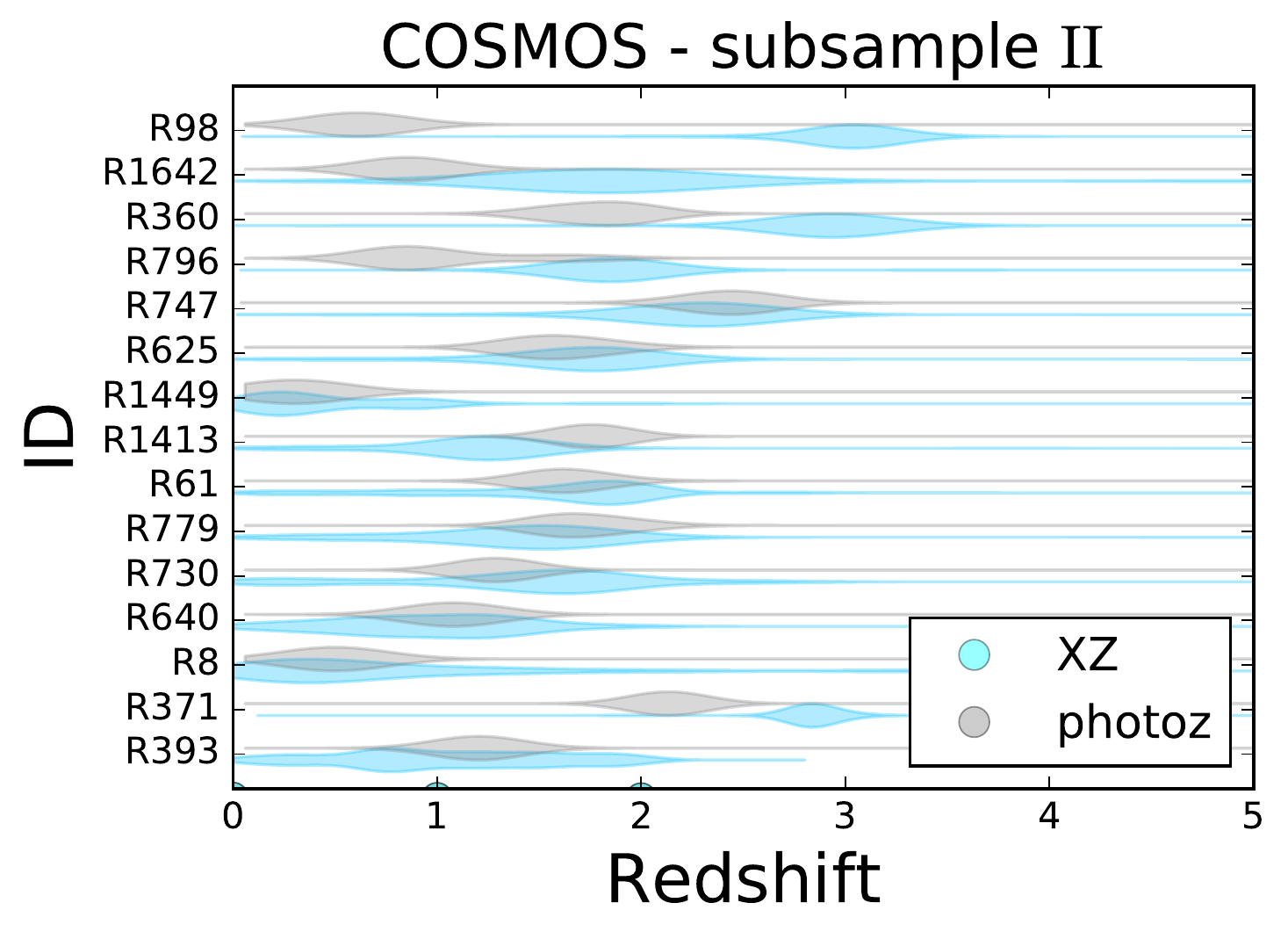}
\caption{XZ (cyan violin plots) and photoz (gray violin plots) probability distributions for subsample II (IG $\geq$ 2 $bits$).}
   \label{fig:cosmosII}
\end{figure}

\begin{figure}
\centering
\includegraphics[width=0.5\textwidth,origin=l]{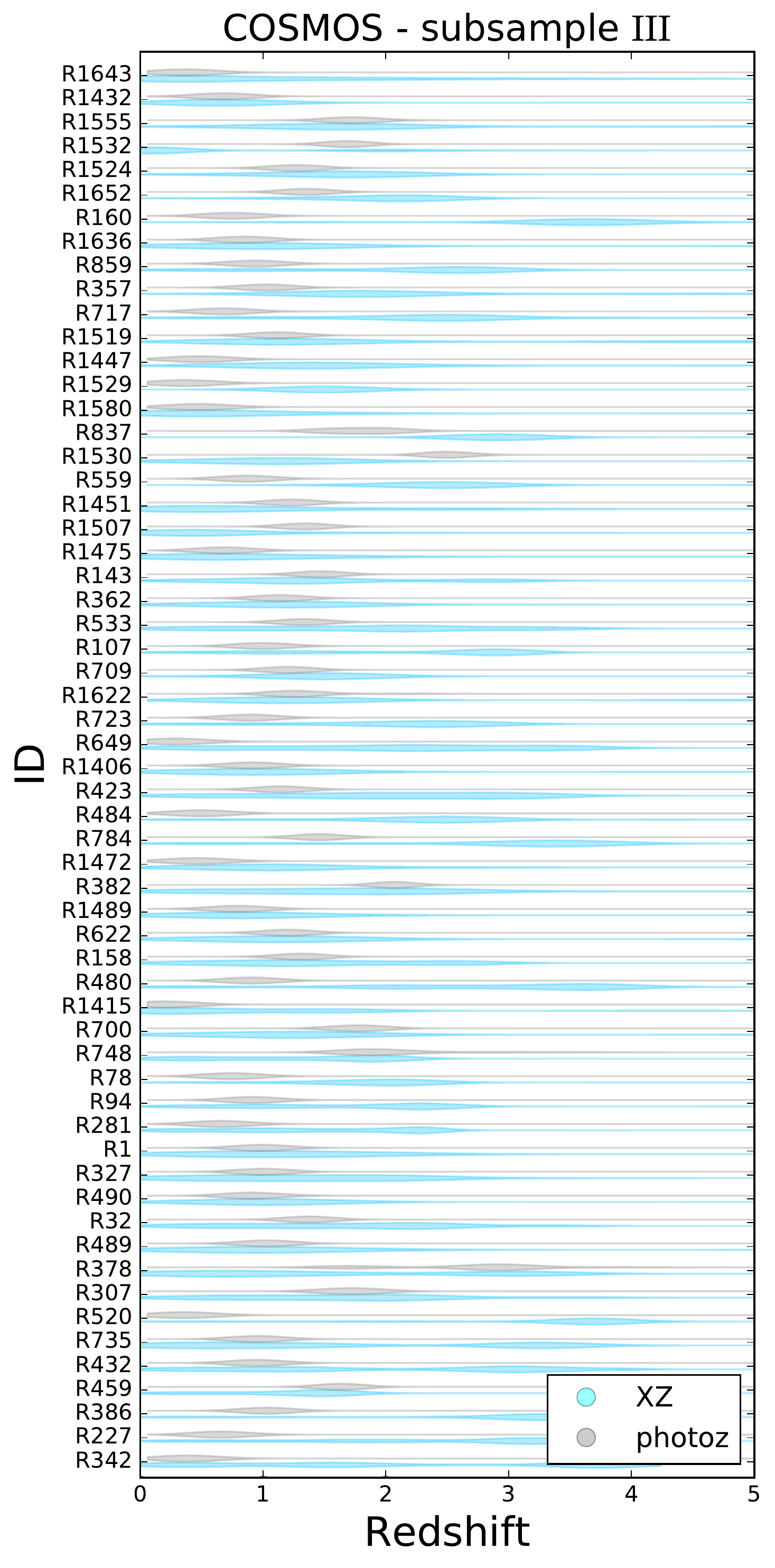}
\caption{XZ (cyan violin plots) and photoz (gray violin plots) probability distributions for subsample III (IG $\geq$ 2 $bits$).}
   \label{fig:cosmosIII}
\end{figure}

\begin{figure}
\centering
\includegraphics[width=0.5\textwidth,origin=l]{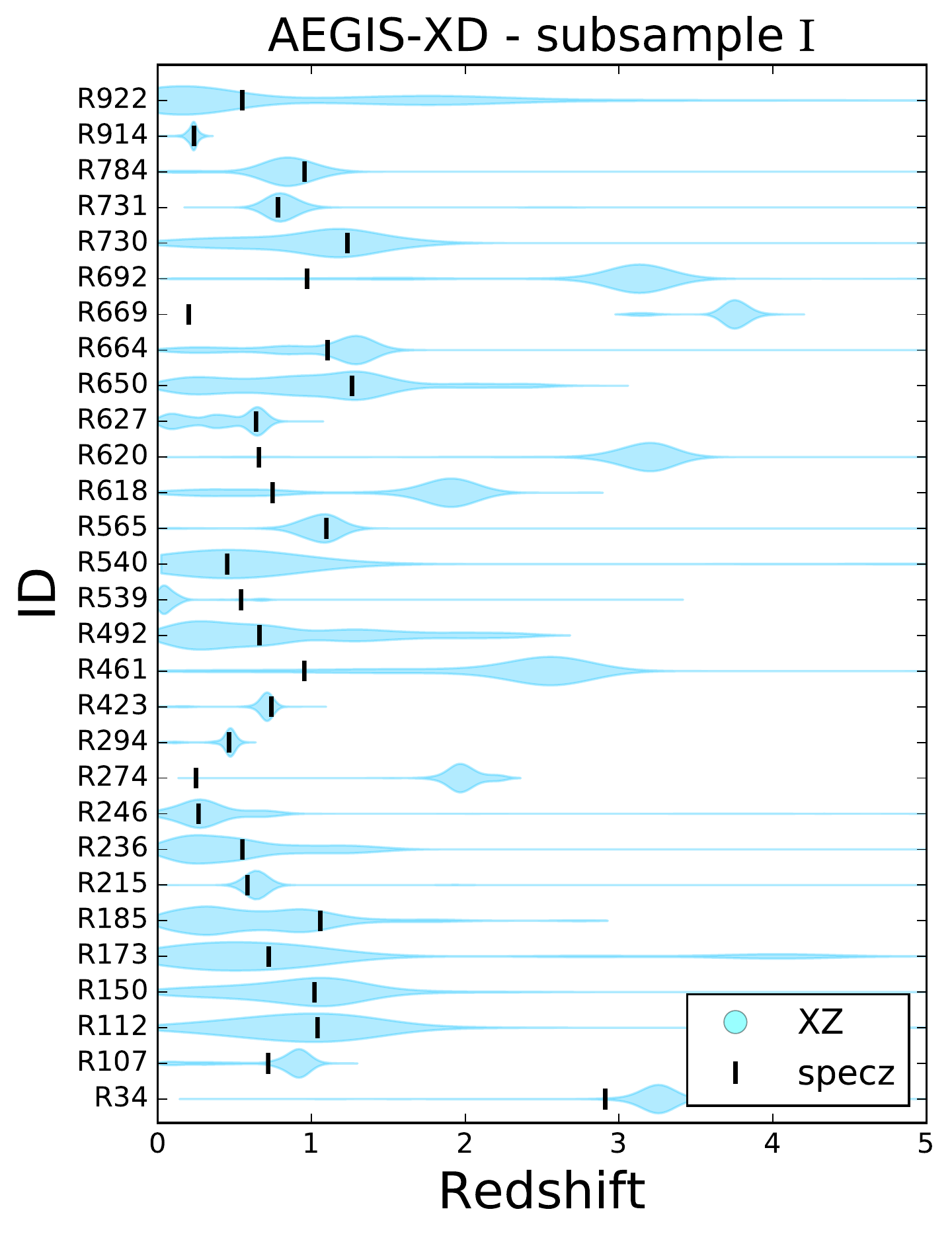}
\caption{XZ (cyan violin plots) and specz (black vertical lines) probability distributions for subsample I (IG $\geq$ 2 $bits$).}
   \label{fig:aegisI}
\end{figure}

\begin{figure}
\centering
\includegraphics[width=0.5\textwidth,origin=l]{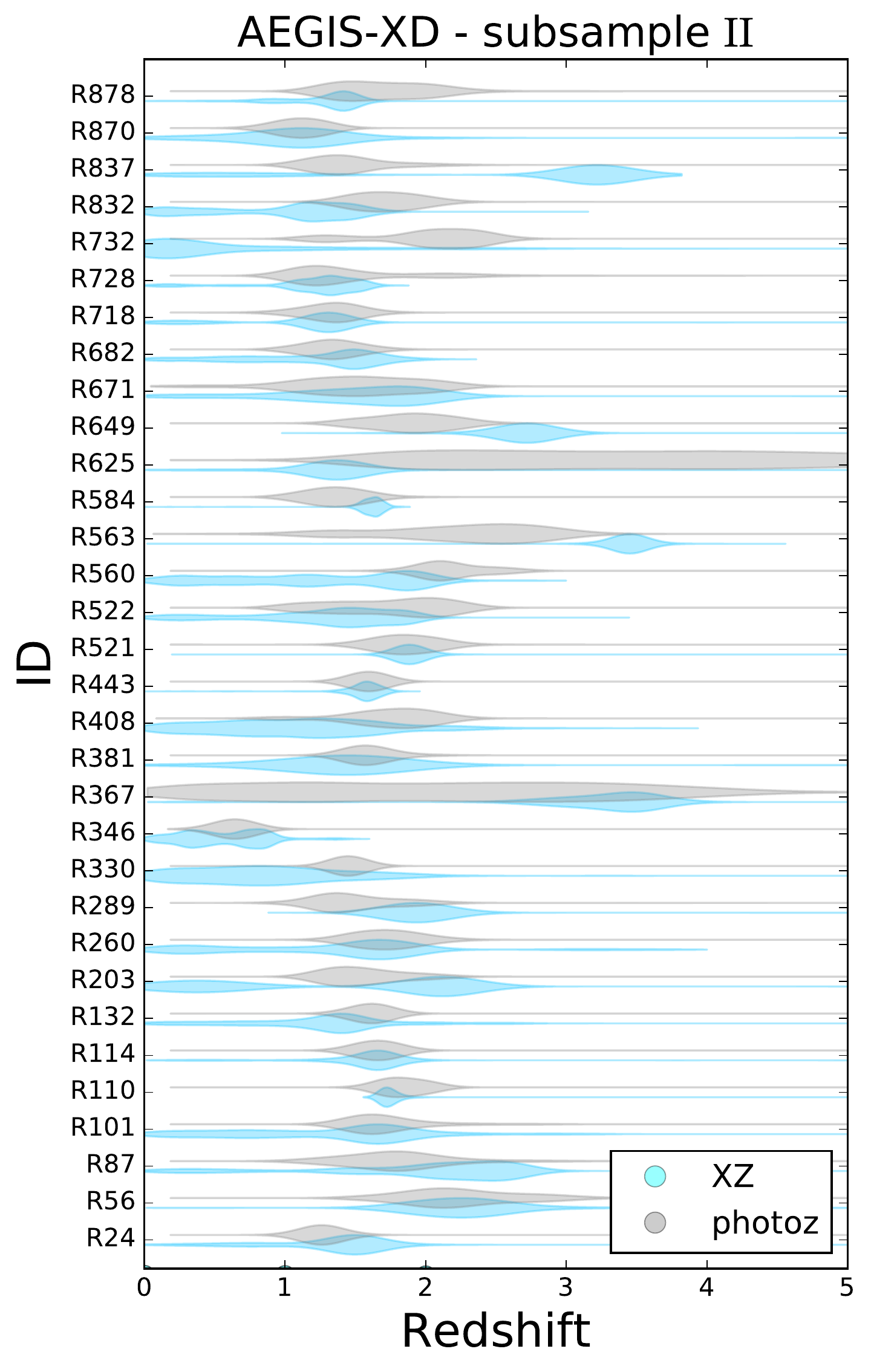}
\caption{XZ (cyan violin plots) and photoz (gray violin plots) probability distributions for subsample II (IG $\geq$ 2 $bits$).}
   \label{fig:aegisII}
\end{figure}

\begin{figure}
\centering
\includegraphics[width=0.5\textwidth,origin=l]{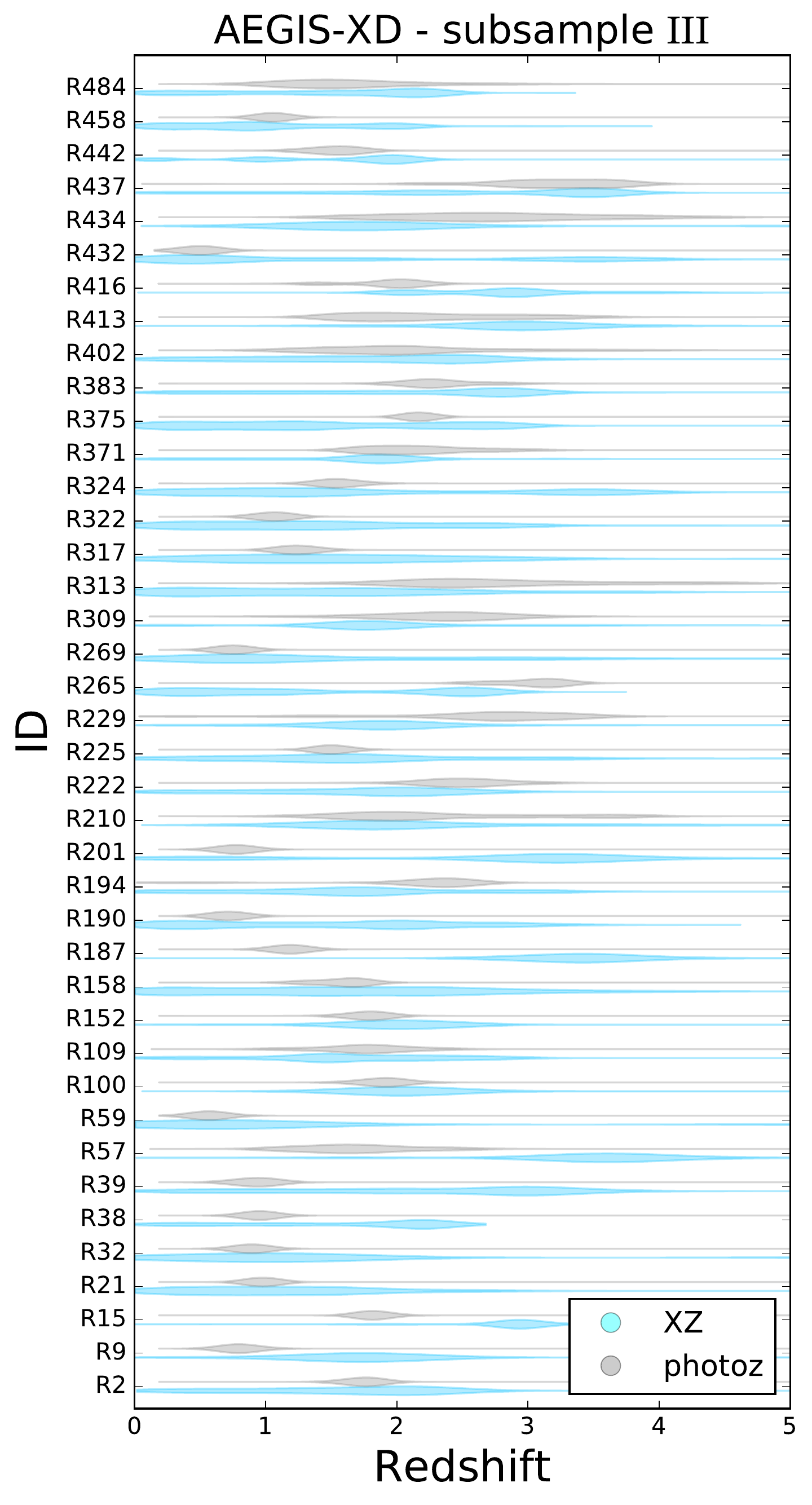}
\caption{XZ (cyan violin plots) and photoz (gray violin plots) probability distributions for subsample III (IG $\geq$ 2 $bits$).}
   \label{fig:aegisIIIa}
\end{figure}

\begin{figure}
\centering
\includegraphics[width=0.5\textwidth,origin=l]{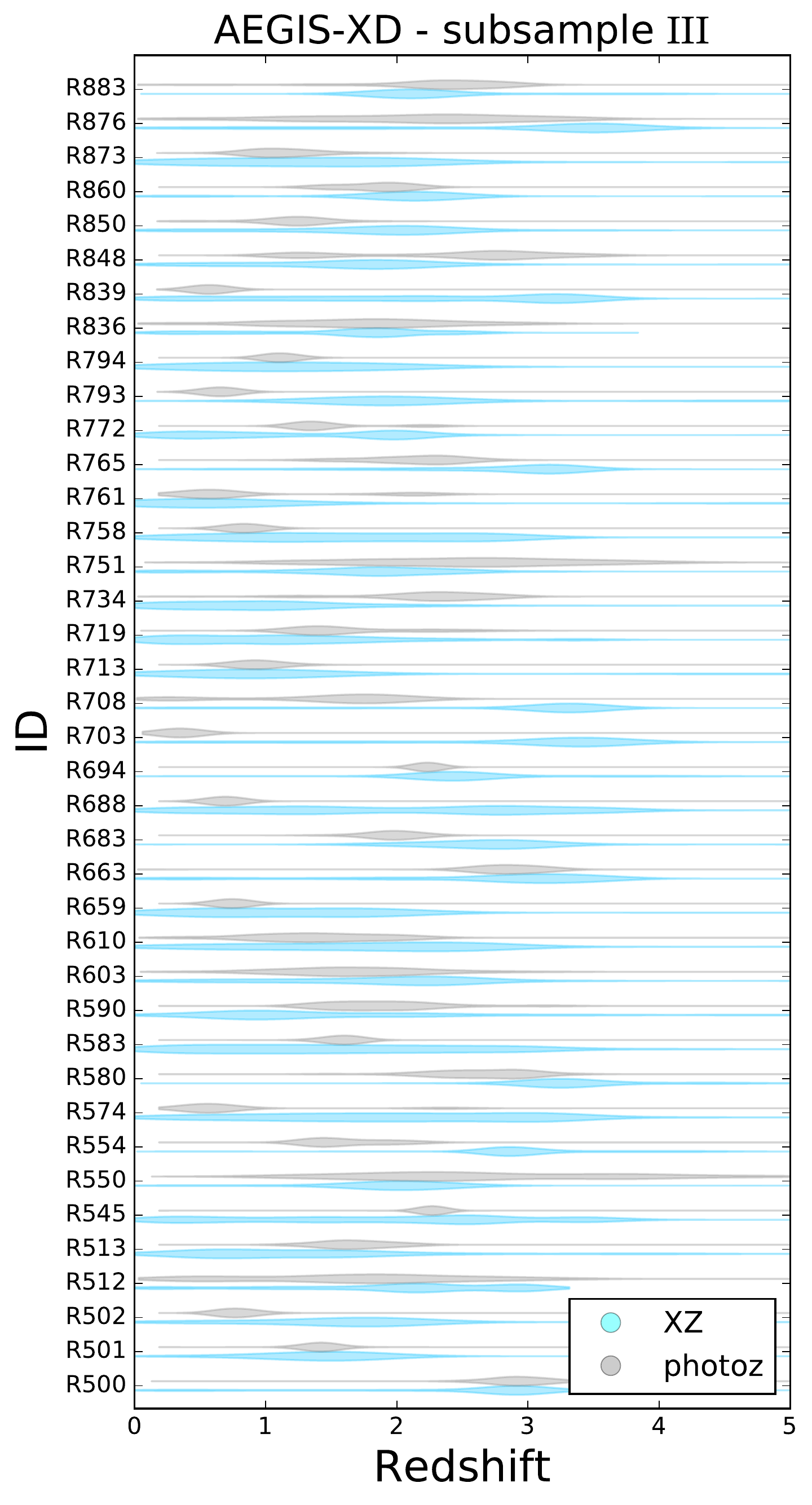}
\caption{Figure \ref{fig:aegisIIIa} cont.}
   \label{fig:aegisIIIb}
\end{figure}

Similar to the CDF-S results, we plot the XZ constraints in three subsamples divided by information gain. Subsample~I contains sources with specz and XZ-constrained redshifts (Figures~\ref{fig:cosmosI} and \ref{fig:aegisI}). Subsample~II contains sources without specz and compares XZ-constrained redshifts to photoz (Figures~\ref{fig:cosmosII} and \ref{fig:aegisII}).
Subsample~III contains sources without specz and compares XZ with some redshift information to photoz (Figures~\ref{fig:cosmosIII}, \ref{fig:aegisIIIa} and \ref{fig:aegisIIIb}).


\begin{table*}
{\small
\hfill{}
\tiny
   \begin{center}
   \caption[]{Properties of COSMOS Sources with IG $\geq$ 1 $bit$)}
       \begin{tabular}{ccccccccc}
           	\hline
           	\noalign{\smallskip}
           	ID & Counts & RA & Dec & XZ & photoz & log$_{10}$N$_{\rm H}$ & specz & IG [$bits$]\\
            \noalign{\smallskip}
           	\hline
           	\noalign{\smallskip}
 			342 & 1128 & 149.7008 & 2.40256 & 1.68 (+2.09,-1.31) & 0.39 (+0.75,-0.04) & 22.71 (+0.70,-0.72) & 0.375 & 1.43 \\ 
227 & 1016 & 150.24247 & 1.76643 & 2.42 (+0.86,-1.58) & 0.66 (+0.48,-0.04) & 22.67 (+0.30,-0.67) & 0.623 & 1.37 \\ 
437 & 254 & 149.99335 & 2.37723 & 0.46 (+0.58,-0.31) & 0.93 (+0.21,-0.05) & 22.94 (+0.40,-0.27) & 0.931 & 2.84 \\ 
393 & 331 & 150.47461 & 2.09414 & 1.01 (+0.66,-0.54) & 1.22 (+0.16,-0.12) & 22.89 (+0.34,-0.35) & -99.0 & 2.21 \\ 
726 & 248 & 150.2325 & 1.94266 & 0.19 (+0.02,-0.08) & 0.18 (+0.96,-0.05) & 23.46 (+0.08,-0.08) & 0.219 & 4.32 \\ 
386 & 425 & 150.10557 & 2.12631 & 3.17 (+0.25,-2.10) & 1.03 (+0.12,-0.04) & 23.42 (+0.24,-0.97) & -99.0 & 1.33 \\ 
664 & 211 & 150.01203 & 2.52562 & 1.01 (+0.40,-0.64) & 0.87 (+0.27,-0.05) & 23.15 (+0.24,-0.45) & 1.028 & 2.22 \\ 
376 & 201 & 149.88424 & 2.33814 & 0.41 (+0.70,-0.29) & 1.03 (+0.11,-0.08) & 22.89 (+0.49,-0.25) & 1.022 & 2.64 \\ 
336 & 279 & 149.834 & 2.30884 & 1.21 (+2.99,-0.02) & 1.1 (+0.09,-0.04) & 25.11 (+0.63,-2.16) & 1.162 & 2.72 \\ 
459 & 163 & 149.87645 & 2.56878 & 1.38 (+0.26,-0.95) & 1.65 (+0.13,-0.13) & 23.03 (+0.19,-0.54) & -99.0 & 1.94 \\ 

			\noalign{\smallskip}
           	\hline
       	\end{tabular}
     \tablefoot{{\it Col. 1:} identification number of \cite{Elvis2009}. {\it Col. 2:} total amount of detected source counts. {\it Cols. 3,4:} J2000 right ascension and declination in degrees. {\it Col. 5:} redshift (XZ). {\it Col. 6:} redshift (photoz) with one $\sigma$ errors from \cite{Salvato2009}. {\it Col. 7:} neutral hydrogen column in units of cm$^{-2}$ from XZ fit. {\it Col. 8:} reported specz values from \cite{Civano2012}. {\it Col. 9:} information gained in units of bits. All quoted errors are 1$\sigma$.
     }
\label{table:COSMOStable}
   \end{center}
   }
   \end{table*}

\begin{table*}
{\small
\hfill{}
\tiny
   \begin{center}
   \caption[]{Properties of AEGIS-XD Sources with IG $\geq$ 1 $bit$)}
       \begin{tabular}{ccccccccc}
           	\hline
           	\noalign{\smallskip}
           	ID & Counts & RA & Dec & XZ & photoz & log$_{10}$N$_{\rm H}$ & specz & IG [$bits$]\\
            \noalign{\smallskip}
           	\hline
           	\noalign{\smallskip}
 			2 & 292 & 214.5296 & 52.4231 & 1.68 (+0.62,-1.05) & 1.76 (+0.15,-0.19) & 22.96 (+0.28,-0.57) & -99.0 & 1.59 \\ 
9 & 280 & 214.6242 & 52.45268 & 1.75 (+0.51,-0.34) & 0.79 (+0.16,-0.10) & 23.59 (+1.24,-0.29) & -99.0 & 1.54 \\ 
15 & 386 & 214.538 & 52.46775 & 2.9 (+0.13,-0.91) & 1.82 (+0.16,-0.14) & 23.56 (+0.10,-0.33) & -99.0 & 1.85 \\ 
21 & 903 & 214.4244 & 52.47308 & 1.1 (+0.85,-0.79) & 0.98 (+0.10,-0.09) & 22.18 (+0.38,-0.48) & 1.14825 & 1.6 \\ 
24 & 196 & 214.4684 & 52.48146 & 1.47 (+0.10,-0.66) & 1.26 (+0.17,-0.15) & 23.63 (+0.20,-0.42) & -99.0 & 2.35 \\ 
32 & 201 & 214.4394 & 52.49756 & 1.3 (+5.56,-0.81) & 0.9 (+0.10,-0.12) & 23.59 (+0.77,-0.53) & 0.872816 & 1.19 \\ 
34 & 184 & 214.5468 & 52.50349 & 3.25 (+0.07,-0.10) & 2.9 (+0.19,-0.22) & 25.33 (+0.46,-0.41) & 2.91 & 2.89 \\ 
38 & 934 & 214.4415 & 52.50898 & 1.86 (+0.43,-1.35) & 0.96 (+0.10,-0.10) & 23.09 (+0.19,-0.70) & 0.985102 & 1.77 \\ 
39 & 159 & 214.4748 & 52.5095 & 2.38 (+0.83,-1.52) & 0.94 (+0.16,-0.20) & 23.54 (+0.41,-0.78) & -99.0 & 1.14 \\ 
56 & 94 & 214.5373 & 52.53091 & 2.32 (+0.20,-0.34) & 2.2 (+0.69,-0.46) & 24.75 (+0.71,-0.36) & -99.0 & 2.04 \\ 

			\noalign{\smallskip}
           	\hline
       	\end{tabular}
     \tablefoot{{\it Col. 1:} identification number of \cite{Nandra2015}. {\it Col. 2:} total detected source counts. {\it Cols. 3,4:} J2000 right ascension and declination in degrees. {\it Col. 5:} redshift (XZ). {\it Col. 6:} redshift (photoz) from \cite{Nandra2015}. {\it Col. 7:} neutral hydrogen column in units of cm$^{-2}$ from XZ fit. {\it Col. 8:} Reported specz values from \cite{Nandra2015}. {\it Col. 9:} information gained in units of bits. All quoted errors are 1$\sigma$.
     }
\label{table:AEGIS}
   \end{center}
   }
   \end{table*}

\newpage
\clearpage 

\end{appendix}

\end{document}